%% file: notes.tex
\documentclass[preprint, 10pt]{elsarticle}

\usepackage{graphics}
\usepackage{graphicx}
\usepackage{amsfonts}
\usepackage{amssymb}
\usepackage{amsmath}
\usepackage{color}
\usepackage{hyperref}
\usepackage{subcaption}
\usepackage{float}
\usepackage{natbib}

\input{def}

\begin{document}

\begin{frontmatter}

\title{Reduced Wiener Chaos representation of random
  fields via basis adaptation and projection}
\date{}

\author[label1]{Panagiotis Tsilifis}
\ead{tsilifis@usc.edu}
\author[label2]{Roger G. Ghanem \corref{cor1}} 
\ead{ghanem@usc.edu}

\cortext[cor1]{Corresponding author}
\address[label1]{Department of Mathematics, University of Southern
  California, Los Angeles, CA 90089, USA}
\address[label2]{Department of Civil Engineering, University of Southern
California, Los Angeles, CA 90089, USA}

%\begin{keyword}
%Bayesian experimental design \sep Expected information gain \sep
%Stochastic optimization \sep Polynomial Chaos \sep Two-phase transport
%\end{keyword}

%\maketitle

\input{abstract}

\end{frontmatter}

\input{intro}

\input{adaptation_metho}

\input{example_elliptic}

\input{example_geometric}

\input{conclusions}

\appendix

\input{appendix_A}

\input{appendix_B}

\section*{}
\bibliographystyle{plain}
\bibliography{references}

\end{document}

%% file: def.tex
\newcommand{\bx}{\mathbf{x}}
\newcommand{\by}{\mathbf{y}}
\newcommand{\bz}{\mathbf{z}}

\newcommand{\bh}{\mathbf{h}}

\newcommand{\bv}{\mathbf{v}}

\newcommand{\bw}{\mathbf{w}}
\newcommand{\ba}{\mathbf{a}}

\newcommand{\bn}{\mathbf{n}}
\newcommand{\be}{\mathbf{e}}

\newcommand{\bI}{\mathbf{I}}
\newcommand{\bA}{\mathbf{A}}
\newcommand{\bS}{\mathbf{S}}
\newcommand{\bD}{\mathbf{D}}

\newcommand{\bU}{\mathbf{U}}
\newcommand{\bC}{\mathbf{C}}

\newcommand{\bepsilon}{\boldsymbol{\epsilon}}

\newcommand{\bxi}{\boldsymbol{\xi}}
\newcommand{\bk}{\boldsymbol{\kappa}}

\newcommand{\balpha}{\boldsymbol{\alpha}}
\newcommand{\bbeta}{\boldsymbol{\beta}}
\newcommand{\bgamma}{\boldsymbol{\gamma}}
\newcommand{\bfeta}{\boldsymbol{\eta}}

\newcommand{\bzeta}{\boldsymbol{\zeta}}

\newcommand{\calA}{\mathcal{A}}
\newcommand{\calB}{\mathcal{B}}
\newcommand{\calG}{\mathcal{G}}

\newcommand{\calN}{\mathcal{N}}

\newcommand{\calF}{\mathcal{F}}

\newcommand{\calI}{\mathcal{I}}

\newcommand{\calJ}{\mathcal{J}}

\newcommand{\calX}{\mathcal{X}}
\newcommand{\calR}{\mathcal{R}}

\newcommand{\R}{\mathbb{R}}
\newcommand{\N}{\mathbb{N}}
\newcommand{\E}{\mathbb{E}}
\newcommand{\Prob}{\mathbb{P}}

\newcommand{\lang}{\Bigl\langle}
\newcommand{\rang}{\Bigr\rangle}

\newcommand{\wick}{\diamond}

%% file: abstract.tex
\begin{abstract}

A new characterization of random fields appearing in physical models
is presented that is based on their well-known Homogeneous Chaos expansions.
We take advantage of the adaptation capabilities of these expansions
where the core idea is to rotate the basis of the underlying Gaussian Hilbert space,
in order to achieve reduced functional representations that 
concentrate the induced probability measure in a lower dimensional
subspace. For a smooth family of rotations along
the domain of interest, the uncorrelated Gaussian inputs are
transformed into a Gaussian process, thus introducing a mesoscale that
captures intermediate characteristics of the quantity of interest. 
\end{abstract}

%% file: intro.tex
\section{Introduction}

Modeling, characterizing and propagating uncertainties in complex physical
systems have been extensively explored in recent years as they
straddle engineering and the physical, computational, and
mathematical sciences.  
The computational burden associated with a
probabilistic representation of these uncertainties is a persistent
related challenge.
One class of approaches to this challenge has been
to seek proper functional representations of the quantities of
interest (QoI) under investigation that will be consistent with the
observed reality as well as with the
mathematical formulation of the underlying physical system, which for
instance, is characterized within the context of partial differential
equations with stochastic parameters. Additionally, these
representations are equipped to serve as accurate propagators useful for prediction 
or statistical inference purposes. Among the
criteria that make such a functional representation a successful
candidate, are often the ability to provide a parametric interpretation
of the uncertainties involved in a
subscale level of the governing physics, as well as its quality as an
approximation of what is assumed to be the reality and its discrepancy
from it, in terms of several modes of convergence such
as distributional, almost sure or functional ($L^2$). 

The Homogeneous (Wiener) Chaos \cite{wiener} representation of random
processes has provided a convenient way to characterize solutions of
systems of equations that describe physical phenomena as was
demonstrated in \cite{ghanem_spanos} and further applied to a wide
range of engineering problems \cite{ghanem, matthies_bucher,
  ghanem_wrr, ghanem_dham, ghanem_redhorse}. Generalization of these
representations beyond the Gaussian white noise
\cite{xiu_karniadakis,soize_physical} provided the foundation
for a multi-purpose tool for uncertainty characterization and
propagation \cite{le_maitre_etal, najm, xiu_fluid}, 
statistical updating \cite{saad_ghanem,marzouk_etal, marzouk_najm} and design
\cite{huan, tsilifis} or as a generic mathematical model in order to
characterize uncertainties using maximum likelihood techniques
\cite{desceliers_etal, ghanem_doostan_redhorse}, Bayesian inference
\cite{ghanem_doostan, arnst_ghanem} or maximum entropy
\cite{das_etal}. Despite its wide applicability which has resulted in
significant gains, including but not limited to 
computational efficiencies, its use can still easily become prohibitive
with the increase of the dimensionality of the stochastic
input. Several attempts using sparse representations
\cite{doostan_owhadi, doostan_icc} have only
partially managed to sidestep the issue which still remains a major
drawback.  Recently, a new method for adapted Chaos
expansions in Homogeneous Chaos spaces has shown some promising
potential as a generic dimensionality reduction technique \cite{tipireddy}. The core
idea is based on rotating the independent Gaussian inputs through a
suitable isometry to form a new
basis such that the new expansion expressed in terms of that basis
concentrates its probability measure in a lower dimensional
subspace, consequently, the basis terms of the Homogeneous Chaos
spaces that lie outside that subspace can be filtered out via a
projection procedure. Several special cases along with intrusive and
non-intrusive computational algorithms were suggested which result in
significant model reduction while maintaining high fidelity in the probabilistic
characterization of the scalar QoIs.

It is the main objective of the present paper to extend further the basis
adaptation technique from simple scalar quantities of
interest to random fields or vector valued quantities that admit a
polynomial chaos expansion. Such random fields emerge, for instance, as
solutions of partial differential equations with random parameters and can
be found to have different degree of dependence on the stochastic
inputs at different spatio-temporal locations, therefore their adapted
representations and the corresponsing adapted basis should be expected
to exhibit such a spatio-temporal dependence. We
provide a general framework where a family of isometries are indexed
by the same topological space used for indexing the random field of
interest. Several
important properties are proved for the new adapted expansion, namely
the new stochastic input is no longer a vector of standard normal
variables but a Gaussian random field that admits a Karhunen-Loeve
\cite{karhunen, loeve}
expansion with respect to those variables. This new quantity essentially
merges uncertainties into a new basis that varies at different
locations, thus introducing a new way of upscaling uncertainties with
localized information about the quantity of interest. In addition, new
explicit formulas are derived that allow the transformation of an
existing chaos expansion to a new expansion with respect to any chosen
basis. One major benefit of this capability is that, once a chaos expansion
is available, any suitable adaptation can be achieved without further relying
on intrusive and non-intrusive methods that would require additional (repeated) evaluations
of the mathematical model, thus delivering us from further
computational costs.

This paper is organized as follows: First we introduce the basis
adaptation framework for Homogeneous Chaos expansions and the
reduction procedure via projection on subspaces of the Hilbert space of square
integrable random fields. Next we demonstrate how the framework
applies when stochasticity is also present in the coefficients of the
chaos expansion and finally we provide the theoretical foundations of
an infinite dimensional perspective of our approach which shows that
our derivations remain consistent and are nothing more but a special case of Hilbert spaces of
arbitrary dimension. Finally, our results are illustrated with two
numerical examples: That of an elliptic PDE with random diffusion
parameter, which explores various ways of obtaining reduced order
expansions that adapt well on the random field of interest and an
explicit chaos expansion where its first order coefficients consist
of a geometric series which allows the comparison of infinite dimensional
adaptations and their truncated versions. 

%% file: adaptation_metho.tex
\section{Basis adaptation in Homogeneous Chaos expansions of random fields}

\subsection{The Homogeneous (Wiener) Chaos}

We consider a probability space $\left(\Omega, \calF, \Prob\right)$ and $\calG$
a $d$-dimensional Gaussian Hilbert space, that is a closed vector
space spanned by a set of $d$ independent standard (zero-mean and
unit-variance) Gaussian random variables $\{\xi_i\}_{i=1}^d$, equipped with the inner
product $\langle \cdot, \cdot \rangle_{\calG}$ defined as $\langle \xi, \zeta \rangle_{\calG}
= \E[\xi \zeta]$ for $\xi, \zeta \in H$, where $\E[\cdot]$ denotes the
mathematical expectation with respect to the probability measure
$\Prob$. For simplicity, throughtout this section we will drop the
index $_\calG$ and simply write
$\langle \cdot , \cdot \rangle$ whenever there is no confusion. Let now
$\calF(\calG)$ be the $\sigma$-algebra generated by the
elements of $\calG$, then since all Gaussian variables have finite second
moments, it follows that $\calG$ is a closed subspace of $L^2(\Omega,
\calF(\calG), \Prob)$. We also define $\calG^{\wick n}$, for $n \in \N \cup \{-1,0\}$ to be the space of
all polynomials of exact order $n$, with the convention $\calG^{\wick -1} :=
\{0\}$. Then clearly $\calG^{\wick 0}$ is the space of constants and
$\calG^{\wick 1} = \calG$ and in fact from the Cameron-Martin theorem \cite{cameron,janson} we have that
$L^2(\Omega, \calF(\calG), \Prob) = \bigoplus_{n=0}^{\infty} \calG^{\wick n}$
which has an orhogonal basis that consists of the multidimensional Hermite polynomials defined as 
\begin{equation}
\bh_{\balpha}(\bxi) = \prod_{i = 1}^d h_{\alpha_i}(\xi_i),
\end{equation}
where $\balpha = (\alpha_1, ..., \alpha_d) \in \calJ := \N^d\cup\{\mathbf{0}\}$
and $h_{\alpha_i}(\xi_i)$ are the $1$-dimensional Hermite
polynomials of order $\alpha_i$. More precisely $\{\bh_{\balpha},
|\balpha| = n\}$ spans $\calG^{\wick n}$, where $|\balpha| = \sum_i \alpha_i$
and by introducing the orthonormal basis that consists of 
\begin{equation}
\psi_{\balpha}(\bxi) = \frac{\bh_{\balpha}(\bxi)}{\sqrt{\balpha !}},  \ \ \
\balpha \in \calJ,
\end{equation} 
any $u \in L^2(\Omega, \calF(\calG), \Prob)$ can be represented by its
Homogeneous Chaos expansion
\begin{equation}
u(\bxi) = \sum_{\alpha \in \calJ} u_{\balpha}\psi_{\balpha}(\bxi),
\end{equation}
where the convergence of the infinite summation is with respect to the
$L^2(\Omega, \calF(\calG), \Prob)$ norm and $\balpha ! = \prod_{i=1}^d \alpha_i !$.

Consider now a real-valued quantity of interest $u(\bx,\bxi)$ where $\bx \in D
\subset \R^k$, $D$ is typically bounded, and assume that $u \in L^2(\Omega \times D, 
\calF(\calG\times D), \mathbb{\Prob} \times \lambda)$, where
$\calF(\calG\times D)$ is the $\sigma$-algebra generated by the
rectangles $A\times B \in \calG\times D$, $\lambda$ is the
Lebesgue measure on $\R^k$ and $ \mathbb{\Prob} \times \lambda$ is
the product measure on $\Omega \times D$. Then is holds that
\begin{eqnarray}
\E\left[||u(\bx,\bxi)||^2_{L^2(D)}\right]  =  \int_{\Omega}
||u(\bx,\bxi)||^2_{L^2(D)} p(\bxi) d\bxi = \int_{\Omega} \int_{D}
|u(\bx, \bxi)|^2 p(\bxi) d\lambda(\bx) d\bxi < +\infty.
\end{eqnarray} 
Then, for each $\bx \in D$ we have $u(\bx, \bxi) \in L^2(\Omega, \calF(\calG),
\mathbb{\Prob})$ and as above it admits a
representation in terms of its orthogonal basis, that is the Hermite polynomials,
\begin{equation}
u(\bx, \bxi) = \sum_{\alpha\in \calJ} u_{\alpha}(\bx)
\psi_{\alpha}(\bxi),
\end{equation}
and the above square-integrability condition
($||u(\bx,\bxi)||_{L^2(D)} < +\infty$, for $\bxi$ a.s.) implies that $||u_{\alpha}(\bx) ||_{L^2(D)}
< +\infty$ for all $\alpha \in \calJ$, a condition that will be useful below.

\subsection{Change of basis for random fields}

In what follows we work with a truncated representation of $u$, that is
we assume that only a finite
number of terms of order up to $p \in \N$ are present 
\begin{equation}
u(\bx, \bxi) = \sum_{\balpha \in \calJ_p} u_{\alpha}(\bx) \psi_{\alpha}(\bxi),
\end{equation}
with $\calJ_p = \{\balpha\in \calJ :  |\balpha| \leq p\}$. The change
of basis framework \cite{tipireddy} presented below can easily be
generalized for the case of an infinite series. Namely, we consider an
isometry $\bA : \R^d \to
\R^d$ and we observe that $\bfeta := \bA \bxi$ is a basis in $\calG$ if
and only if $\bxi$ is. Since the Cameron-Martin theorem applies for
any basis in $\calG$, then $u$ can also be written as 
\begin{equation}
\label{eq:adapt_series}
u^{\bA}(\bx,\bxi) := u(\bx, \bfeta) = \sum_{\bbeta, |\bbeta| \leq p}
u_{\bbeta}^{\bA}(\bx) \psi_{\bbeta}(\bfeta),
\end{equation}
and by denoting $\psi^{\bA}_{\bbeta}(\bxi) := \psi_{\bbeta}(\bfeta) = \psi_{\bbeta}(\bA\bxi)$
and using the orthogonality between the polynomials we
can write the new coefficients as
\begin{equation}
\label{eq:adapt_coeffs}
u^{\bA}_{\bbeta}(\bx) = \sum_{\balpha} u_{\balpha}(\bx)
\lang\psi_{\balpha}, \psi^{\bA}_{\bbeta} \rang,\qquad
\forall \bx.
\end{equation}
This can be seen as a pointwise convergence in $L^2(D)$ but in fact
a stronger result is true: For the new expansion we still have that
$||u_{\bbeta}^{\bA}(\bx)||_{L^2(D)} < +\infty$ so the series actually
converges in $L^2(D)$.

For the above it is clear that once a Homogeneous Chaos series of
$u(\bx, \bxi)$ is available, then given any isometry $\bA$,
Eq. (\ref{eq:adapt_coeffs}) gives the coefficients of the series
expansion with respect to the new basis, as a function of the
initial coefficients and the entries of $\bA$. Although this expession in the
current form is computationally cumbersome, using properties of the Wick
product \cite{janson} we are able to derive analytic formulae with
respect to the entries of $\bA$ that, to the best of our knowledge
have not been presented before. Derivations of these formulae can be found in
\ref{sec:coeff_formulas}.

Note here that since the above expressions hold for any isometry $\bA$ and
all $\bx \in D$, one might consider choosing different $\bA$'s for
various choices of $\bx$. To illustrate this dependence of $\bA :=
\bA(\bx)$ on $\bx$ we take for instance the Gaussian and the
quadratic adaptation \cite{tipireddy}. For the Gaussian case, the first
row of $\bA$ is defined, for each $\bx$, through the mapping $\bxi \to \eta_1$
given as 
\begin{equation}
\label{eq:gauss_adapt}
\eta_1 (\bx) = \frac{1}{\left(\sum_{i=1}^du_{\epsilon_i}^2(\bx)\right)^{1/2}}\sum_{i=1}^d u_{\epsilon_i}(\bx) \xi_i,
\end{equation}
where $\epsilon_i = (0,...,1,...,0)$ is the multi-index with $1$ in the
$i$th location and zeros elsewhere. This represents the (normalized) centered Gaussian part of
$u(\bx)$. Similarly, for the quadratic case, the matrix $\bA$ is the
unitary matrix that satisfies, for each $\bx$,
\begin{equation}
\label{eq:quadr_iso}
\bS(\bx) = \bA^T \bD \bA
\end{equation}
where $\bS$ has entries  $\frac{u_{2\epsilon_i}}{\sqrt{2}}$ along the
diagonal and $\frac{u_{\epsilon_{ij}}}{\sqrt{2}}$ elsewhere.

As these cases indicate, the isometry $\bA$ can depend on $\bx$ and as a consequence,
$\bfeta$ will also depend on $\bx$ which implies
that for each $\bx$, $\bxi$ is transformed to a different basis
$\bfeta(\bx)$. By construction, each component $\eta_i(\bx)$ of the
adapted bases is a Gaussian process with covariance kernel
\begin{eqnarray}
\label{eq:kernel}
k_i(\bx,\by) = \E[\eta_i(\bx)\eta_i(\by)] = \sum_{j,k=1}^d
a_{ij}(\bx)a_{ik}(\by) \E[\xi_j \xi_k] = \ba_i(\bx) \ba_i(\by)^T
\end{eqnarray} 
where for convenience we denote by $\ba_i(\bx) = (a_{i1}(\bx), ...,
a_{id}(\bx))$ the $i$th row of $\bA(\bx)$.  In fact, for the case where the dependence is such that
the entries $a_{ij}(\bx)$ are square integrable, the following result holds:

\textbf{Theorem 1.} \emph{Provided that the entries of
  $\ba_i(\bx)$ are square-integrable, the
function $k_i(\cdot, \cdot) : D\times D \to \R$ defined in
eq.~(\ref{eq:kernel}) is a Hilbert-Schmidt kernel.}

\textbf{Proof. } Detailed proof in \ref{sec:theorem_1}. $\square$

\textbf{Remark 1.} For an example, in the case of linear adaptation, the
square-integrability of $u_{\epsilon_i}(\bx)$ as mentioned in the
previous subsection suffices to show that
$||a_{1j}||_{L^2(D)} < +\infty$, therefore $k_1(\bx, \by)$ is Hilbert-Schmidt.

\textbf{Remark 2.} In fact, we will see below that $k_i(\bx, \by)$ has
at most $d$ positive eigenvalues and the decomposition
(\ref{eq:kernel}) is the one that follows by Mercer's theorem \cite{mercer}.

\subsection{Reduced adapted decompositions via projection}

Next, it is of interest to consider a projection of the above
expansion on a subspace of $V_{\calI} \subset L^2(\Omega \times D)$
with $V_{\calI}$ being the space spanned by $\{\psi_{\bbeta}: \bbeta
\in \calI\}$ for some $\calI \subset \calJ_p$, resulting in
\begin{eqnarray}
u^{\bA,\calI}(\bx,\bxi) := u^{\calI}(\bx, \bfeta) &=& \sum_{\bbeta \in \calI}
u_{\bbeta}^{\bA}(\bx) \psi_{\bbeta}(\bfeta)\nonumber\\ &=& \sum_{\bbeta \in \calI}
\sum_{\balpha\in \calI_p} u_{\balpha}(\bx) \lang\psi_{\balpha}, \psi^{\bA}_{\bbeta} \rang \psi_{\bbeta}(\bfeta).
\end{eqnarray}

Such projections introduce an error that can be described as the
difference $u - u^{\bA,\calI}$. Trivially in the case where $\calI = \calJ_p$,
this difference is zero. Futhermore one can write
$u(\bx, \bfeta)$ as a series of
$\{\psi_{\balpha}(\bxi)\}_{\balpha\in \calJ_p}$
\begin{equation}
u(\bx, \bfeta) = \sum_{\bgamma\in \calJ_p} u_{\bgamma}(\bx) \psi_{\bgamma}(\bxi),
\end{equation}
which gives 
\begin{equation}
u_{\bgamma}(\bx) = \sum_{\bbeta\in \calJ_p} \sum_{\balpha \in \calJ_p}
u_{\balpha}(\bx) \lang\psi_{\balpha}, \psi^{\bA}_{\bbeta}
\rang \lang \psi^{\bA}_{\bbeta}, \psi_{\bgamma}\rang
\end{equation}
and in the case of a projection on some $\calI$, the sum over $\bbeta$
is simply taken in $\calI$ instead of $\calJ_p$. We denote by
$\bw(\bx)$ and $\bw^{\bA,\calI}(\bx)$ the vectors with entries the coefficients
$\{u_{\balpha}(\bx)\}_{\balpha\in\calJ_p}$ and
$\{u_{\bgamma}(\bx)\}_{\calJ_p}$ respectively and with $\vert \calJ
\vert$ the cardinality of a set $\calJ$. By introducing the
$\vert \calJ_p \vert \times \vert \calJ_p \vert$
Grammian matrix $\bC$ with entries $\bC_{\balpha, \bbeta} = \lang
\psi_{\balpha} , \psi_{\bbeta}^{\bA} \rang$ for $\bbeta
\in \calI$ and $0$ otherwise, we can write the error associated with a
projection $\calI$ as 
\begin{equation}
\label{eq:projection_error}
\bw(\bx) - \bw^{\bA,\calI}(\bx) = \left(\bI - \bC \bC^T \right) \bw(\bx),
\end{equation}
which depends solely on $\calI$ and $\bA$. Note that as mentioned previously, as $\calI$
approaches $\calJ_p$ the error becomes zero independently of
$\bA$. However, for $\calI$ being a strict subset of $\calJ_p$ the
error can vary as a function of the entries of $\bA$. A closer look,
using Proposition $2$ from \ref{sec:coeff_formulas}, indicates that $\bC$
is a block diagonal matrix and so is $\bC\bC^T$. Furthermore for the
case of $n$-dimensional adaptations ($n < d$), each block matrix of the diagonal
has only $n$ non-zero columns. 

Several options are available for exploration of the error of a particular adaptation
procedure. For instance, for each $\bx \in D$ and a fixed projection space $\calI$ one might
wish to minimize, with respect to $\bA$, an appropriately chosen norm of 
$\bw - \bw^{\bA,\calI}$ in order to
locally adapt the chaos expansion of $u(\bx)$ at the point of interest
$\bx$. Alternatively for a global adaptation one can also minimize an
$L^2(D)$ norm of $\bw - \bw^{\calI}$, that is 
\begin{equation}
\label{eq:projection_error_norm}
||\bw(\bx) - \bw^{\calI}(\bx)||_{L^2(D)} = \left(\int_D ||\bw(\bx) - \bw^{\calI}(\bx)||^2d\bx\right)^{1/2}.
\end{equation}
Further investigation of the interrelation between the error and the
choice of $\bA$ falls beyond the scope of the present paper and can be the
subject of future work.

\subsection{Basis adaptation of Chaos expansions with random
  coefficients}

In this subsection we consider the case where the coefficients of the
chaos expansion are themselves taken to be random variables. We adopt
the formulation presented in \cite{soize_reduced} where the random
coefficients can be thought of as the result of a reduced
decomposition. More specifically, let two orthonormal bases $\bxi \in \calG_1$ and $\bzeta \in
\calG_2$ with $\calG_1$, $\calG_2$ being $d_1$- and $d_2$-dimensional Gaussian
Hilbert spaces respectively, that are statistically independent and let
$\calG = \overline{\calG_1 \times \calG_2}$ the closure of the product space
$\calG_1\times \calG_2$. Then it is known \cite{soize_physical} that any $u(\bx, \bxi,\bzeta) \in L^2(\Omega\times D,
\calF(\calG), \Prob)$ admits an expansion 
\begin{equation}
u(\bx, \bxi, \bzeta) = \sum_{\balpha\in\calJ^{d_1}} \sum_{\bbeta \in
  \calJ^{d_2} } u_{\balpha, \bbeta}(\bx) \psi_{\balpha}(\bxi) \psi_{\bbeta}(\bzeta),
\end{equation}
where $\calJ^{d_i} := \N^{d_1} \cup \{\mathbf{0}\}$, $i = 1,2$. The
above expansion can be rearranged in the form,
\begin{equation}
\label{eq:reduced_chaos}
u(\bx, \bxi, \bzeta) = \sum_{\balpha\in \calJ^{d_1}} \bU_{\balpha}(\bx,
\bzeta) \psi_{\balpha}(\bxi),
\end{equation}
where 
\begin{equation}
\bU_{\balpha}(\bx, \bzeta) = \sum_{\bbeta \in \calJ^{d_2}} u_{\balpha, \bbeta}(\bx) \psi_{\beta}(\bzeta).
\end{equation}
Thus, $u(\bx, \bxi, \bzeta)$ can be written as a polynomial chaos
expansion with respect to $\bxi$ with random coefficients that depend
on $\bzeta$ and are independent of the basis functions
$\{\psi_{\balpha}(\bxi)\}_{\balpha\in \calJ^{d_1}}$. In order to
proceed, we consider again the truncated
series 
\begin{equation}
\label{eq:reduced_chaos_trunc}
u^{\calJ_p}(\bx, \bxi, \bzeta) = \sum_{\balpha\in \calJ_p^{d_1}} \bU_{\balpha}(\bx,
\bzeta) \psi_{\balpha}(\bxi),
\end{equation}
and
\begin{equation}
\label{eq:random_coeffs}
\bU_{\balpha}(\bx, \bzeta) = \sum_{\bbeta \in \calJ_p^{d_2}} u_{\balpha,
  \bbeta}(\bx) \psi_{\beta}(\bzeta),
\end{equation}
where with no loss of generality we take the order of truncation $p$
to be common in both series. Then the extension of the adaptation and
projection procedures presented in the previous subsection is
straightforward. It is clear that for any given isometry $\bA$, the
coefficients $\bU_{\balpha}^{\bA}$ given in
Eq. (\ref{eq:adapt_coeffs}) will also be random since the inner
product used to project $u(\bx, \bxi, \bzeta)$ on the basis functions
$\psi_{\balpha}(\bxi)$ is the merely expectation with respect to
$\bxi$.  Namely,
\begin{equation}
\bU_{\balpha}^{\bA} = \E\left[ u^{\calJ_p}(\bx, \bxi, \bzeta)
  \psi_{\balpha}(\bxi)\right] = \E\left[ u^{\calJ_p}(\bx, \bxi, \bzeta)
  \psi_{\balpha}(\bxi) | \bzeta\right].
\end{equation}
It is also worth
noting that in the case of the standard adaptation schemes (Gaussian,
quadratic), the isometry is itself a random matrix that depends on the
coefficients of the reduced expansion (\ref{eq:reduced_chaos_trunc}) and
more specifically its probability distribution depends on
$\bzeta$. Denote by $\Phi_{\bfeta}(\bf t)$ the characteristic function
of the new basis $\bfeta = \bA(\bzeta)\bxi$.  Then following some
standard manipulations, taking into
account the independence between $\bzeta$ and $\bxi$ and the almost
sure constraint that $\bA(\bzeta)
\bA^T(\bzeta) = \bI_{d_1}$, where $\bI_{d_1}$ is the
unit matrix in $\mathbb{R}^{d_1\times d_1}$, one can evaluate
\begin{equation}
\Phi_{\bfeta}(\bf t) = \E\left[e^{i{\bf t}^T\bfeta}\right] =
e^{-\frac{1}{2}{\bf t}^T{\bf t}}, \quad \bf t\in \R^{d_1}
\end{equation}
thus concluding that the marginal distribution of $\bfeta$ is indeed
$\calN(\mathbf{0},  \bI_{d_1})$ and that the standard Hermite polynomial
chaos expansions remain valid.

\subsection{Extension to infinite-dimensional spaces}
 
In the previous subsections we have developed our basis adaptation
methodology by initially taking the Gaussian Hilbert space $\calG$ to be a finite
dimensional space. In this section we demonstrate that this can be
viewed as a special case of a space $\calG$ that is of arbitraty
dimension (countable or uncountable infinite dimensional). In order to do this, first it is essential to provide 
some further insight on the construction of such spaces. Next we will show
that, for a family of isometries $\{\calA(\bx)\}_{\bx \in D}$, under
suitable topological conditions, the elements of the tranformed basis can be viewed as Gaussian
fields that admit a Karhunen-Loeve type expansion in terms of the initial
basis. 

We start with a necessary definition: 

\textbf{Definition 1.} \emph{For any $H$ real Hilbert space, we say
  that the Gaussian Hilbert space $\calG$ is indexed by $H$ if there is a
linear isometry $\chi \mapsto \xi_\chi$, from $H$ to $\calG$.} 

This definition provides a natural way to construct $\calG$, given
some $H$. Namely,
if $\{\be_i\}_{i \in I}$ is a basis for $H$ and $\{\xi_i\}_{i\in I}$
is a set of uncorrelated standard normal variables with common index set
$I$, then the mapping $\chi := \sum \chi_i \be_i \mapsto \sum \chi_i\xi_i$ is an
isometry and $\calG := span\{\xi_i\}_{i\in I}$ is a Gaussian Hilbert
space indexed by $H$. Of course, in order for the above to make sense
we need the sum $\sum \chi_i\xi_i$ to be defined. In the case where $H$
is finite dimensional, then $\sum \chi_i\xi_i \in\calN(0,
||\chi||^2_H)$ and more generally, if $\bxi = (\xi_1, ..., \xi_d)$ is
$H$-valued, then the map $\chi \mapsto \langle\bxi, \chi\rangle_H \sim
\calN(0, ||\chi||_H^2)$ defines an isometry. For instance, let $H := \R^d$ and $\{\xi_i\}_{i=1}^d$ to
be scalar standard normal variables. This is actually the case upon which our methodology
has been built. 

The main difficulty when $H$ is infinite dimensional is to ensure the existence of some
$\bxi = \{\xi_i\}_{i\in I}$ such that $\langle \bxi, \chi\rangle_H \sim
\calN(0, ||\chi||^2_H)$ for all $\chi \in H$.
Gaussian measures on infinite dimensional spaces are defined in terms
of real measures on their dual space \cite{gelfand,redhorse_2009}. In practice this means that often there
is no $H$-valued Gaussian variable $\bxi$. In the countable case, the construction of
$\calG$ then can be obtained with the following procedure (see
\cite{janson, gelfand} for technical details): A locally convex topological vector space $\calX$ must
be identified such that $H \subset \calX$ for which there is a continuous
inclusion mapping $T : H \mapsto \calX$ which is a Hilbert-Schmidt
operator. Then, we have that $\calX^* \subset H$ and subsequently we obtain the
Gelfand triple $\calX^* \subset H \subset \calX$. It is possible to
choose $\bxi\in \calX$ such that $\langle \bxi, \chi \rangle_{\calX}
\sim \calN(\mathbf{0}, ||\chi||^2_H)$ for any $\chi \in \calX^*$ and we
define the Gaussian Hilbert space as $\calG_0 = \{\langle \bxi, \chi \rangle_{\calX} | \chi \in
\calX\}$. Then the mapping $\bxi \to \langle \bxi, \chi \rangle_{\calX}$
from $\calX^*$ to $\calG_0$ is an isometry and by continuity it can be
extended from $H$ to the closure $\calG = \overline{\calG}_0$. Then
$\calG$ is indexed by $H$. For an
example, let $H := \ell^2(\N)$ the set of real square summable sequences
$\{a_n\}_{n\in \N}$ and take $\bxi = (\xi_1, \xi_2, ...)$ with $\xi_n$
i.i.d. $\calN(0,1)$. Then clearly $\bxi \notin
l^2(\N)$ and we take $\calX := \R^{\infty}$ and $\calX^* = \sum_{i=1}^{\infty}
\R$ and $\sum_{n=1}^{\infty}a_n \xi_n \sim \calN(0, \sum a_n^2)$ where
the sum converges a.s. The basis adaptation procedure here would
consist of selecting an orthonormal basis $\{\be_n\}_{n\in\N}$ on
$\ell^2(\N)$ and then defining the isometry $\calA: \bxi \mapsto \bfeta $ with $\eta_n = \sum e_n^i
\xi_i$. If $\{\xi_n\}_{n\in \N}$ is an
orthonormal basis on $\calG$, then so is $\{\eta_n\}_{n\in\N}$ and any
Wiener chaos expansion of elements in $L^2(\Omega, \calF(\calG),
\Prob)$ can be taken with respect to the new basis. 
A construction of a space $\calG$ for the
uncountably infinite case can be found in \cite{ito}.

At last, motivated by the adaptation schemes presented above, we explore the case where an
isometry is chosen to depend on parameters $\bx$, in a more abstract
setting. For simplicity, we consider the countably infinite
dimensional case, however, all the theorems that 
we recall and prove below are also valid in the uncountably infinite case and we only need
to interpret the infinite sums as limits of nets in $L^2$. 
Let $\calG$ be a Gaussian Hilbert space indexed by a real
Hilbert space $H$ and $D$ be any topological space. Let also $\calB =
\{\{\be_n\}_{n\in\N},\ orthonormal\ basis\ in\ H\}$ the space of all orthonormal bases in $H$ and assume there is a
map $\calA: D \mapsto \calB$ that is continuous and onto. Then for
any basis $\{\be_n\}_{n\in\N}$ there is $\bx \in D$ such that
$\calA(\bx) = \{\be_n\}_{n\in\N}$ and we write $\{\be_n(\bx)\}_{n\in
  \N}$, that is we assume that $D$ is a continuous
parameterization of the space of rotations in $H$, therefore any
basis can be indexed by some $\bx \in D$. Moreover, in order to
maintain the Hilbert-Schmidt structure of the kernels defined below we
will assume that
the entries of $||e^i_n(\bx)||_{L^2(D)} \leq \infty$ for all the entries
of $\be_n(\bx)$. Let $\calX^* \subset H \subset
\calX$ be a Gelfand triple as described above, $\bxi \in \calX$ an
orthonormal basis in $\calG$ and let
\begin{equation}
\label{eq:gp_transformed}
\eta_n(\bx) = \lang \bxi, \be_n(\bx) \rang_H = \sum_{i=1}^{\infty}
e_n^i(\bx) \xi_i, \ \ \ n \in \N.
\end{equation}
In what follows, for the sake of simplicity we drop $n$ and we refer to an abritrary component
$\eta(\bx)$ unless there is a need for further clarification. We have
the following lemma:

\textbf{Lemma 1.} \emph{For $\eta(\bx)$ as above we have that $\calG =
span\{\eta(\bx)\}_{\bx\in D}$.} 

\textbf{Proof.} Detailed proof in \ref{sec:lemma_1}.  $\square$

Now for given $\eta(\bx)$ and for each $\xi \in H$ define the mapping $R : D \mapsto \R$ with 
\begin{equation}
R_{\eta}(\xi)(\bx) = \lang \xi, \eta(\bx) \rang_{\calG} = \E[\xi \eta(\bx)]
\end{equation}
and the Cameron-Martin space, corresponding to $\eta$
\begin{equation}
\label{eq:CM_space}
\calR_{\eta}(\calG) = \{R_{\eta}(\xi): \xi \in H\}
\end{equation} 
which is the space of such mappings. Then we have the following:

\textbf{Theorem 2.} \emph{Let $\eta(\bx)$ defined as in
  Eq. (\ref{eq:gp_transformed}). Then $\{\be^i(\bx)\}_{i\in\N}$ spans the
Cameron-Martin space corresponding to $\eta(\bx)$.}

\textbf{Proof.} Detailed proof in \ref{sec:theorem_2}. $\square$

The above theorem essentially implies that for any $n\in \N$, the 
$\eta_n(\bx)$ obtained after a change of basis transformation throught
the isometries $\calA(\bx)$, $\bx \in D$, are Gaussian processes and
the expression (\ref{eq:gp_transformed}) is their Karhunen-Loeve type
expansion with a number of terms equal to the dimension of
$\calG$. Consequently, in finite dimensional spaces, the expansion
consists of finite terms and their corresponding covariance kernels of
the form (\ref{eq:kernel}) have
at most finitely many positive eigenvalues.

%% file: example_elliptic.tex
\section{Numerical examples}

\subsection{Elliptic PDE}

We consider the following elliptic PDE
\begin{eqnarray}
\label{eq:elliptic}
\begin{array}{c} -\nabla\left(\bk(\bx, \bxi)\cdot\nabla u(\bx, \bxi)\right) = g(\bx), \ \
\ \bx \in D \\ 
\left( \bk(\bx,\bxi)\nabla u(\bx, \bxi)\right) \cdot \bn = 0, \ \ \ \ \ \ \ \ \ \ \ \bx \in \partial D \end{array}
\end{eqnarray}
that can be thought of as the pressure equation in a single flow
problem with no-flux boundary conditions.
The transmissivity tensor $\bk(\bx, \bxi)$ is modeled
as a random process, $g(\bx)$ is a term that describes sinks and
sources and $\bn$ is the unit vector, perpendicular to the
boundary. In addition, the condition 
\begin{equation}
\int_{\partial D}u(\bx) d\bx = 0
\end{equation}
is imposed to ensure well-posedness of the boundary-value problem. In
this 2-dimensional setting we take $D = [0,
400]^2$ which is discretized in a $40\times 40$ rectangular grid and
we place a source and a sink at $\bx_{so} = (0,0)$ and $\bx_{si} = (400,400)$ respectively by taking 
\begin{equation}
g(\bx) = s\exp\left[-\frac{1}{2}\sum_{i=1}^2\frac{(x_i -
    x^i_{so})^2}{l_i^2}\right] - s\exp\left[-\frac{1}{2}\sum_{i=1}^2\frac{(x_i -
    x^i_{si})^2}{\l_i^2}\right]
\end{equation} 
with $s = 0.5$, $l_1 = l_2 = 20$. In what follows, equation
(\ref{eq:elliptic}) is solved using a two-point
flux-approximation finite-volume scheme \cite{aarnes}.

As the prior model of the transmissivity, we take
$\bk= (\bk_{\bx}, \bk_{\by},
\bk_{\bz})$ to be isotropic ($\bk_{\bx} = \bk_{\by} = \bk_{\bz} := \bk_0$)
where the components are a log-normally distributed
process, that is $\bk_0(\bx, \bxi) = \exp\left(G(\bx,\bxi)\right)$ where
$G(\bx,\bxi)$ is a Gaussian field. We parameterize $G(\bx, \bxi)$ by
considering its Karhunen-Loeve (KL) expansion 
\begin{equation}
G(\bx,\bxi) = G_0(\bx) + \sum_{i=1}^{\infty}\sqrt{\lambda_i}\xi_i g_i(\bx)
\end{equation}
where $\{\lambda_i\}_{i\geq 0}$ and $\{g_i(\bx)\}_{i\geq 0}$ are the
eigenvalues and eigenvectors respectively of its covariance kernel,
which is taken to be a squared exponential kernel 
\begin{equation}
k(\bx, \by) = \sigma^2\exp\left[-\frac{1}{2}\sum_{i = 1}^2\frac{(x_i - y_i)^2}{\ell_i^2}\right].
\end{equation}
For the sake of simplicity we take $G_0(\bx) =
0$, whereas the kernel parameters are $\sigma^2 = 0.5$, $\ell_1 = \ell_2 = 80$.
Then we truncate the KL expansion such
that it retains a $97\%$ of the energy. That reduces to a finite
expansion with $20$ terms therefore we have $\bxi \in \R^d$ with
$d = 20$.

Next, a $3$rd-order polynomial chaos
expansion
\begin{equation}
u(\bx,\bxi) = \sum_{\balpha \in \calJ_3} u_{\balpha}(\bx) \psi_{\balpha}(\bxi)
\end{equation} 
of the solution of Eq. (\ref{eq:elliptic}) was contructed. Due to the
relatively high dimensionality of the input, an ensemble of
$N = 10^5$ Monte Carlo samples of the $20$-dimensional Gaussian
input was used in order to estimate the coefficients 
\begin{equation}
u_{\balpha}(\bx) = \lang u(\bx,\bxi)
\psi_{\balpha}(\bxi) \rang \approx \frac{1}{N}\sum_{n=1}^N u(\bx,
\bxi^{(n)}) \psi_{\balpha}(\bxi^{(n)}).
\end{equation}

%\begin{figure}[t]
%\centering
%\includegraphics[width = 0.3\textwidth]{../images/kernel.ps}
%\caption{The structure of the discretized covariance kernel of
 % $\eta_1$. \label{fig:kernel}}
%\end{figure}

\begin{figure}[h]
\centering
\includegraphics[width =
0.19\textwidth]{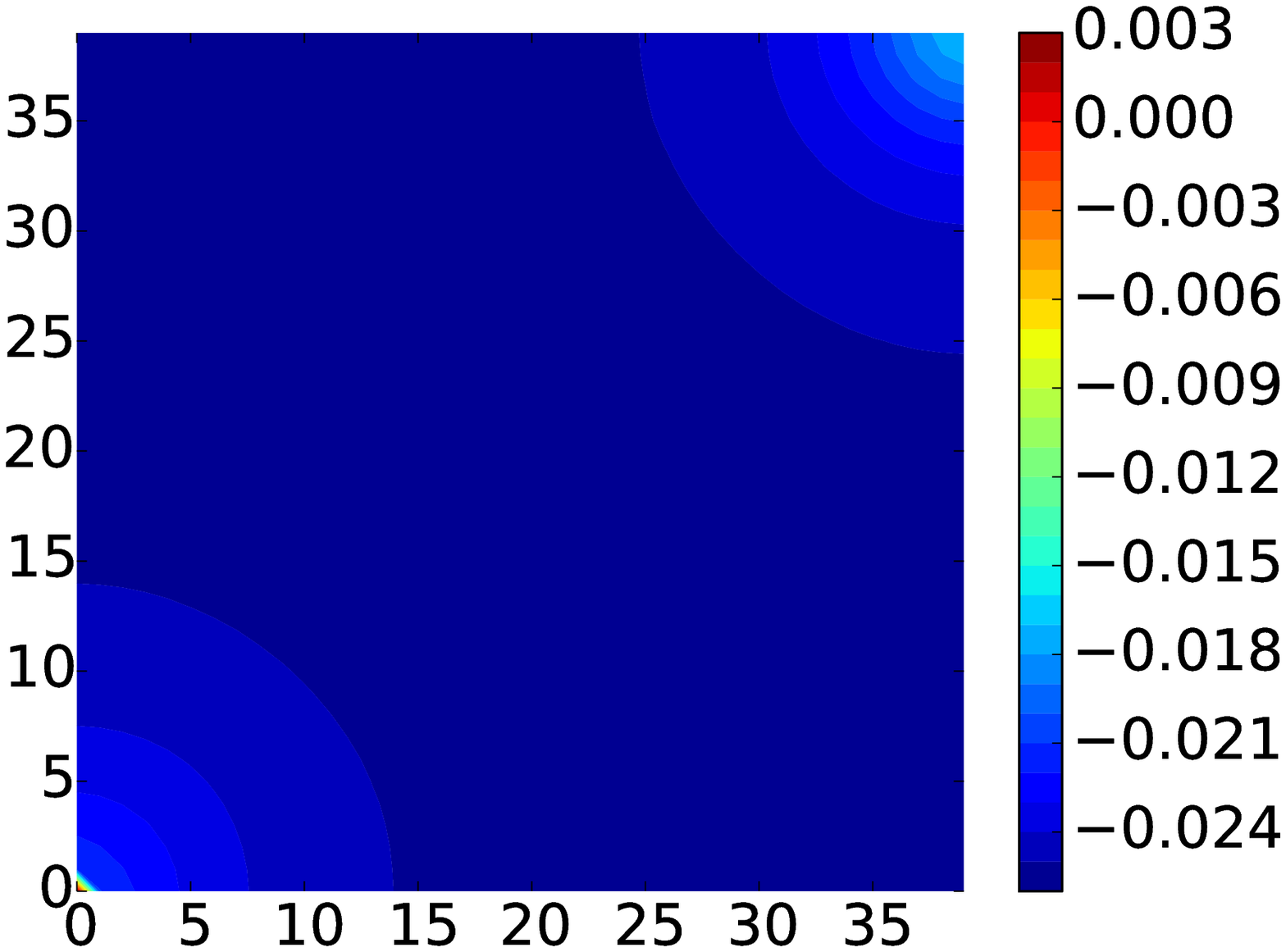}
\includegraphics[width =
0.19\textwidth]{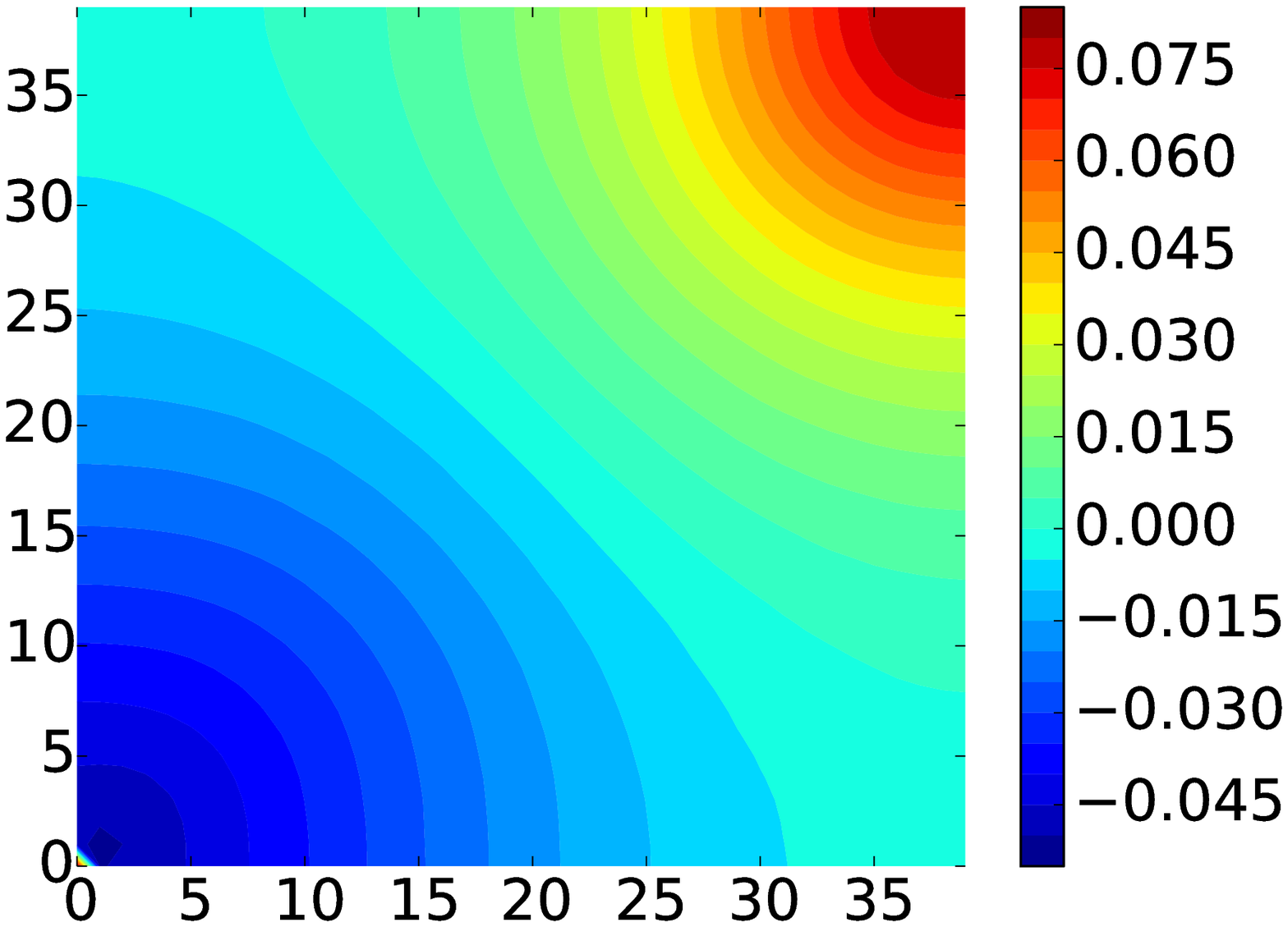}
\includegraphics[width =
0.19\textwidth]{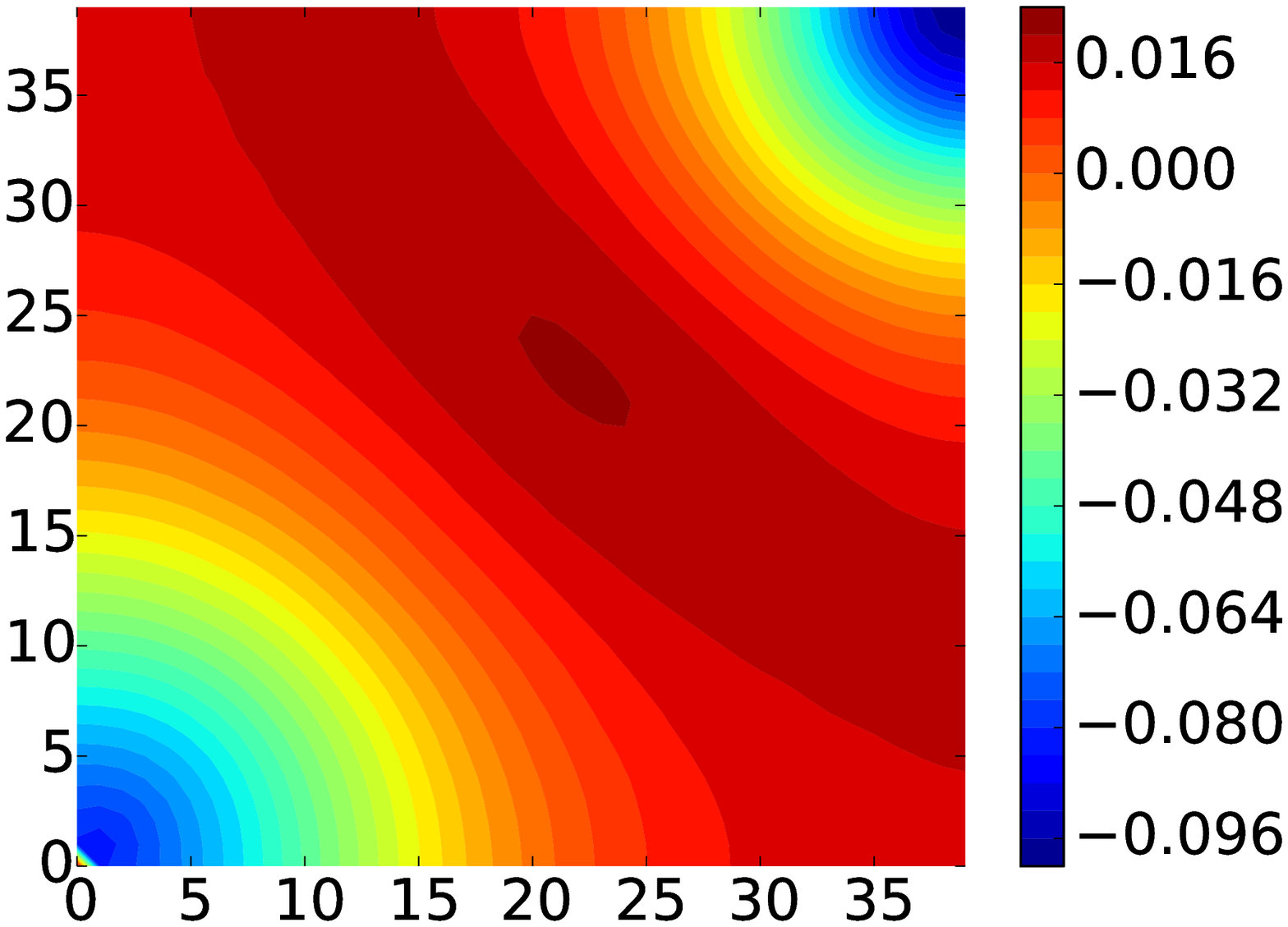}
\includegraphics[width =
0.19\textwidth]{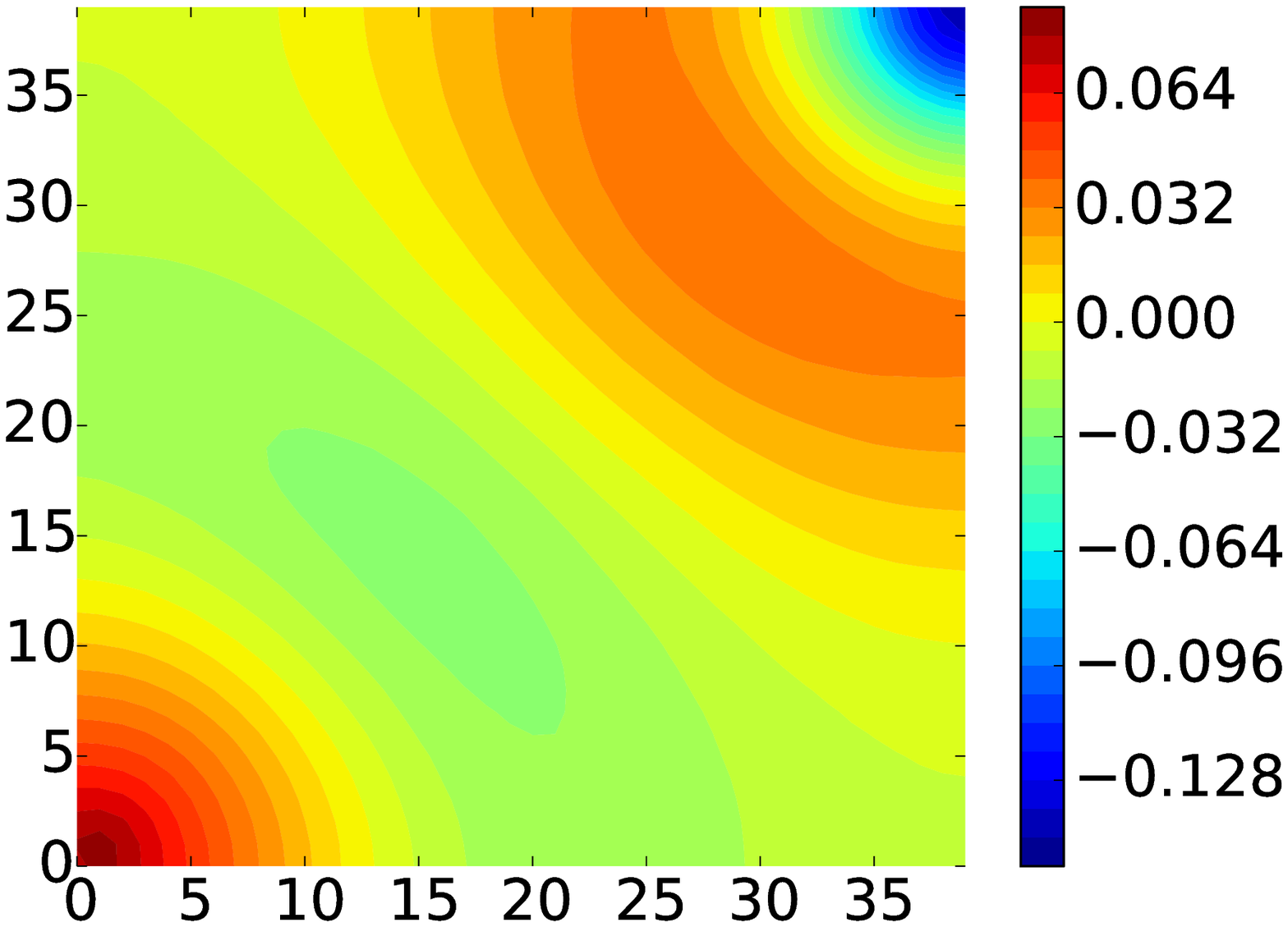}
\includegraphics[width =
0.19\textwidth]{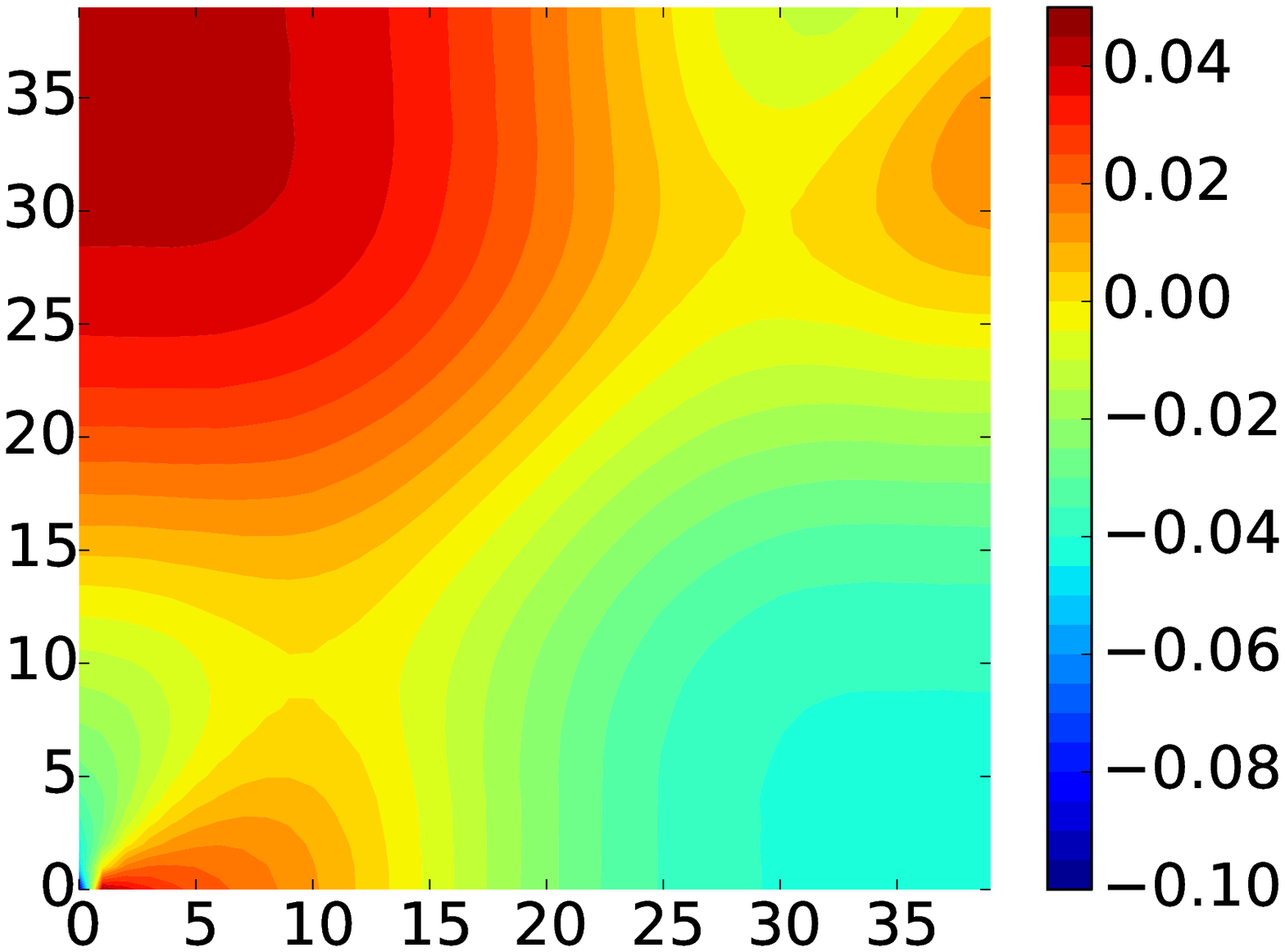}
\includegraphics[width =
0.19\textwidth]{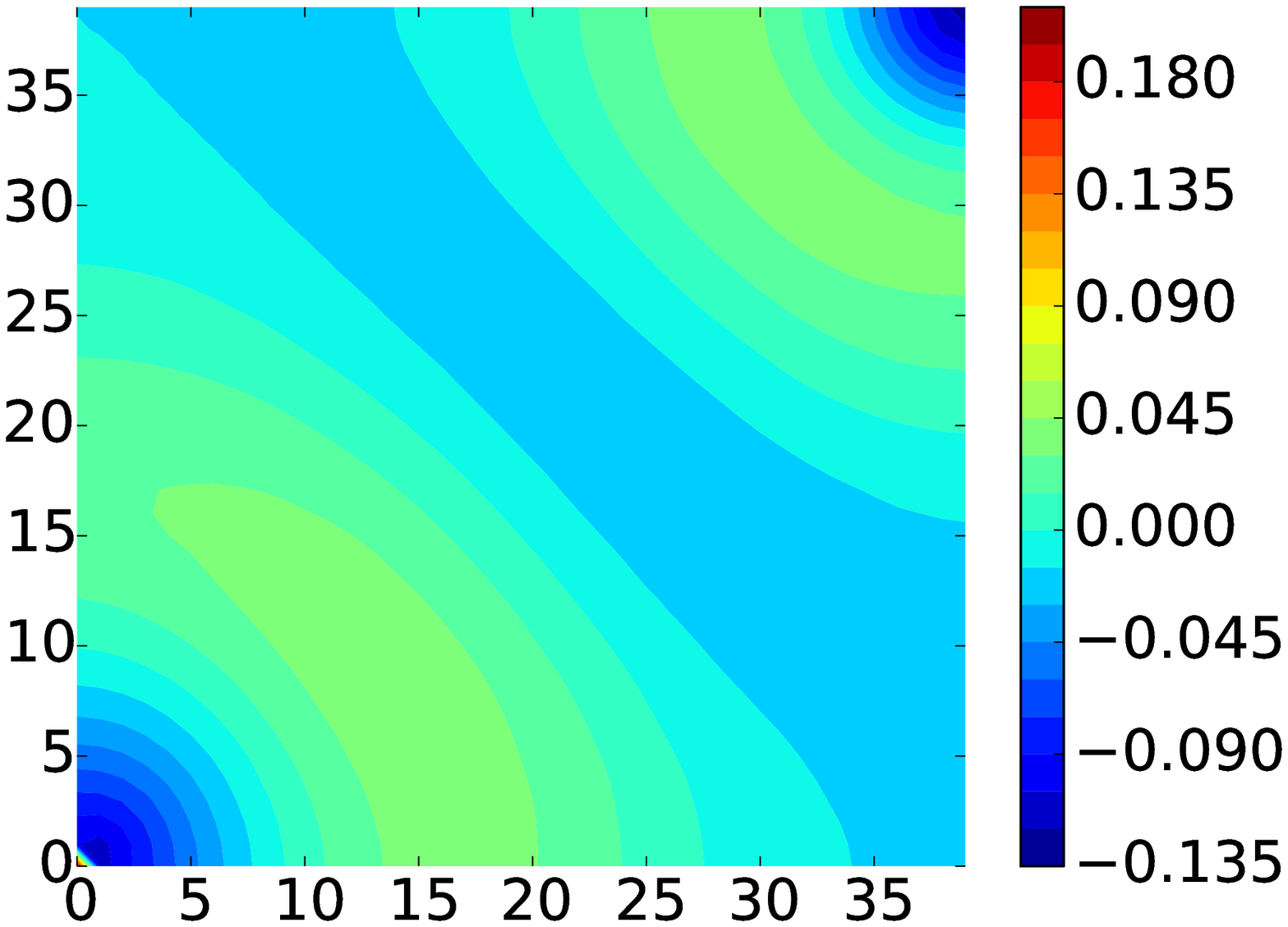}
\includegraphics[width =
0.19\textwidth]{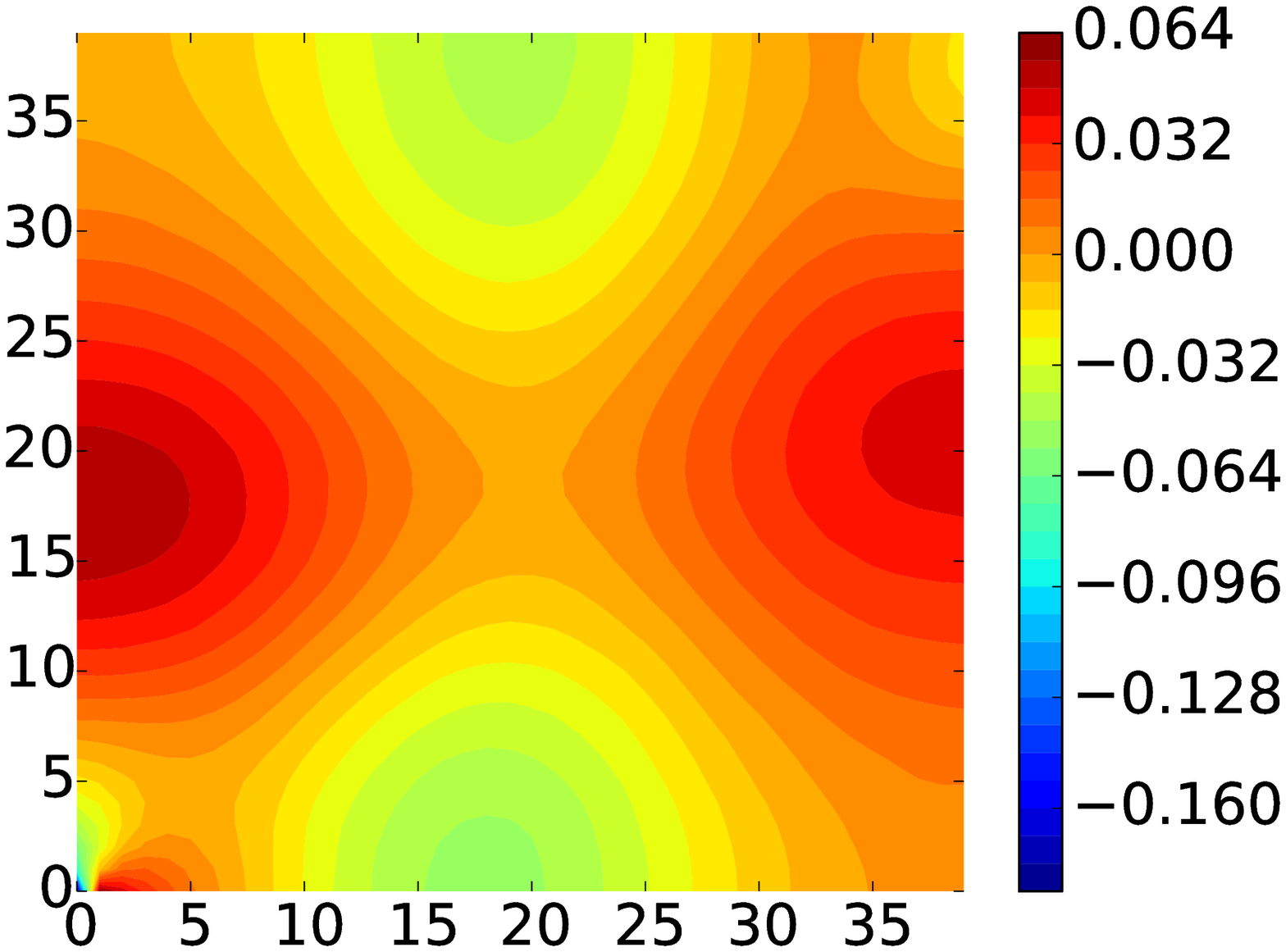}
\includegraphics[width =
0.19\textwidth]{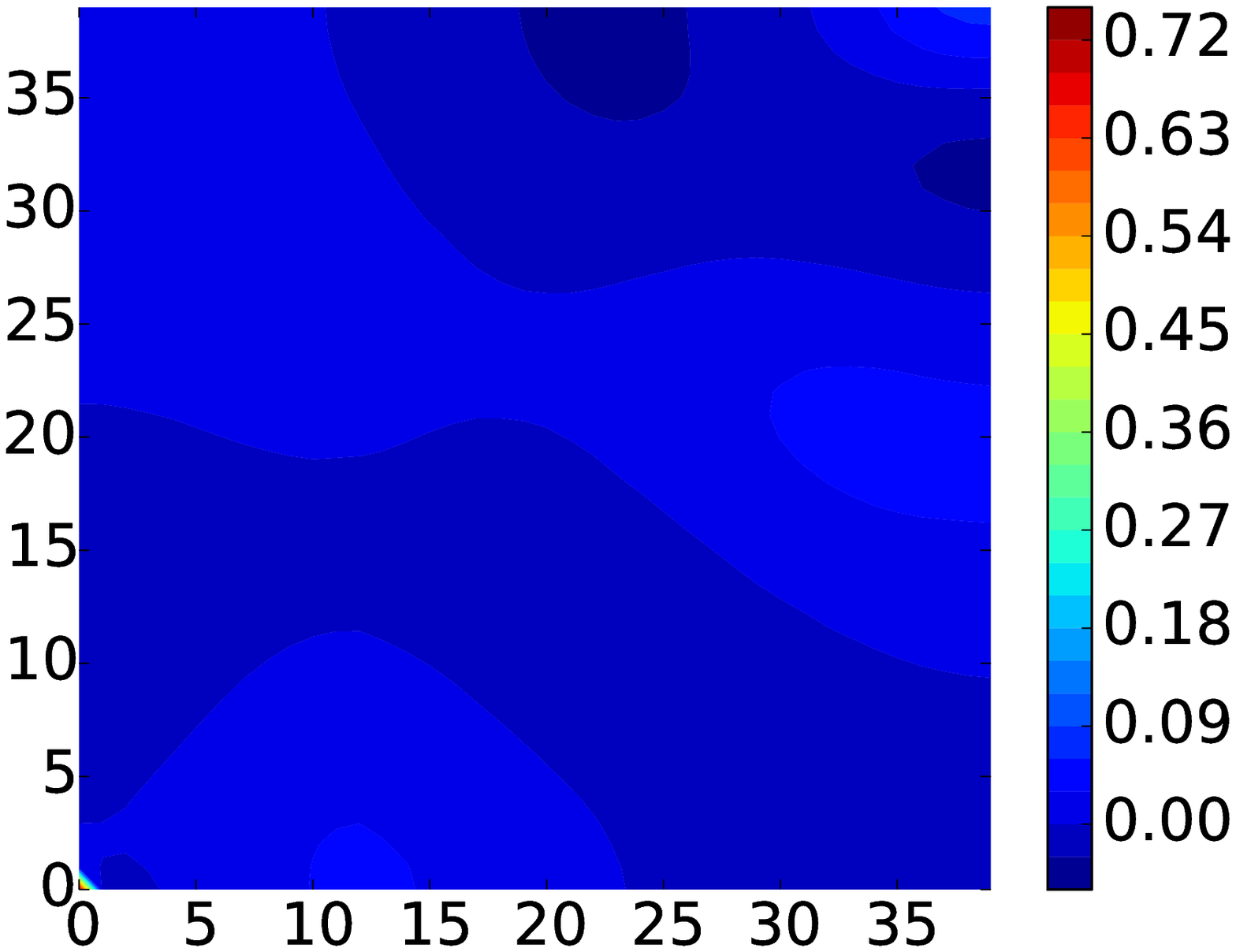}
\includegraphics[width =
0.19\textwidth]{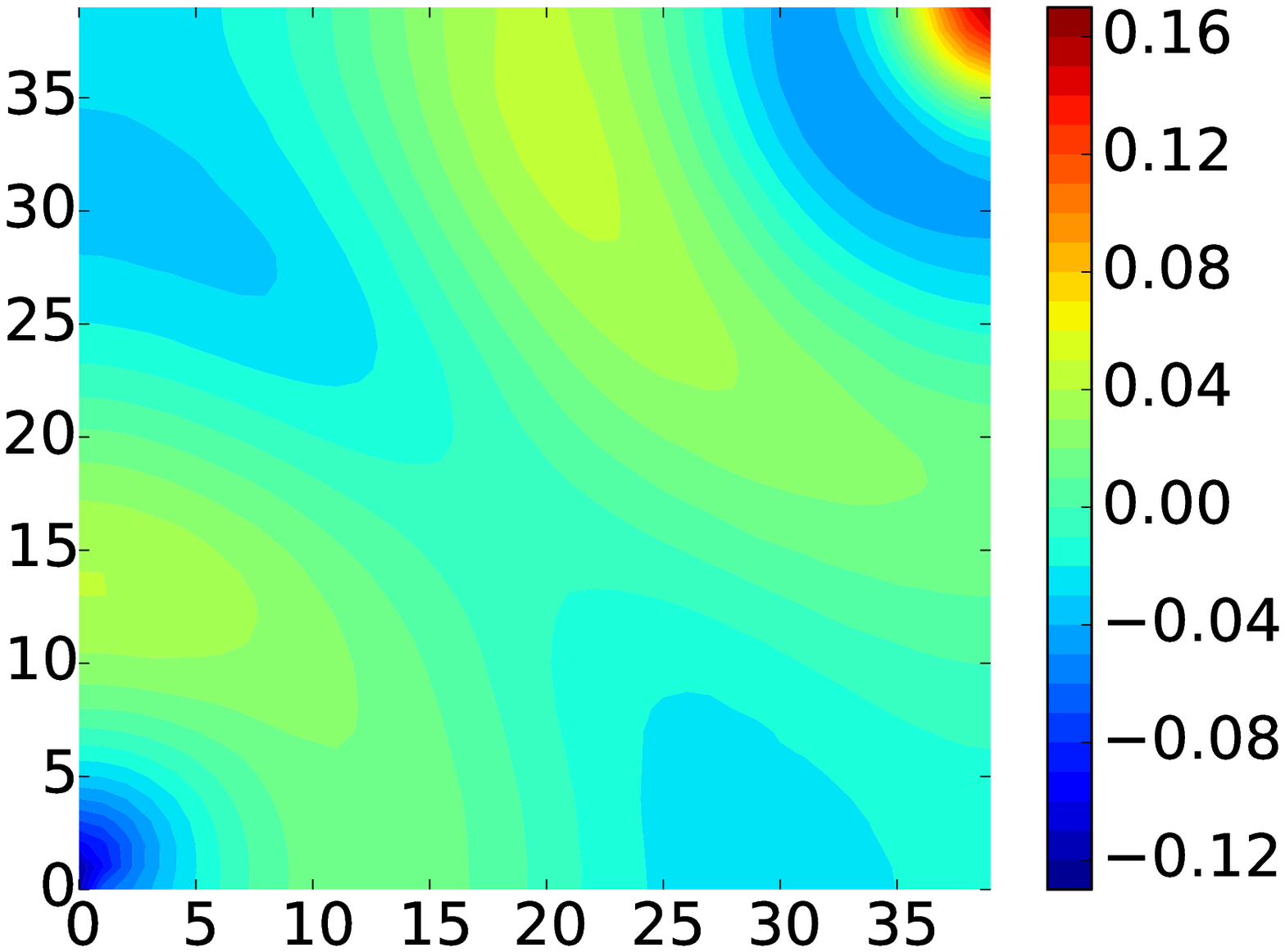}
\includegraphics[width =
0.19\textwidth]{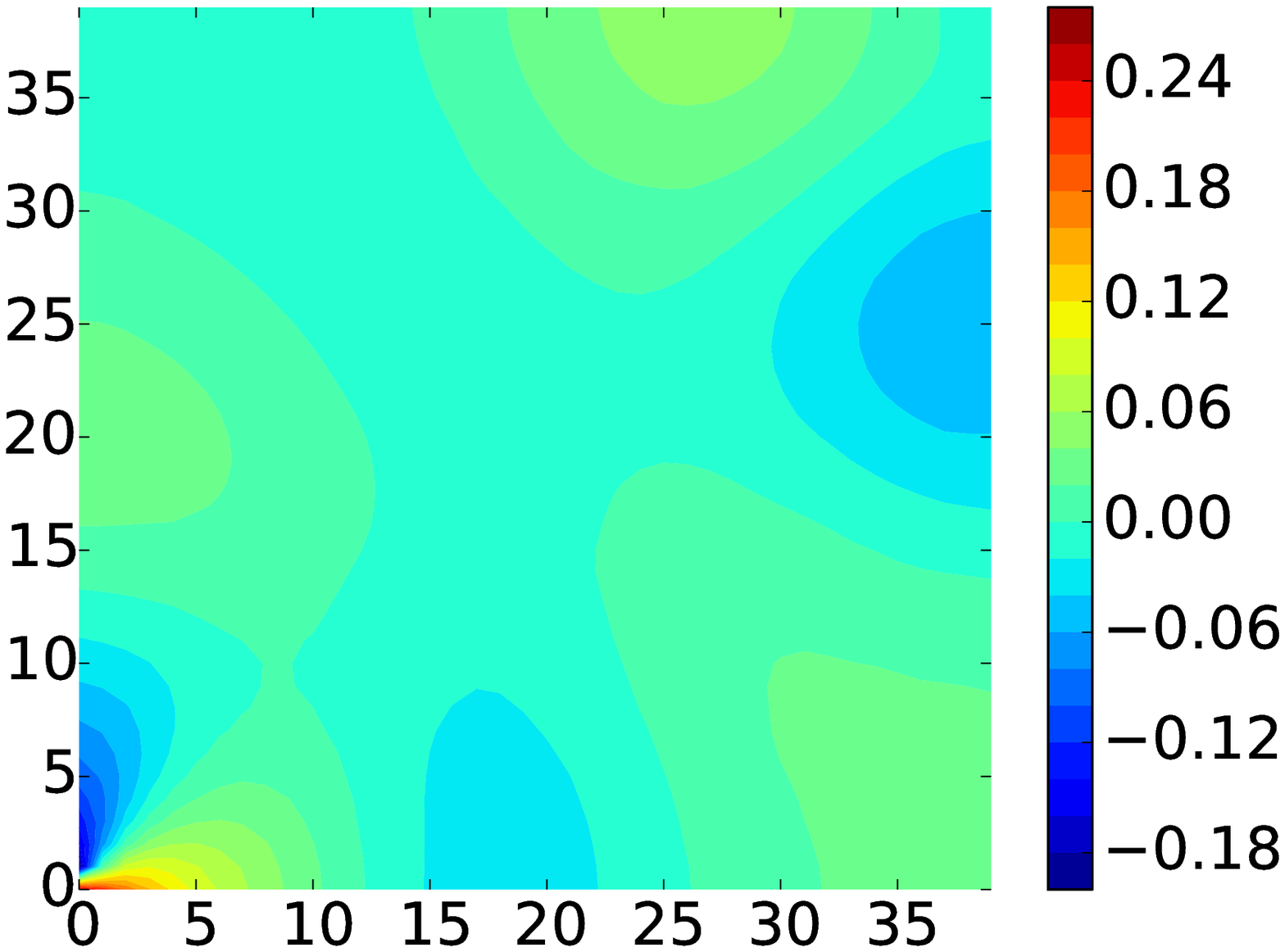}
\includegraphics[width =
0.19\textwidth]{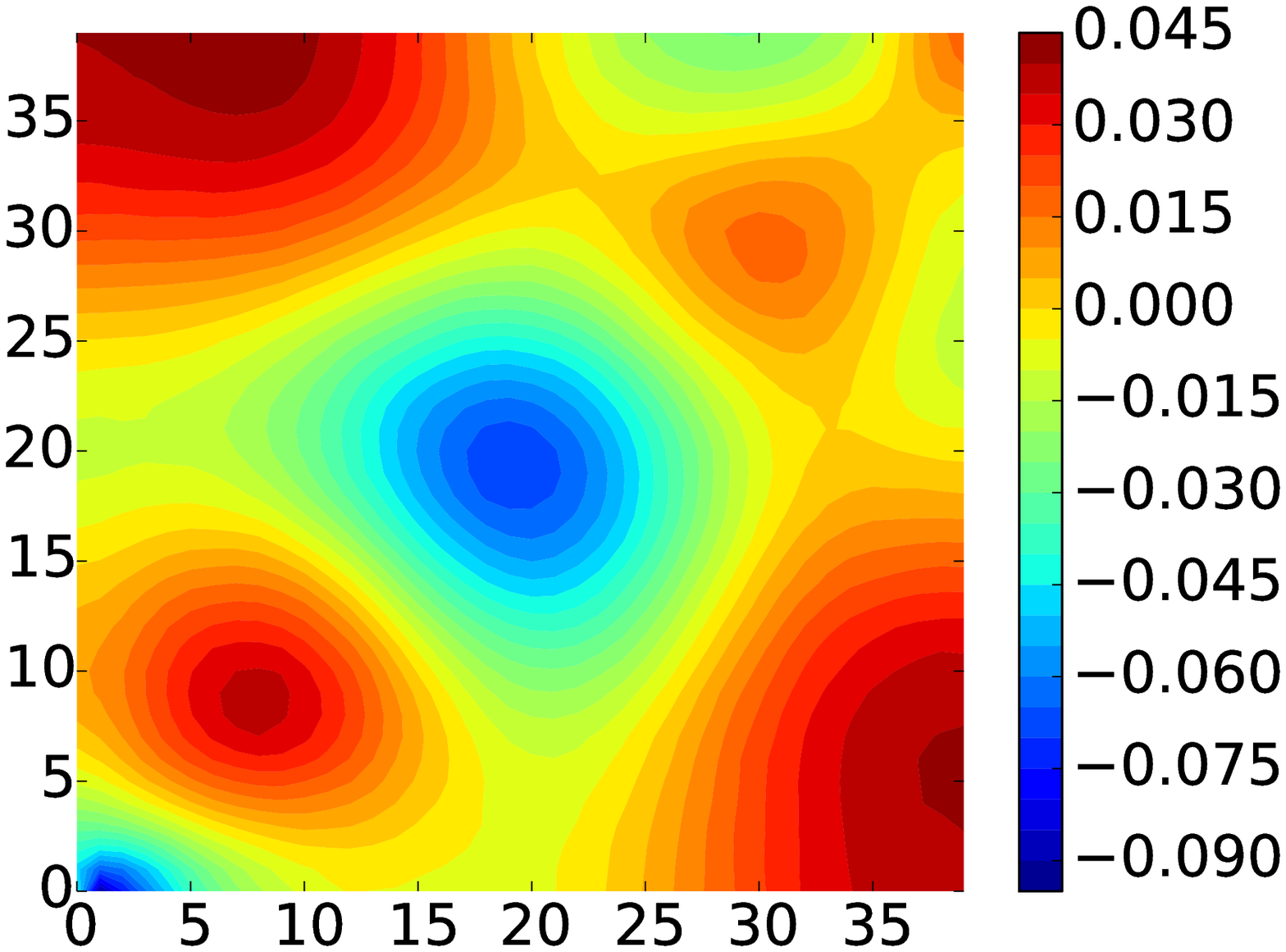}
\includegraphics[width =
0.19\textwidth]{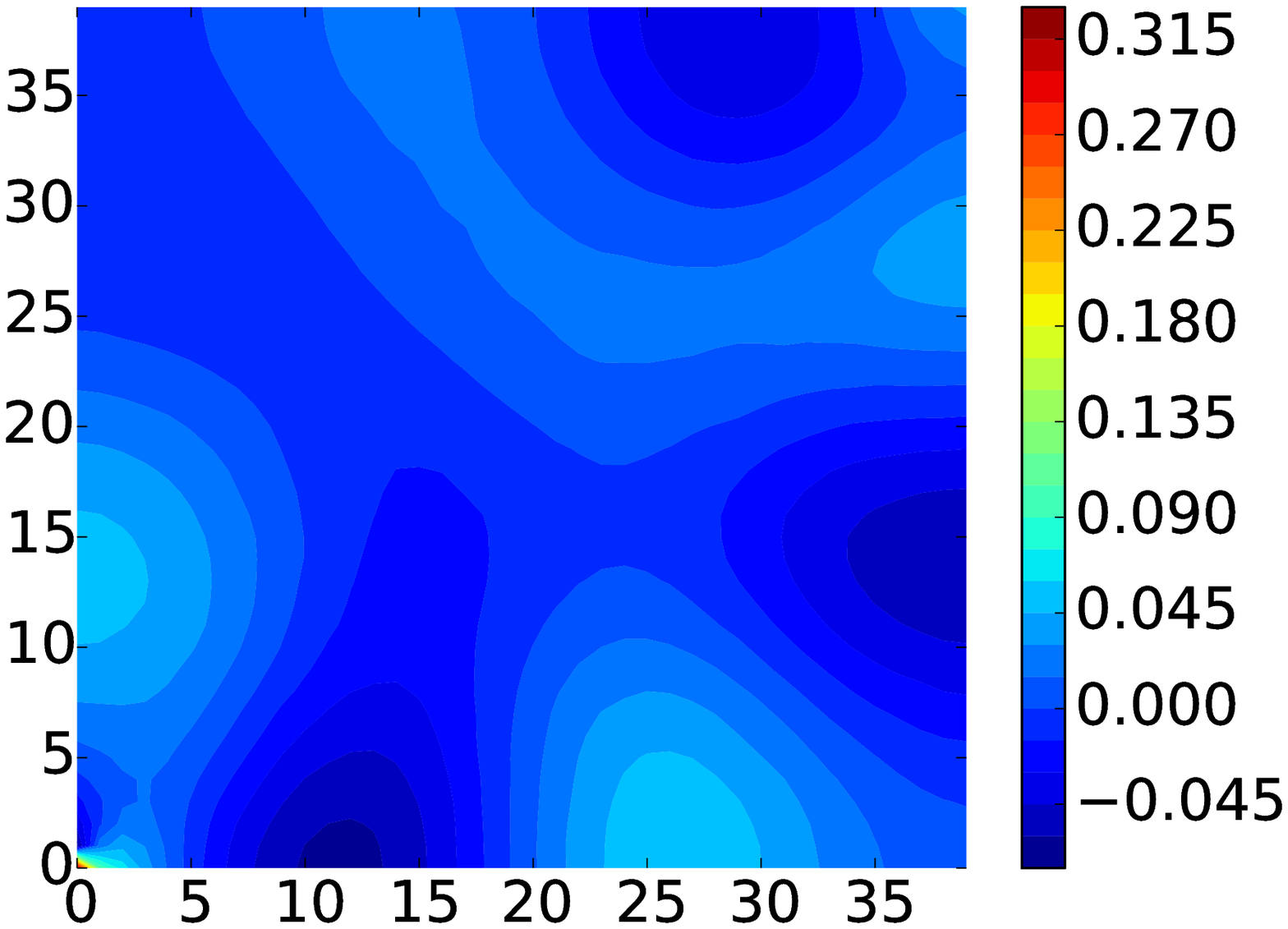}
\includegraphics[width =
0.19\textwidth]{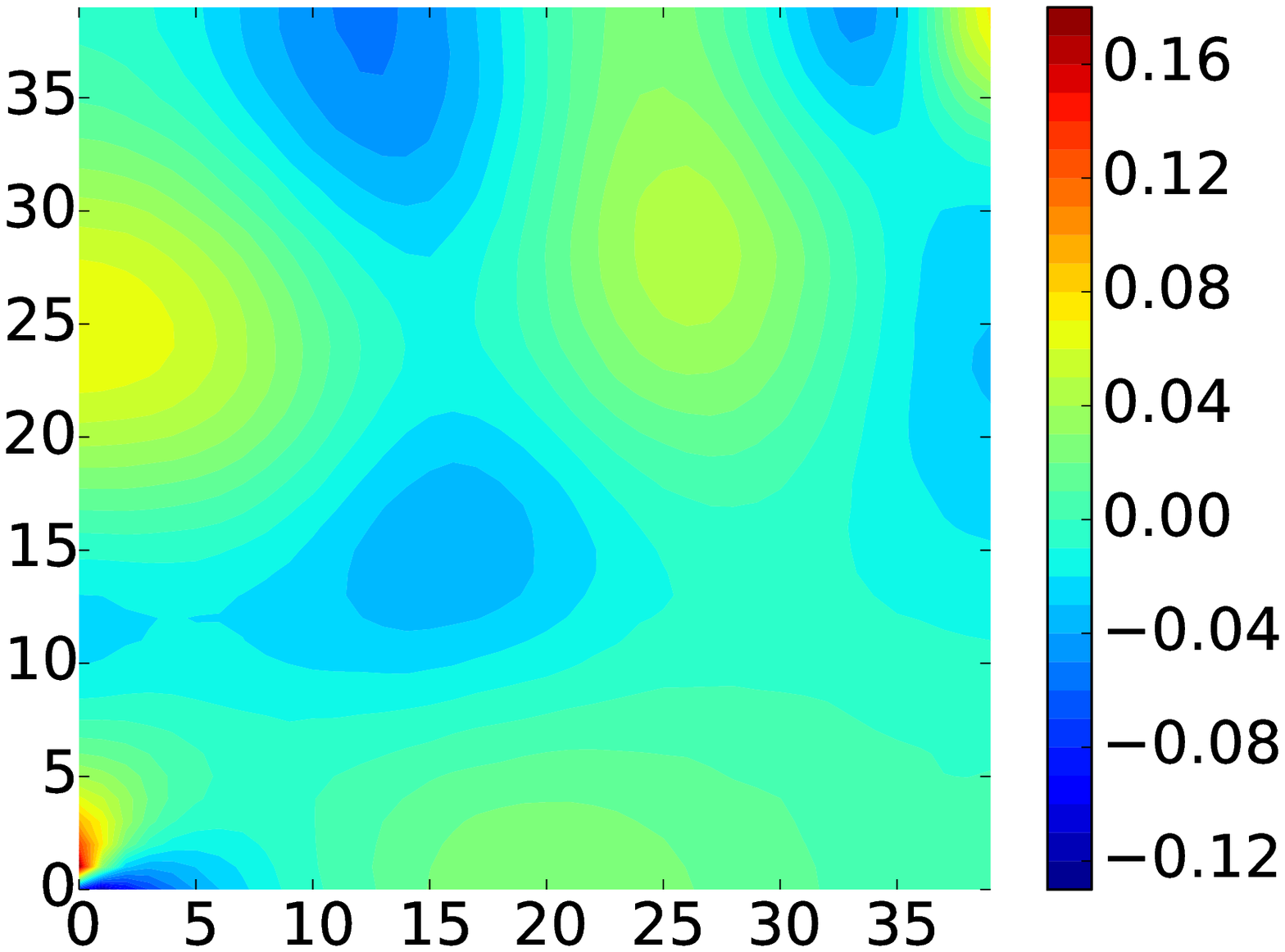}
\includegraphics[width =
0.19\textwidth]{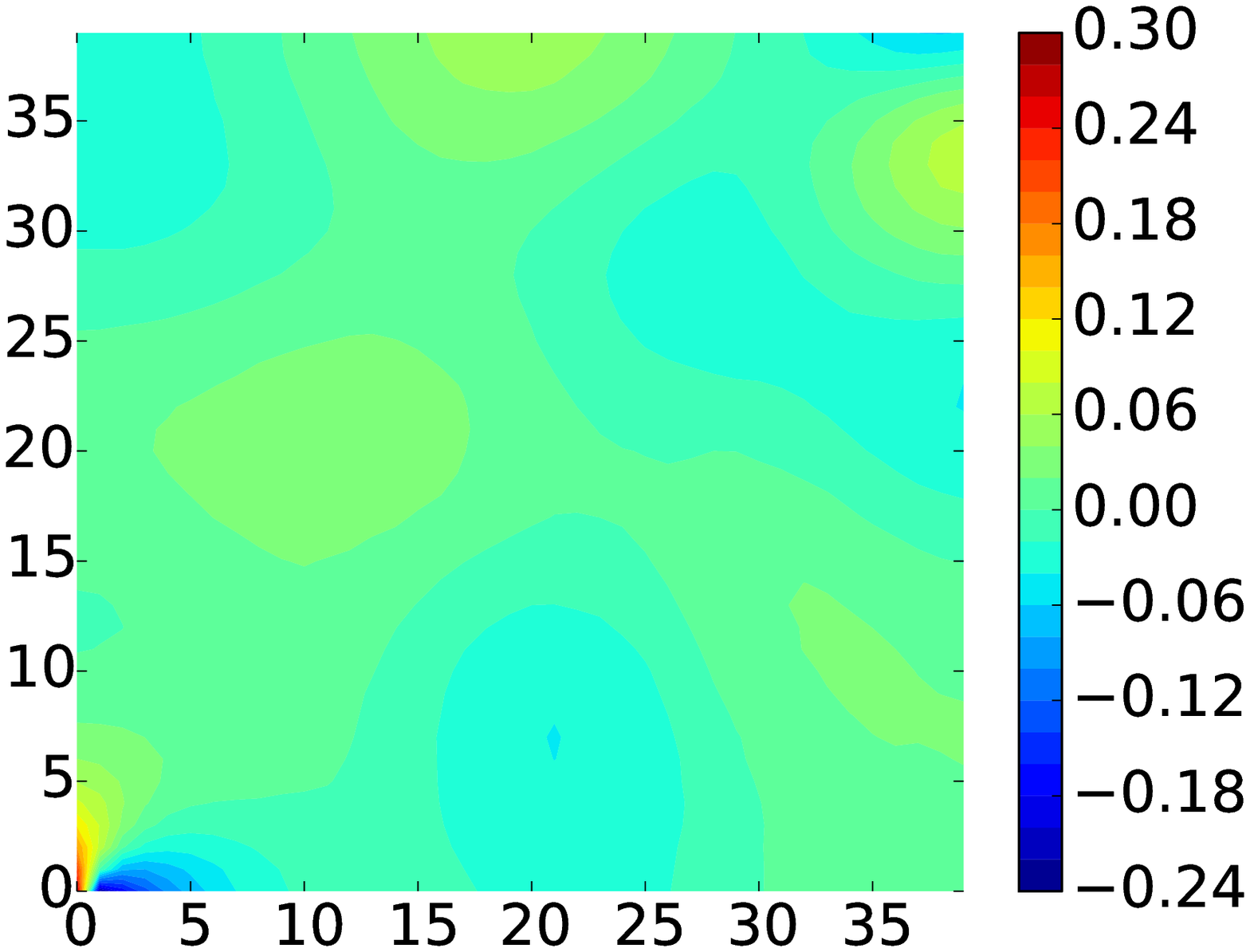}
\includegraphics[width =
0.19\textwidth]{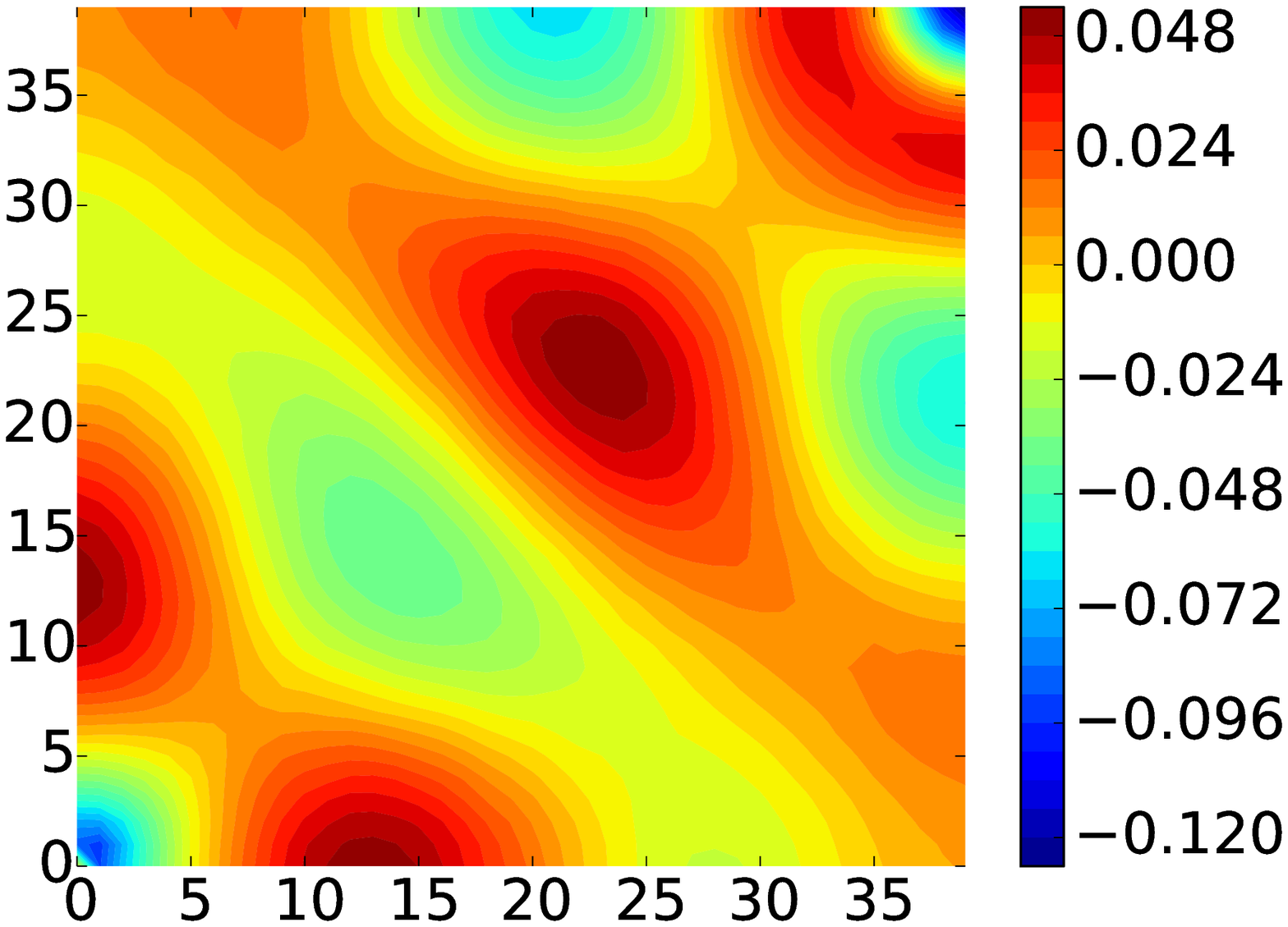}
\includegraphics[width =
0.19\textwidth]{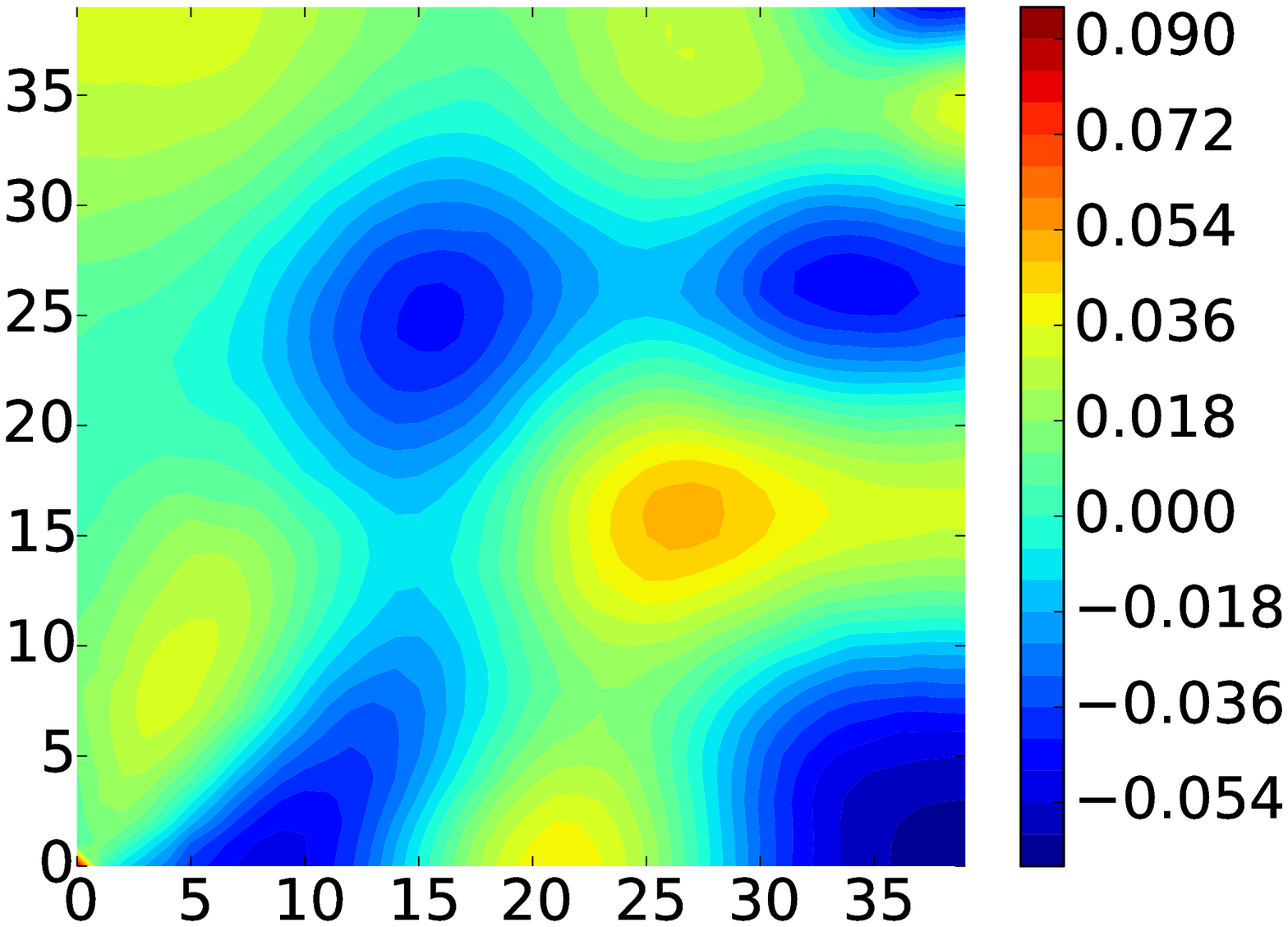}
\includegraphics[width =
0.19\textwidth]{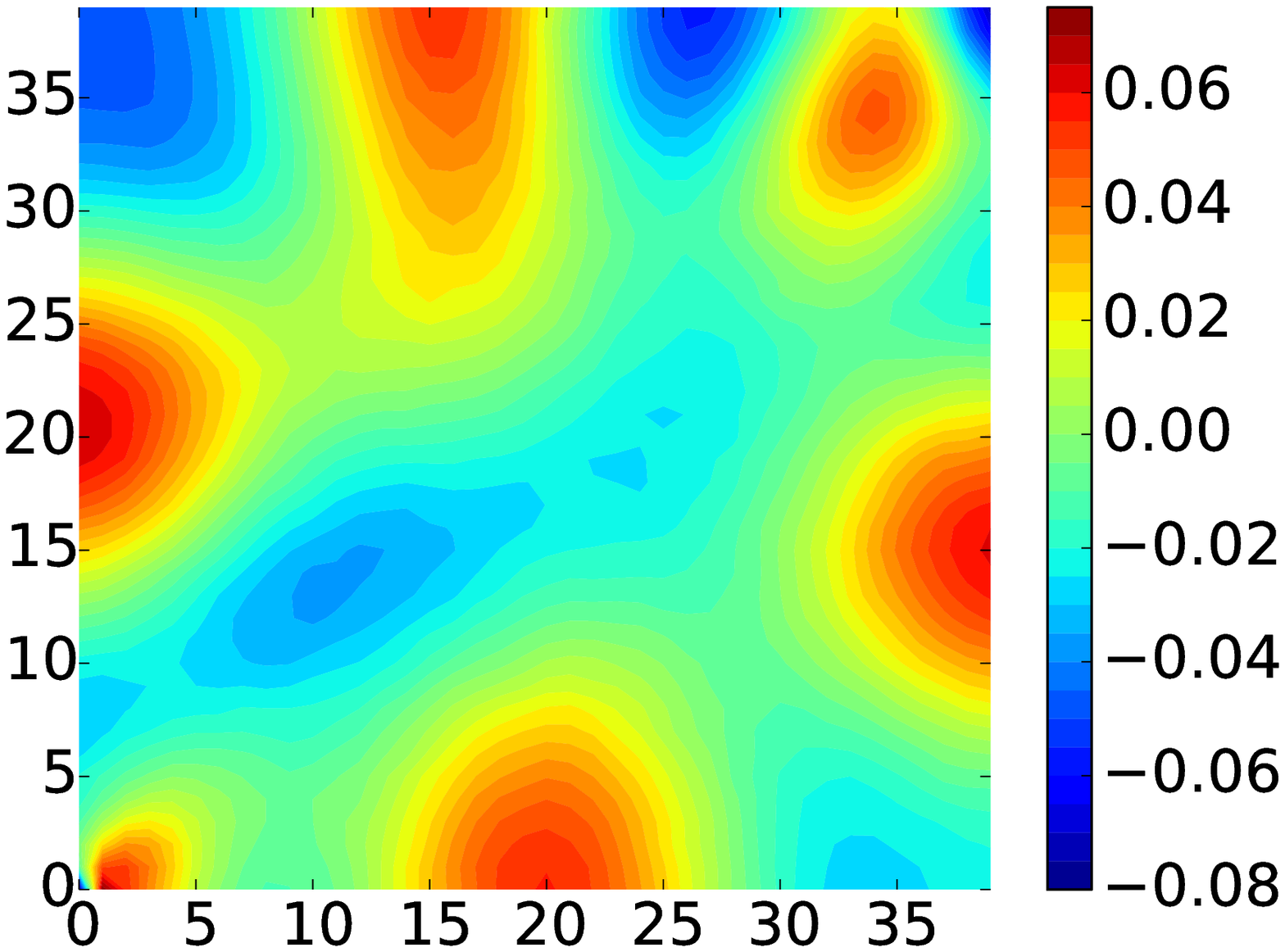}
\includegraphics[width =
0.19\textwidth]{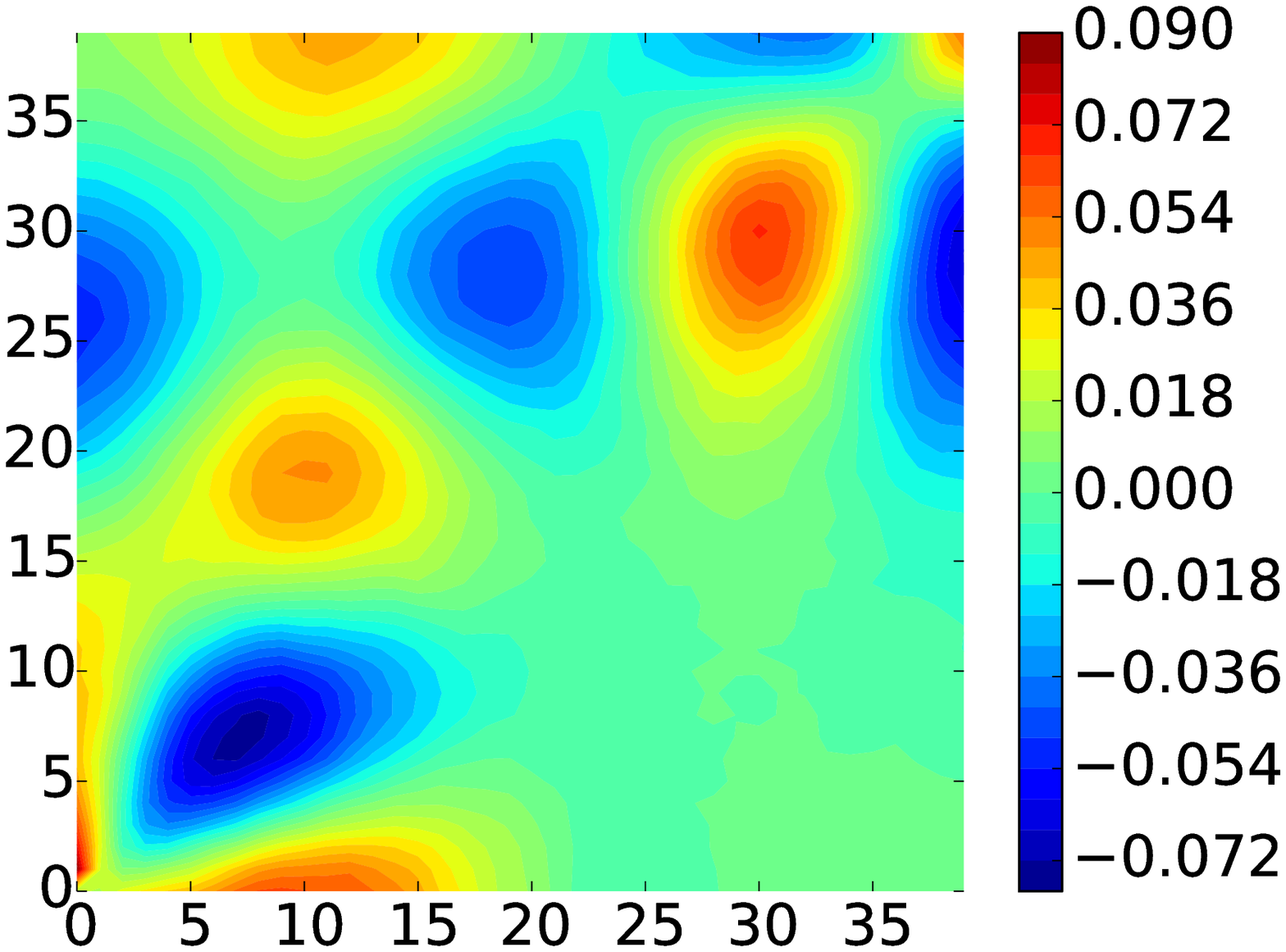}
\includegraphics[width =
0.19\textwidth]{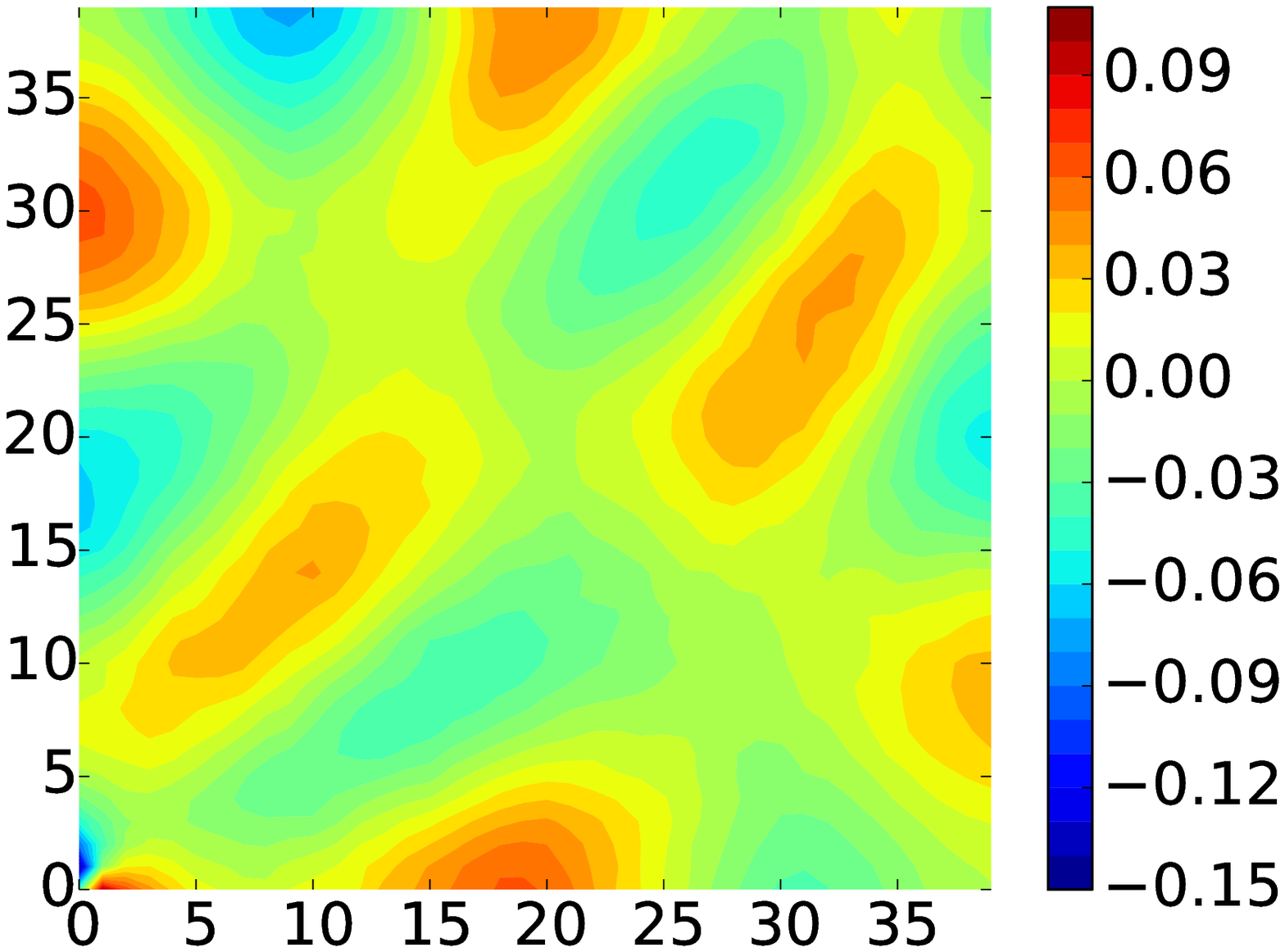}
\includegraphics[width =
0.19\textwidth]{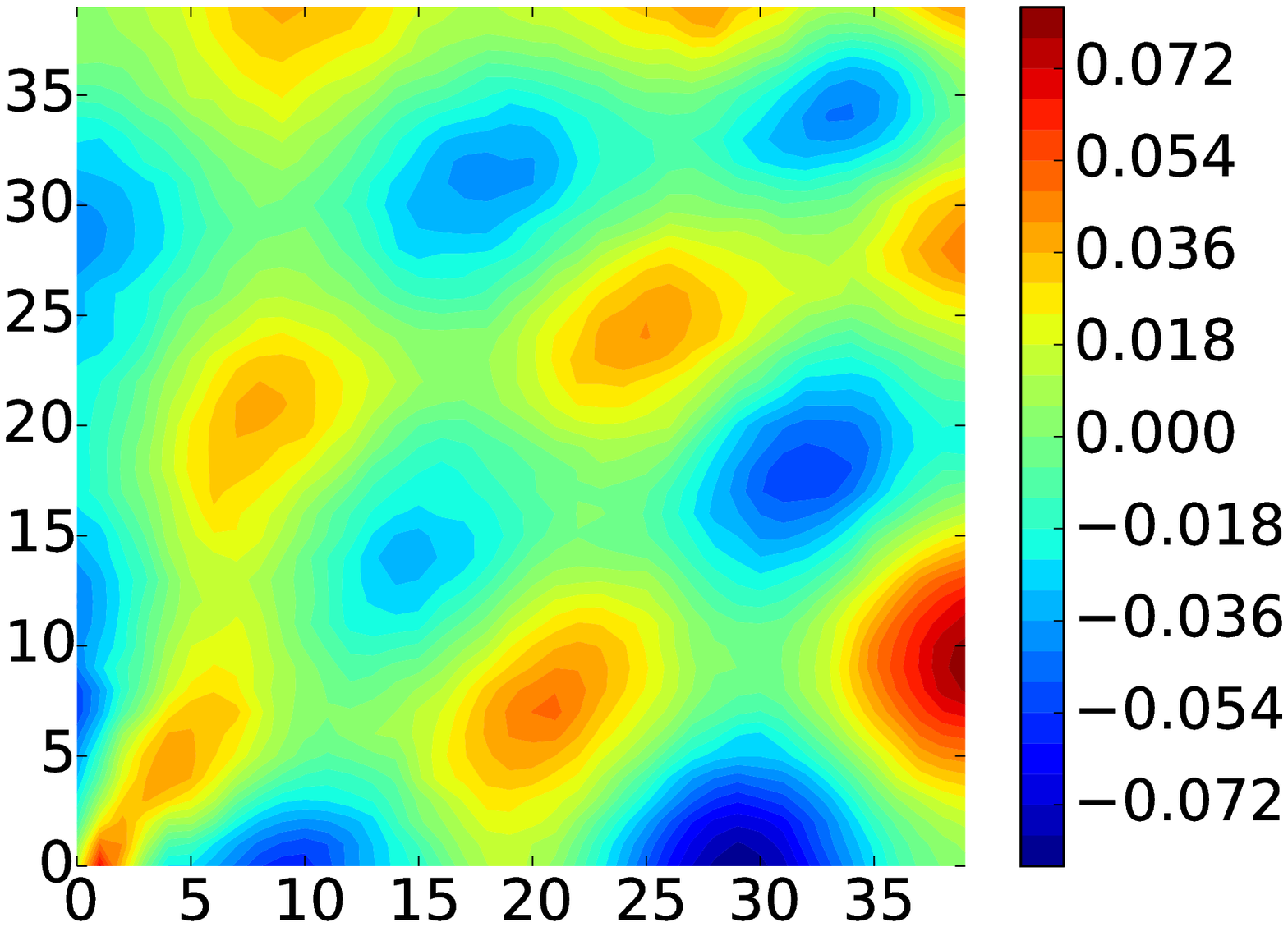}
\caption{The first $20$ eigenvectors of the covariance kernel of
  $\eta_1$. \label{fig:eigvecs}}
\end{figure}

\begin{figure}[H]
\centering
\includegraphics[width =
0.19\textwidth]{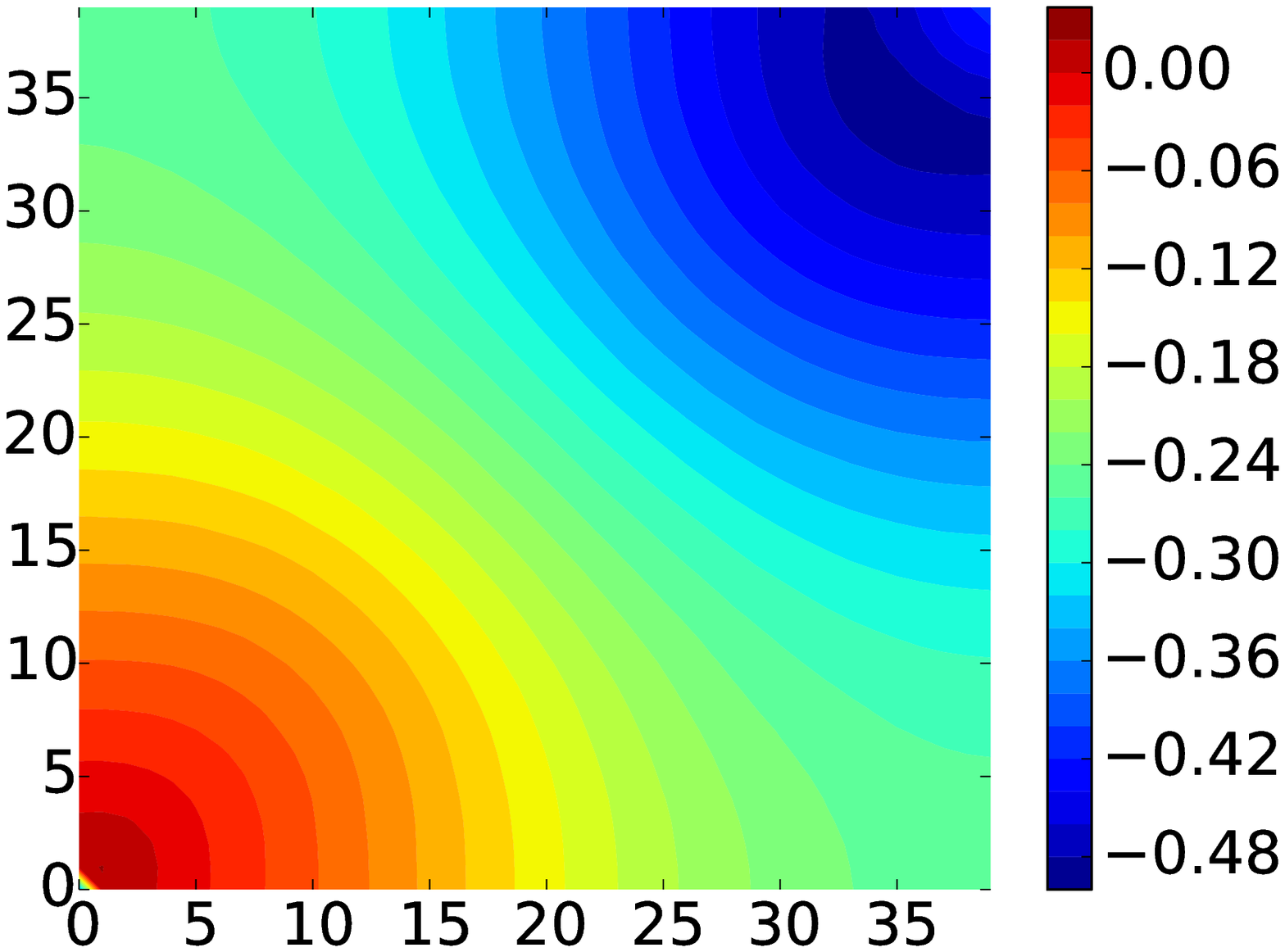}
\includegraphics[width =
0.19\textwidth]{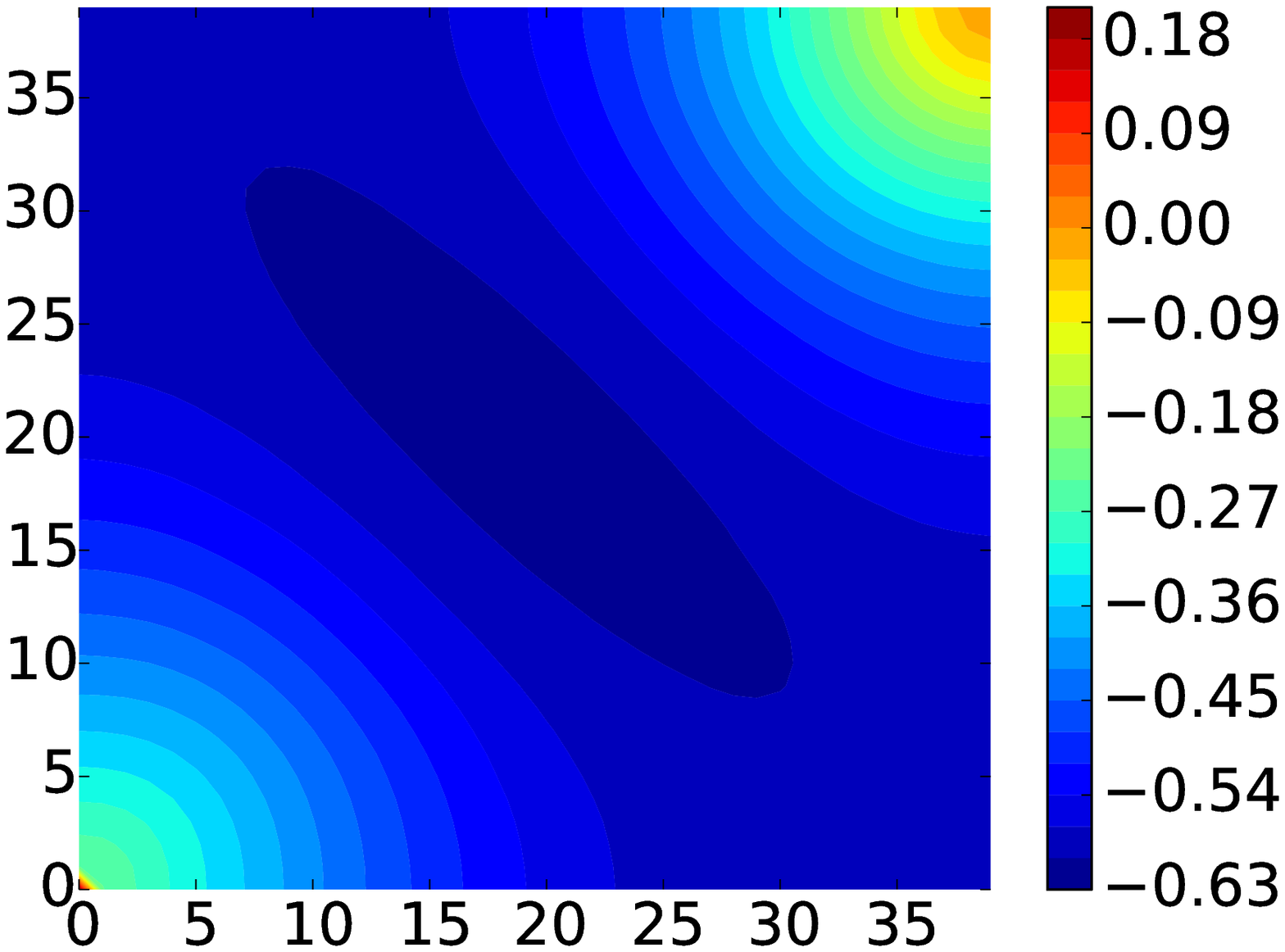}
\includegraphics[width =
0.19\textwidth]{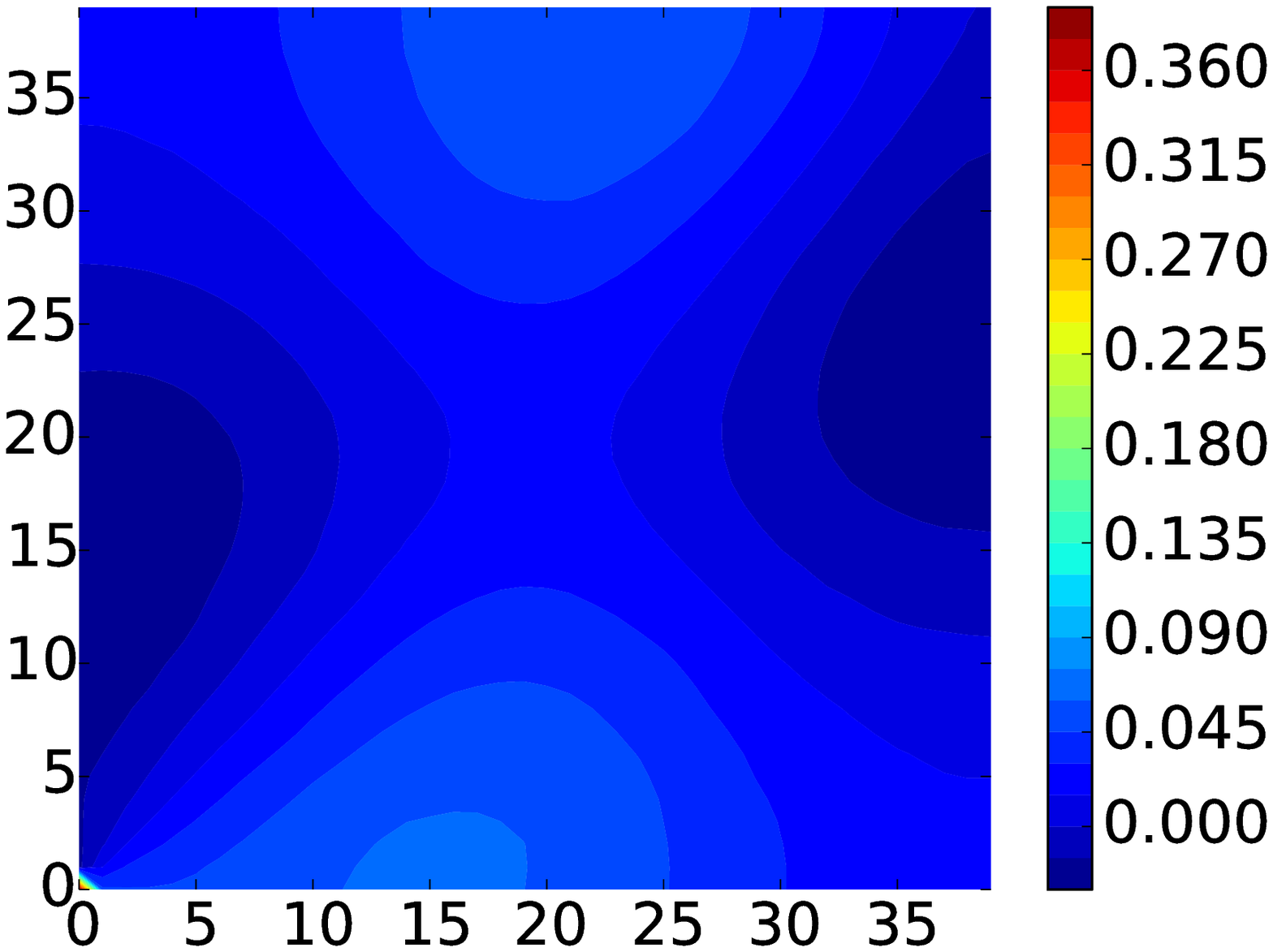}
\includegraphics[width =
0.19\textwidth]{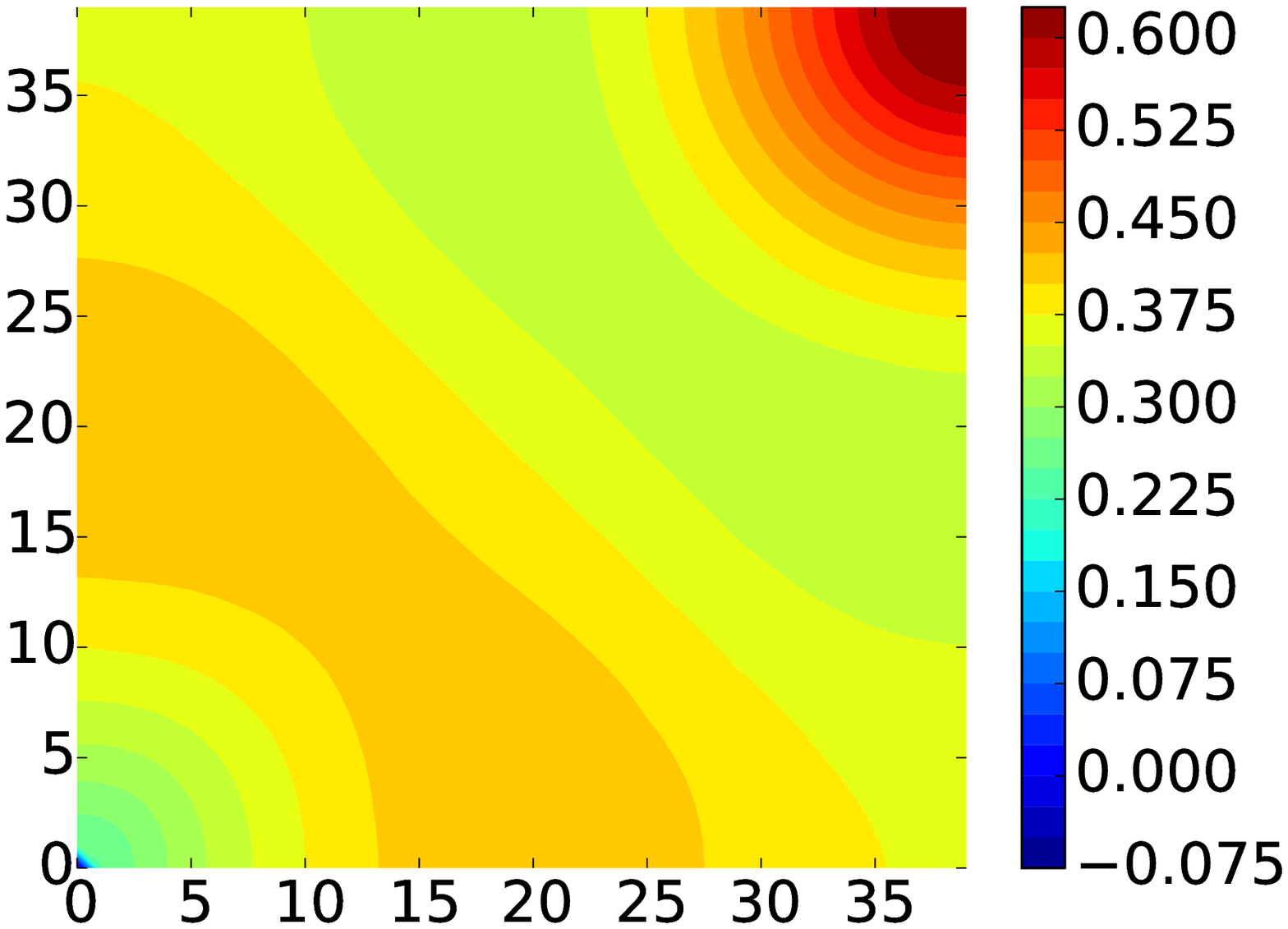}
\includegraphics[width =
0.19\textwidth]{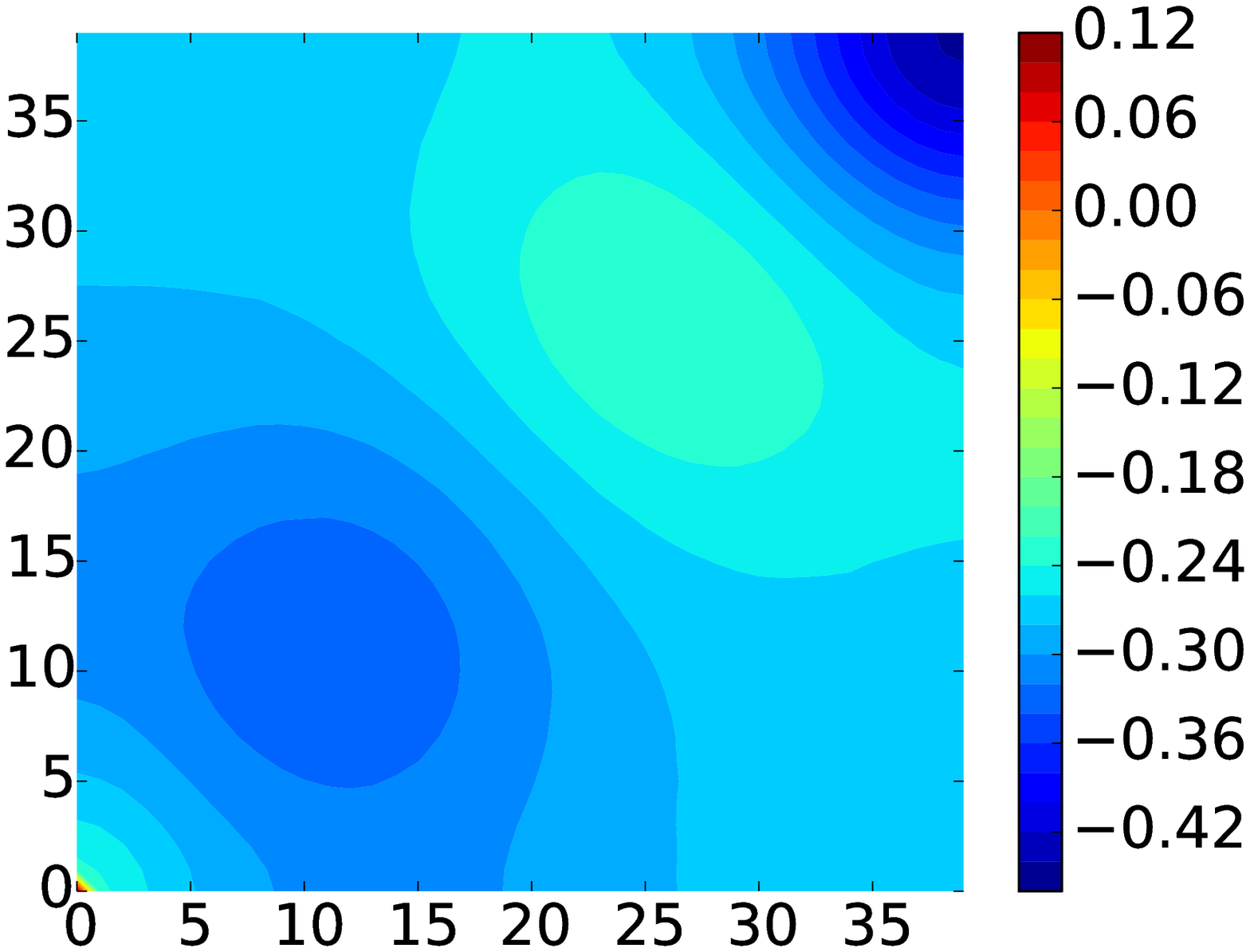}
\includegraphics[width =
0.19\textwidth]{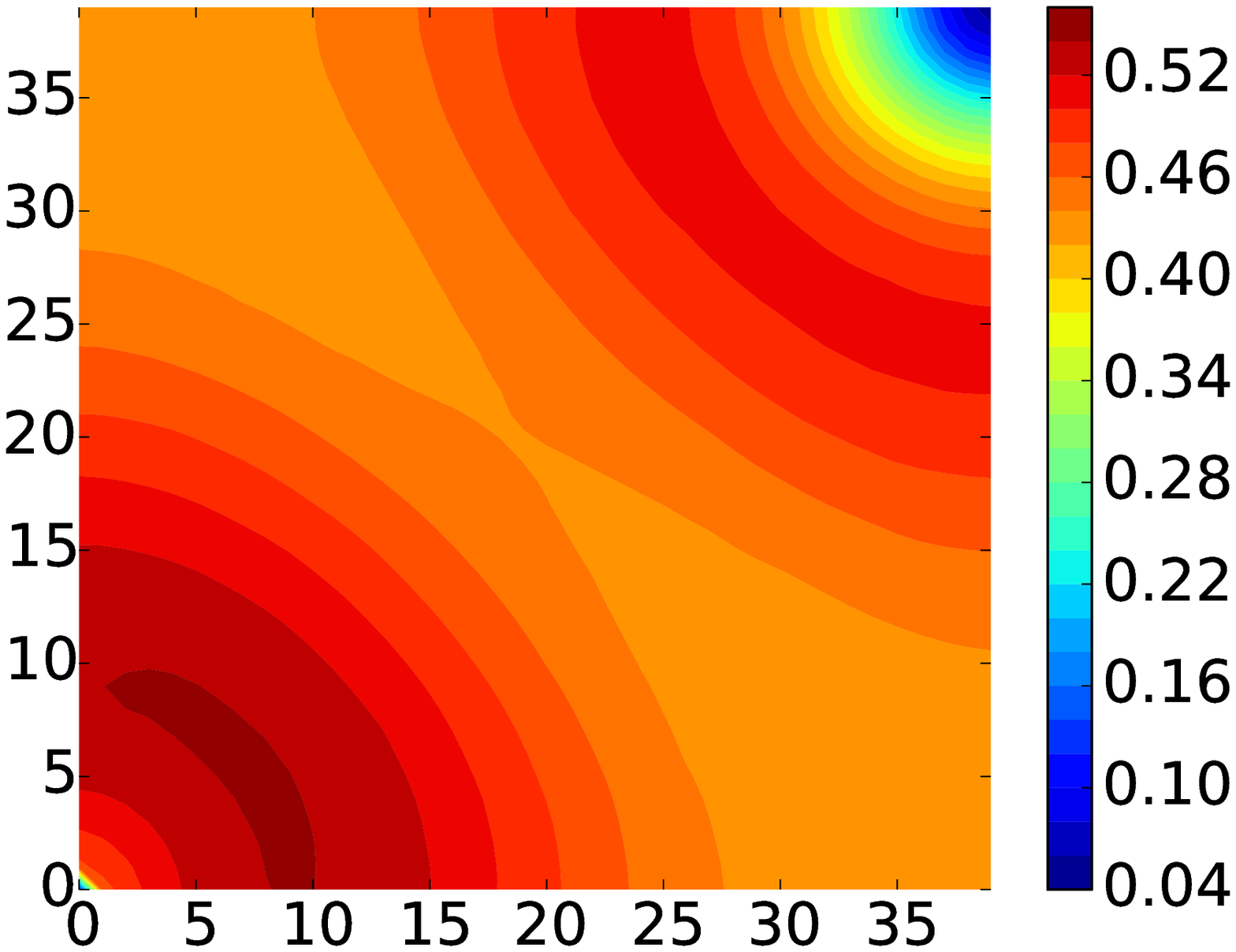}
\includegraphics[width =
0.19\textwidth]{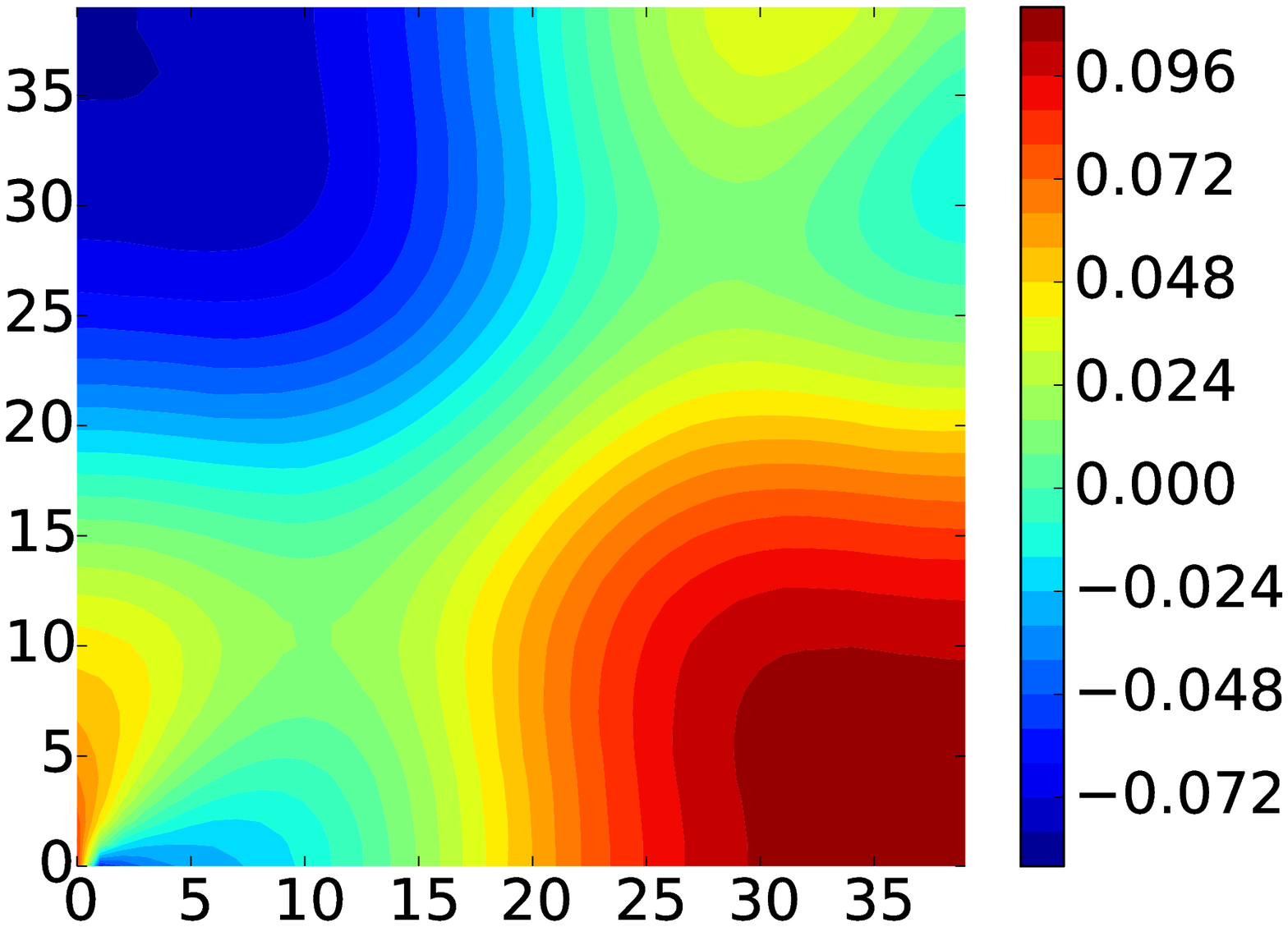}
\includegraphics[width =
0.19\textwidth]{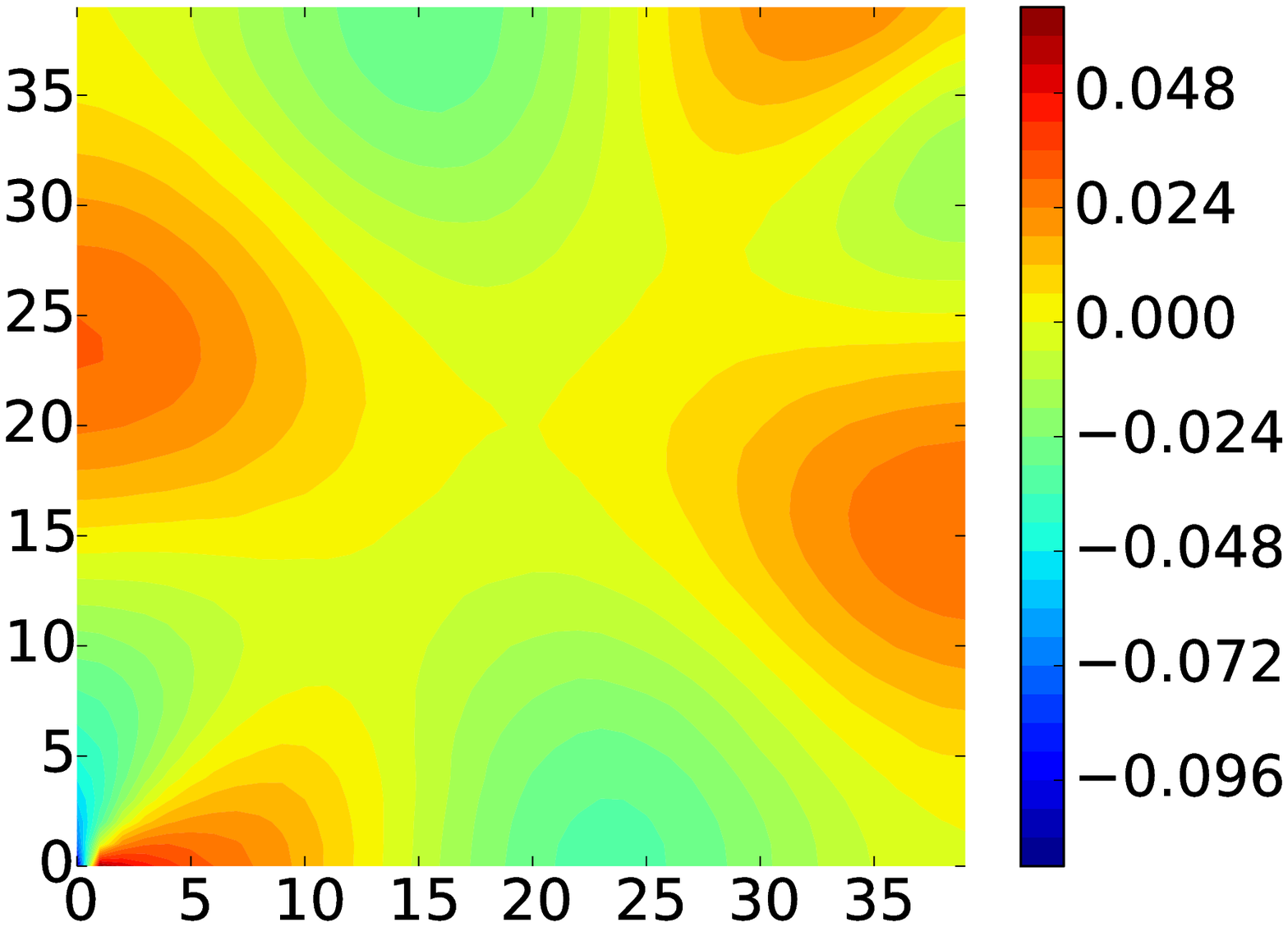}
\includegraphics[width =
0.19\textwidth]{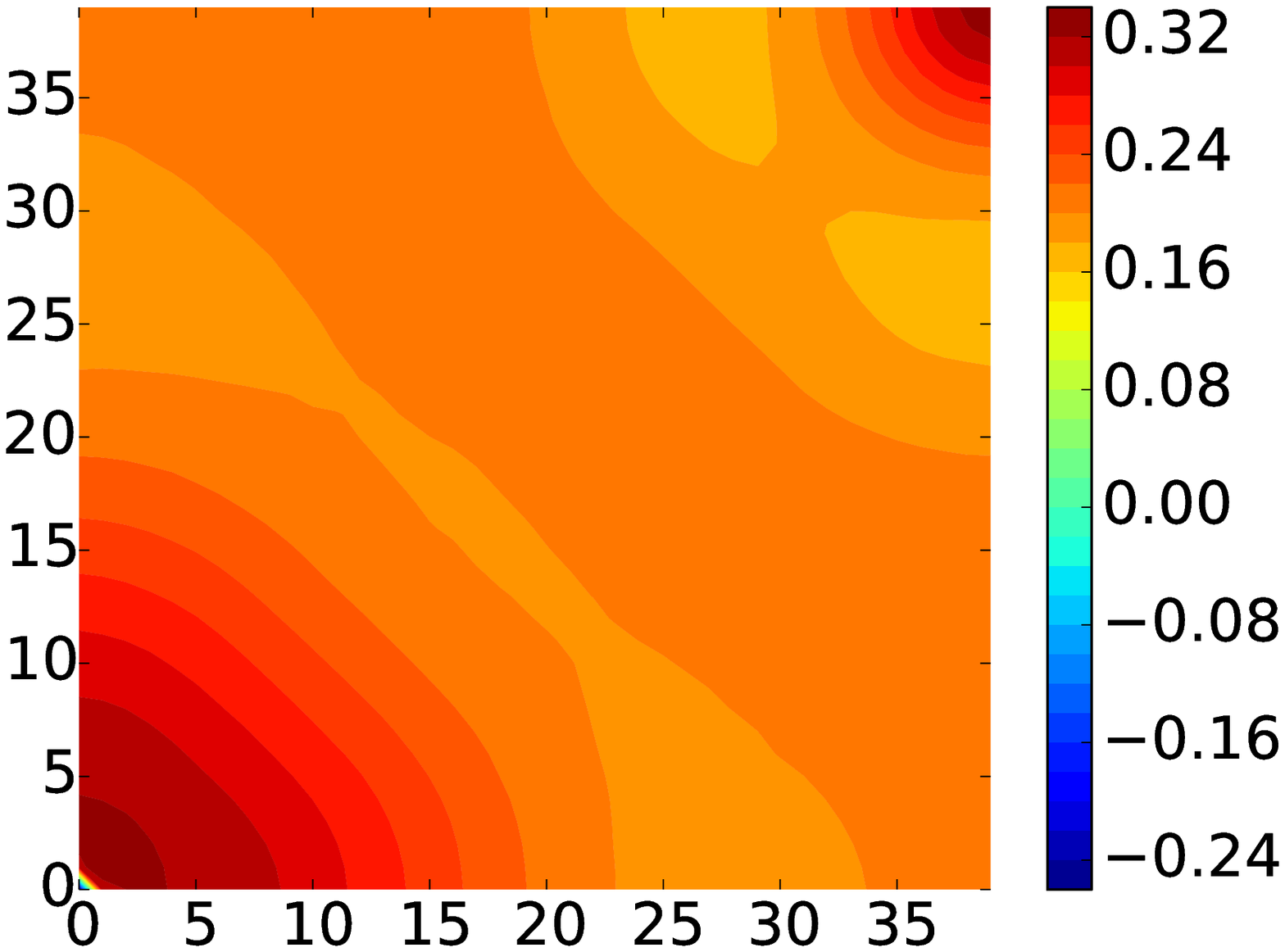}
\includegraphics[width =
0.19\textwidth]{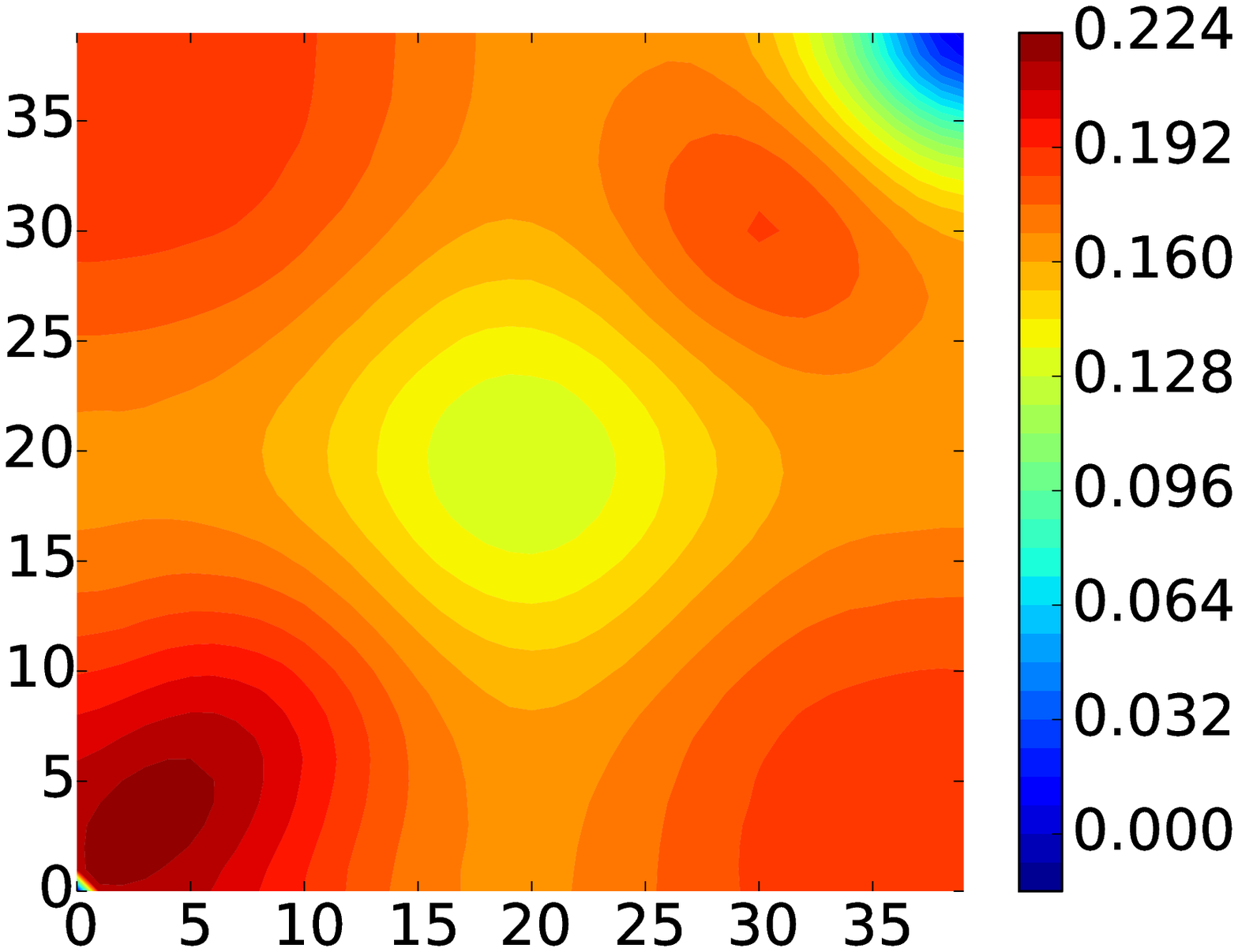}
\includegraphics[width =
0.19\textwidth]{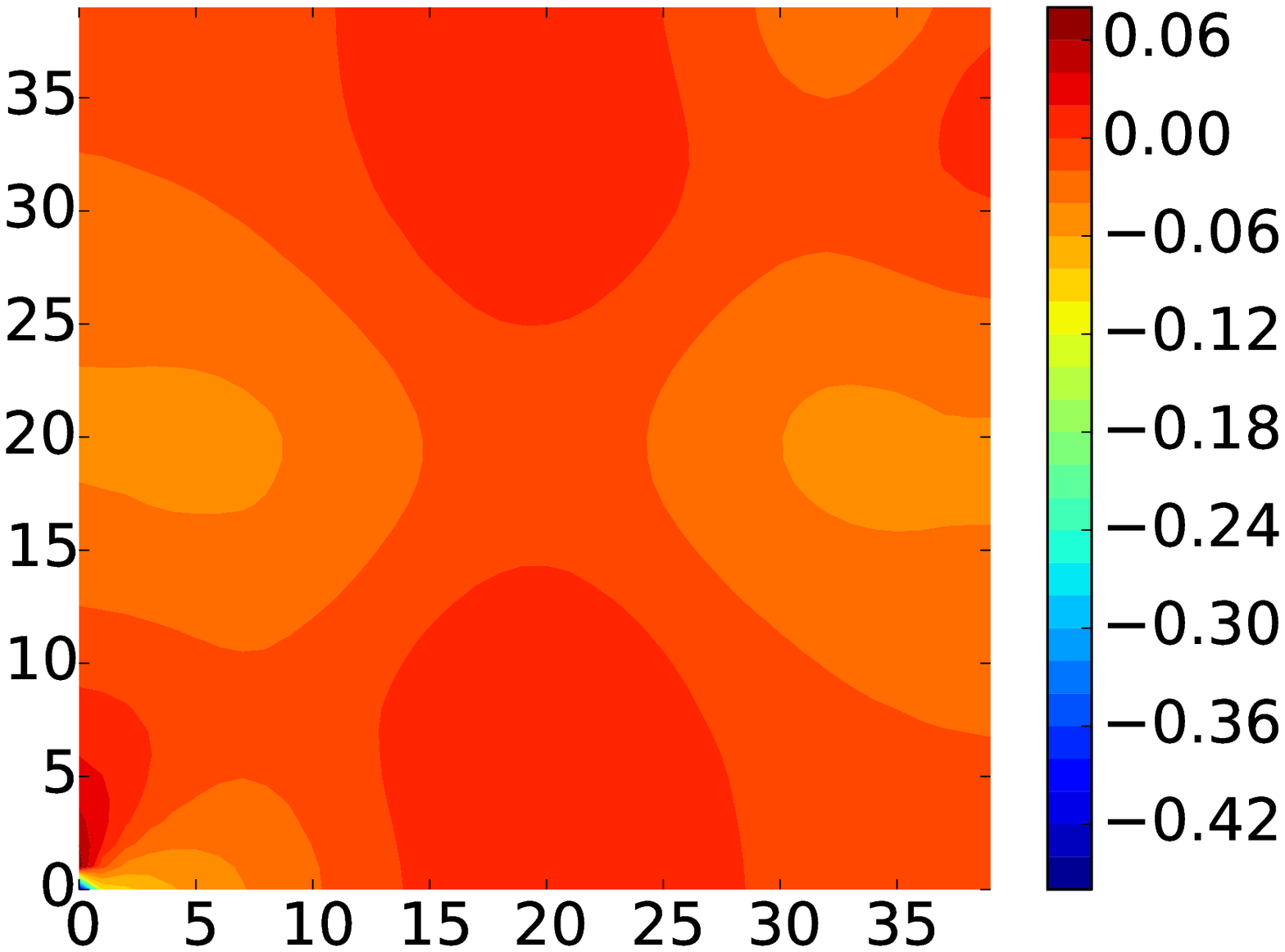}
\includegraphics[width =
0.19\textwidth]{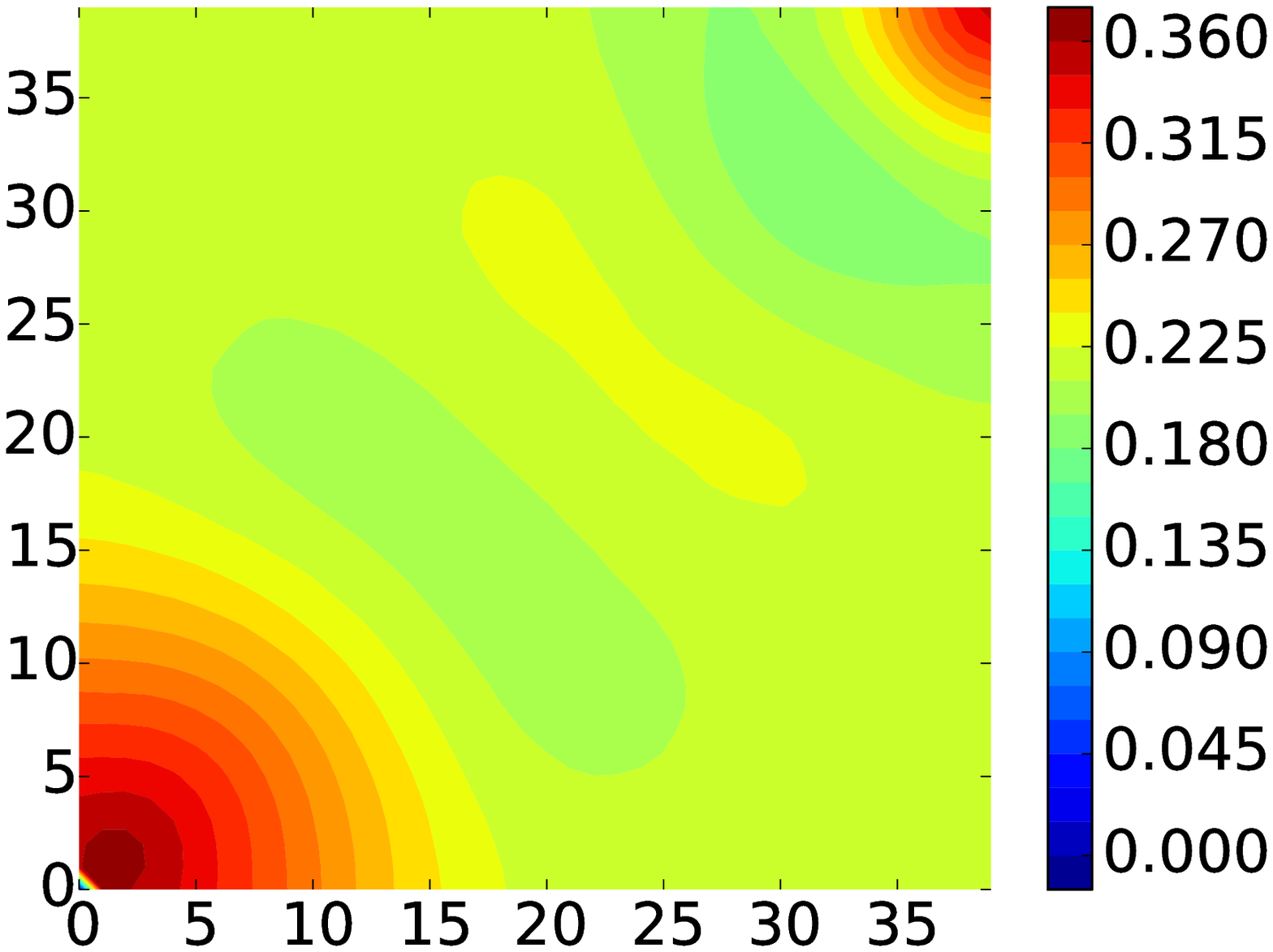}
\includegraphics[width =
0.19\textwidth]{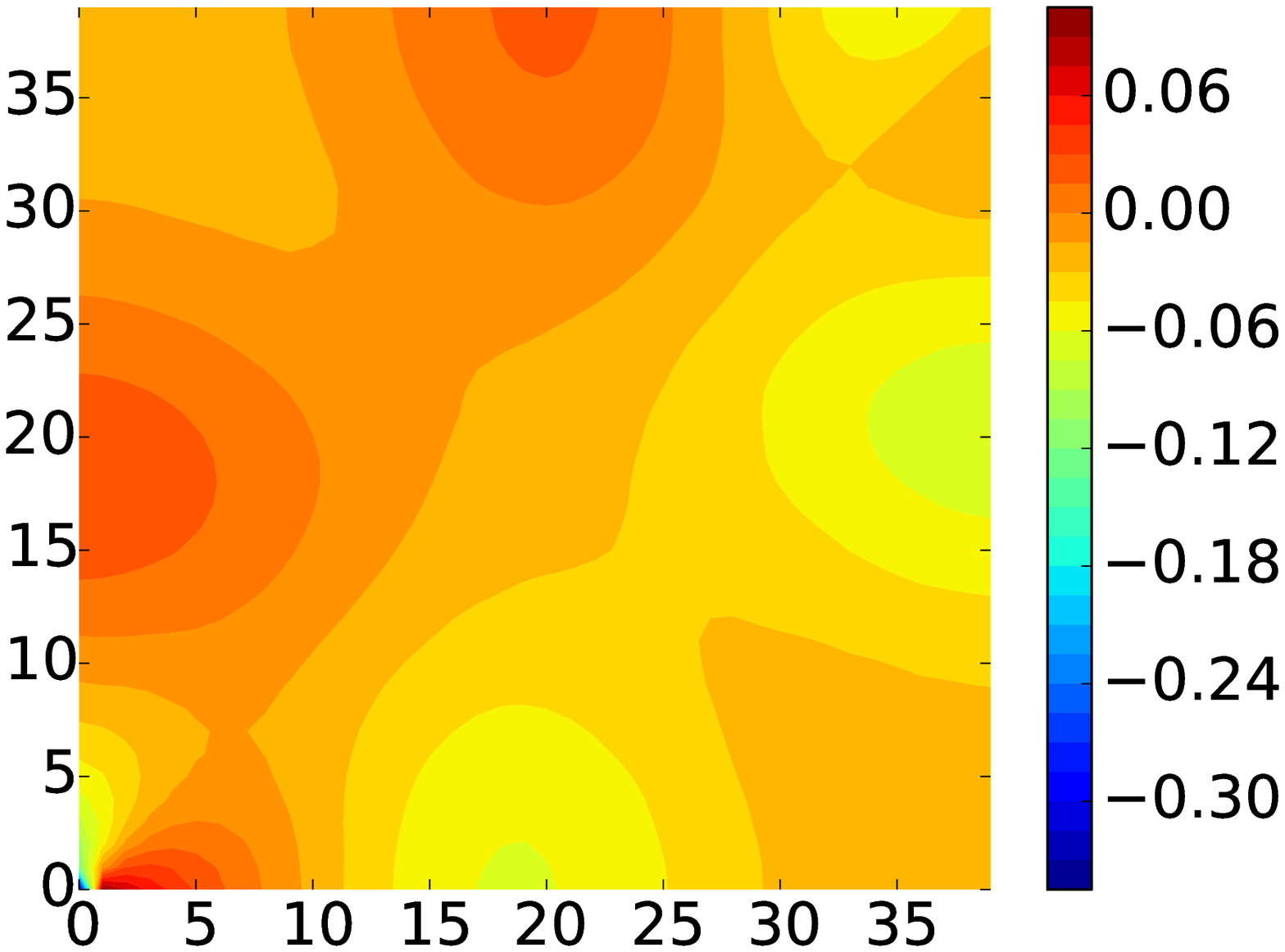}
\includegraphics[width =
0.19\textwidth]{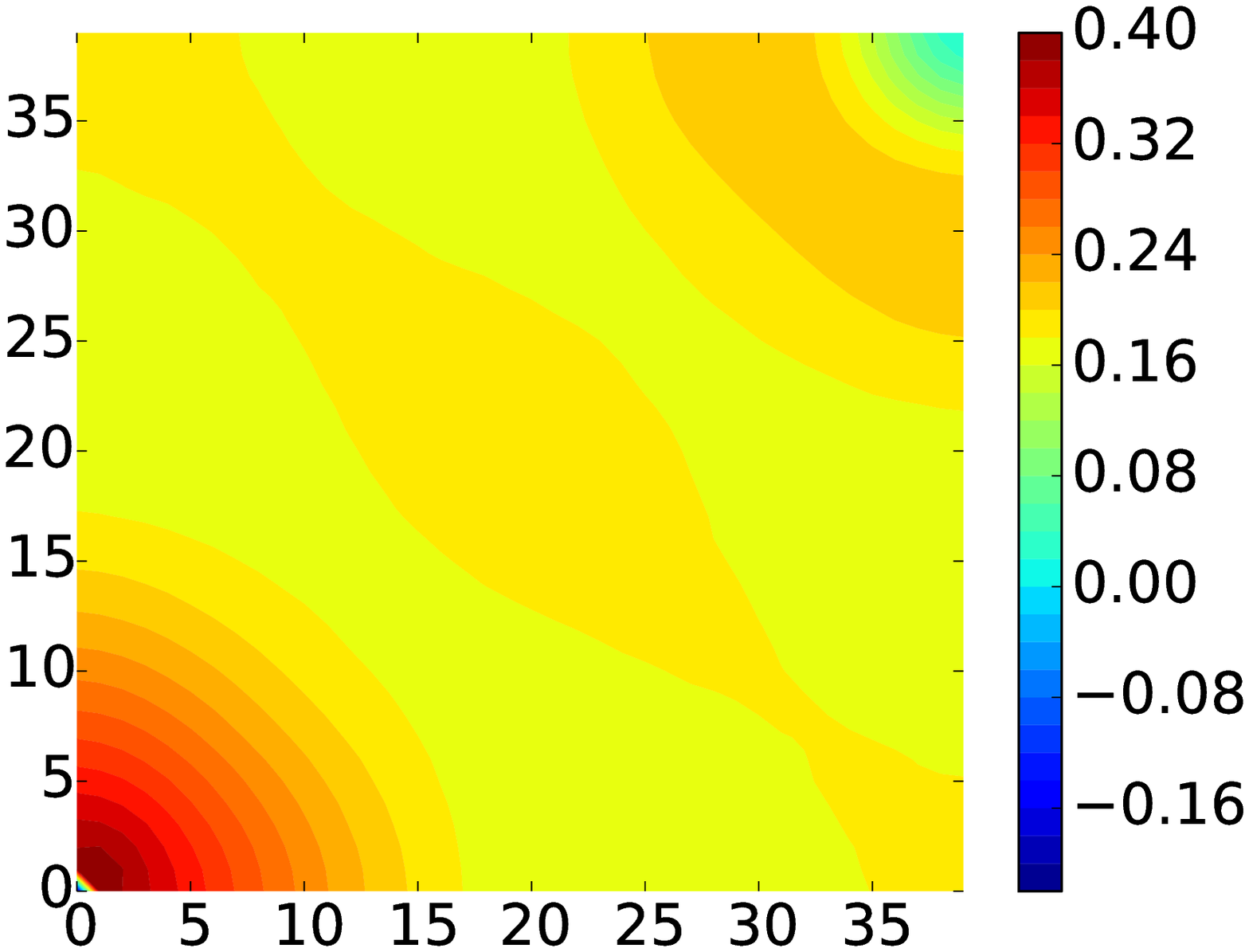}
\includegraphics[width =
0.19\textwidth]{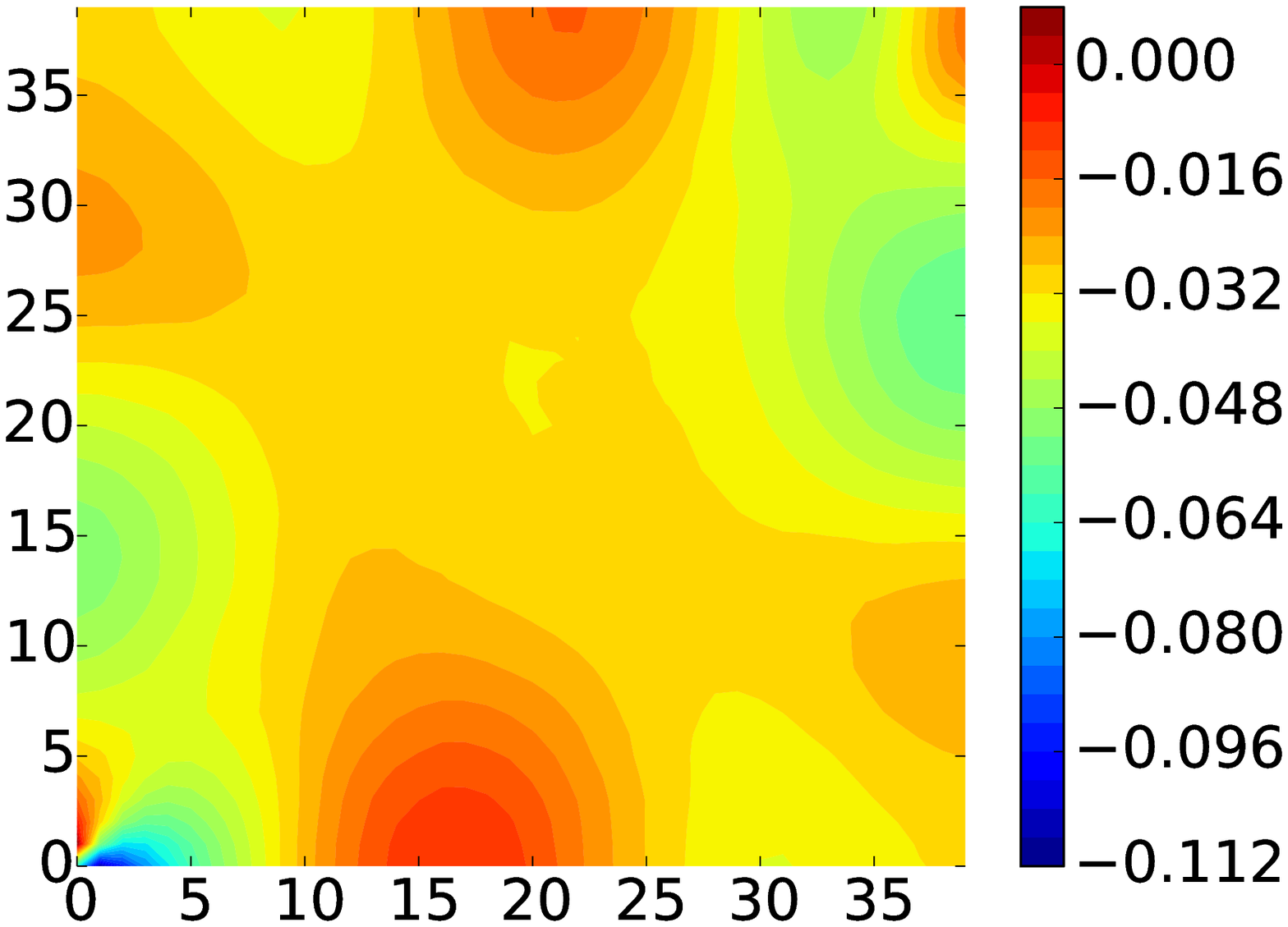}
\includegraphics[width =
0.19\textwidth]{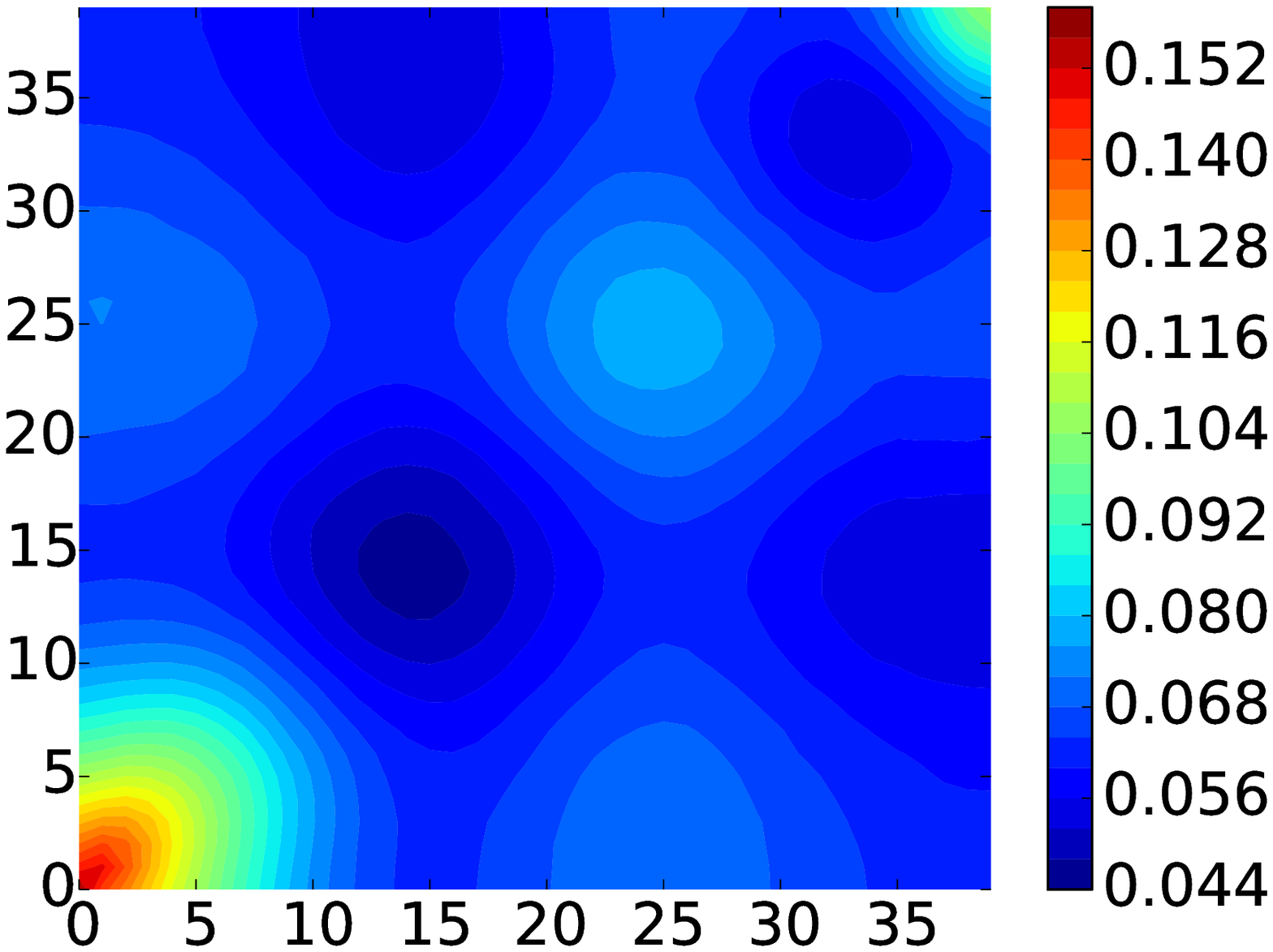}
\includegraphics[width =
0.19\textwidth]{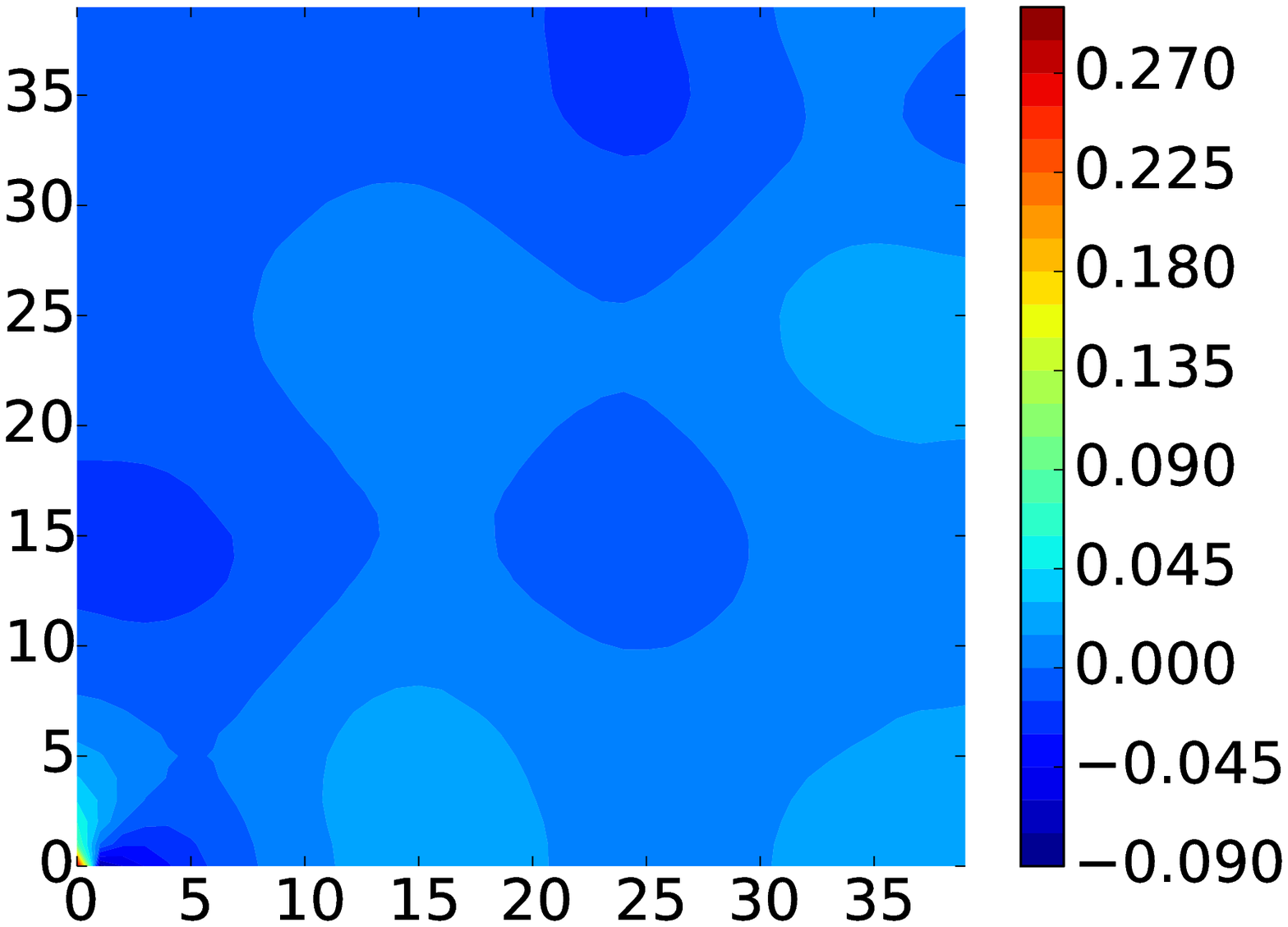}
\includegraphics[width =
0.19\textwidth]{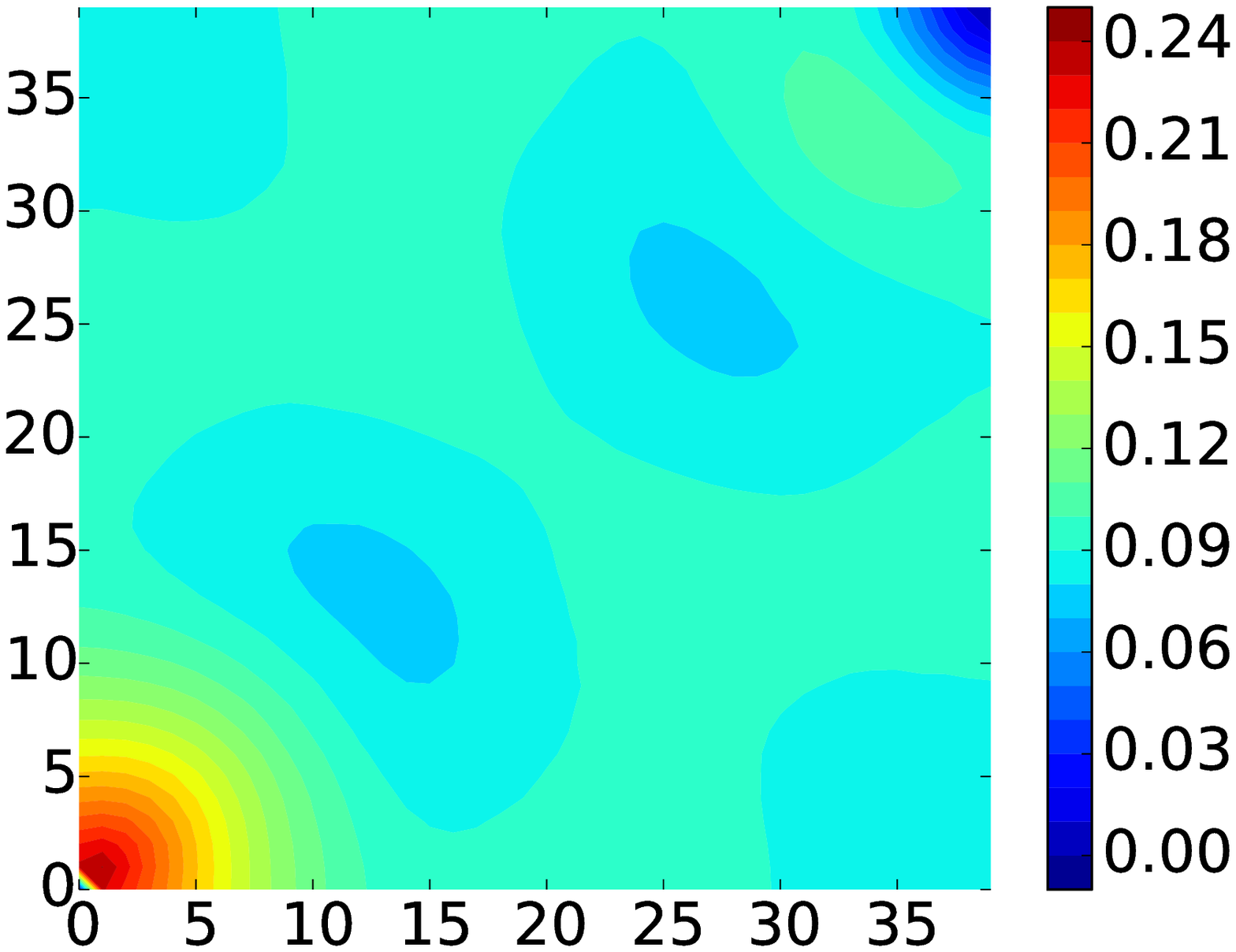}
\includegraphics[width =
0.19\textwidth]{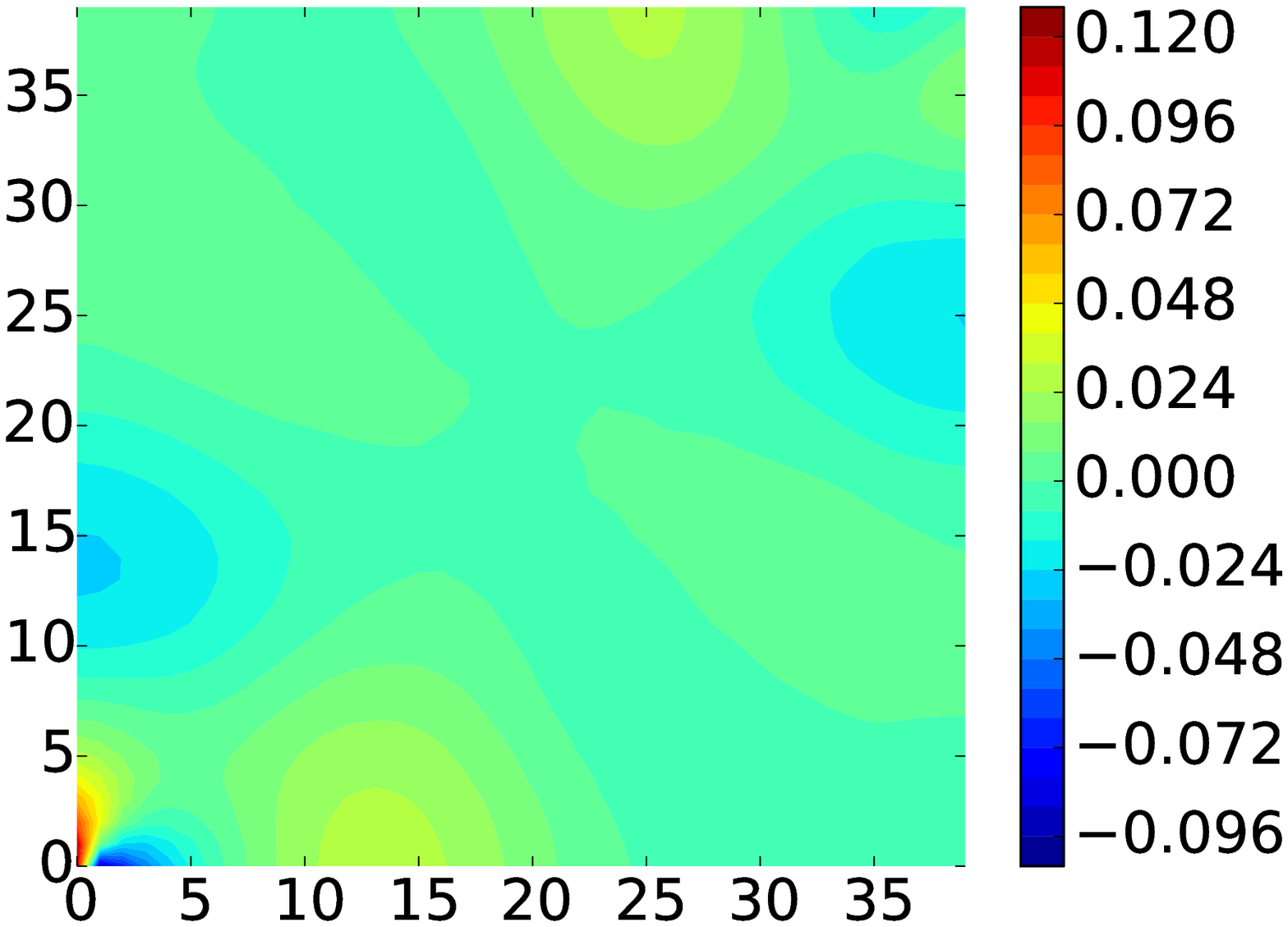}
\includegraphics[width =
0.19\textwidth]{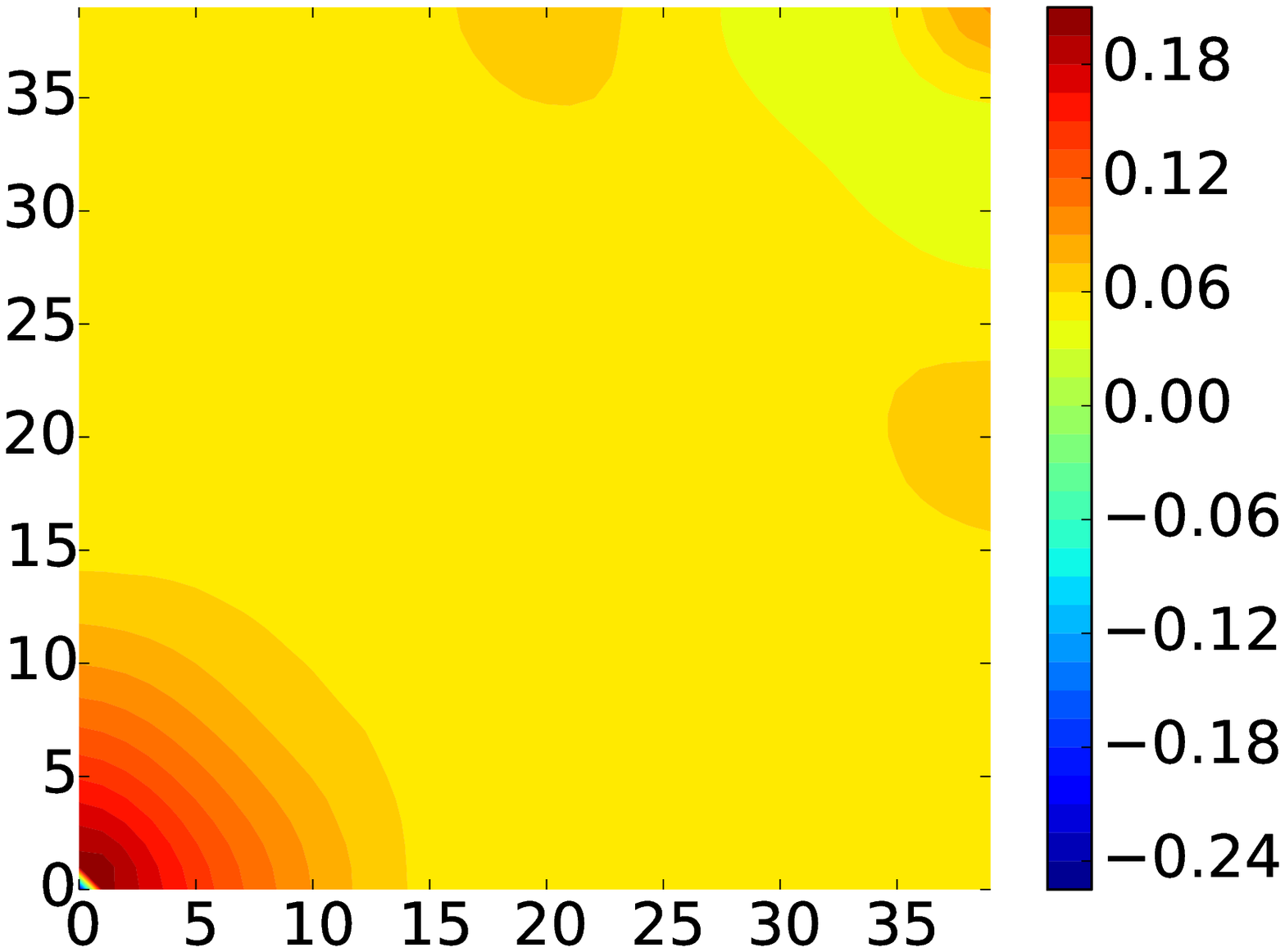}
\caption{The $20$ entries of the first row of $\bA(\bx)$ for the Gaussian adaptation. \label{fig:isometry_entries}}
\end{figure}

\subsubsection{Gaussian adaptation}

First we construct the $1$-dimensional adapted $2$nd-order
series 
\begin{equation}
\label{eq:gauss_pce}
u^{\bA(\bx)}(\eta) = u_0^{\bA(\bx)} + u_1^{\bA(\bx)} \eta
+ u_2^{\bA(\bx)} \frac{\eta^2 - 1}{\sqrt{2}}
\end{equation}
using as $\bA(\bx)$ family of isometries where the first row is defined as in
Eq. (\ref{eq:gauss_adapt}), that is the Gaussian
adaptation. The kernel of the transformed
input $\eta$, that is $k_1(\bx, \by) = \ba_1(\bx)\ba_1(\by)^T$,
has $20$ strictly positive eigenvalues while the rest are zero as
was proved in the previous section. As expected, $\eta$ has unit
variance at each location, $k_1(\bx,\bx) = 1$ and the covariance takes
smaller values elsewhere. Its eigenvectors are shown in Fig. \ref{fig:eigvecs} and the
entries of the first row of $\bA(\bx)$ are shown in
Fig. \ref{fig:isometry_entries}, which are essentially the normalized
coefficients $\{u_{\epsilon_i}(\bx)\}_{i=1}^{20}$ as indicated in Eq. (\ref{eq:gauss_adapt}).

\begin{figure}[h]
\centering
\includegraphics[width = 0.32\textwidth]{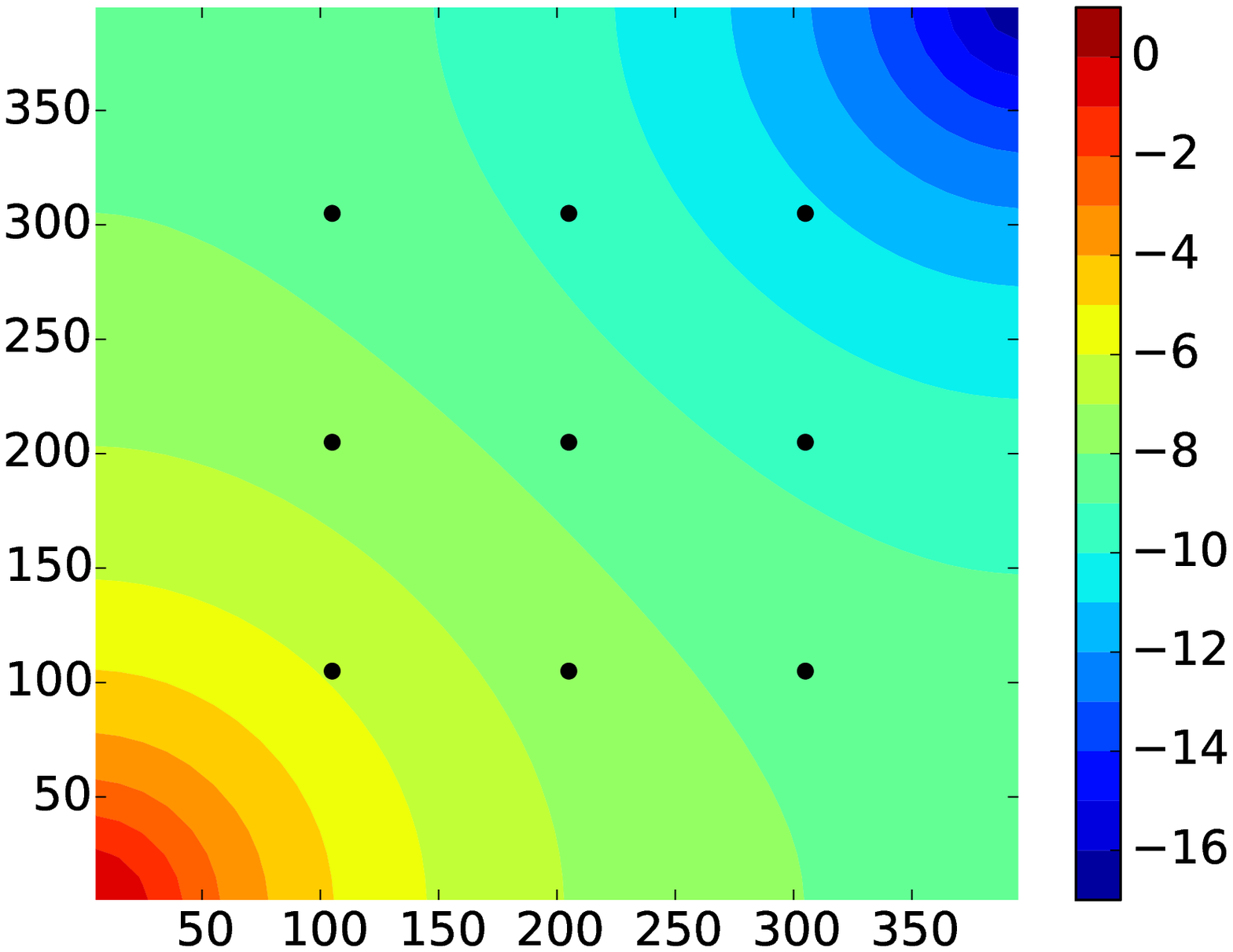}
\includegraphics[width = 0.32\textwidth]{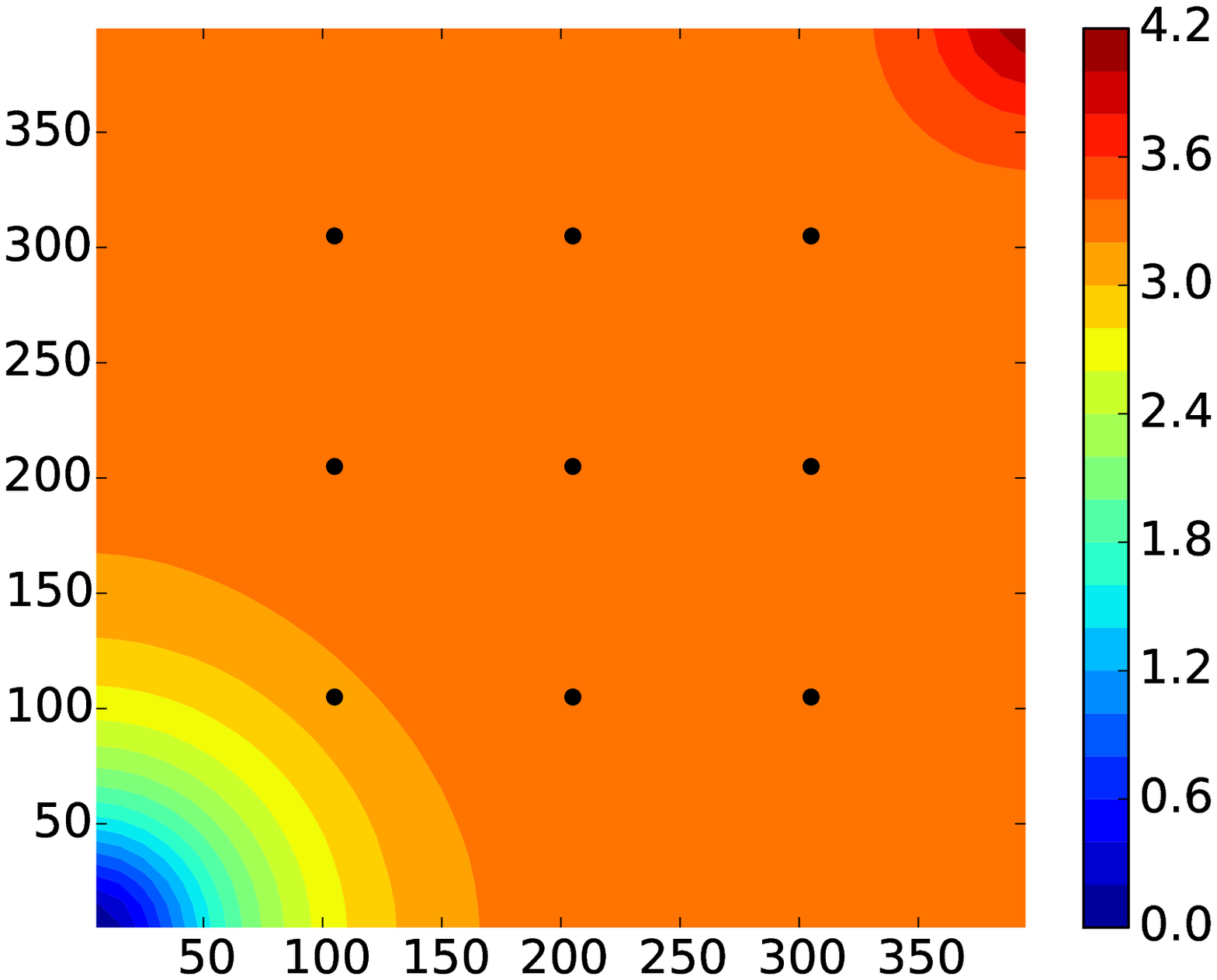}
\includegraphics[width = 0.32\textwidth]{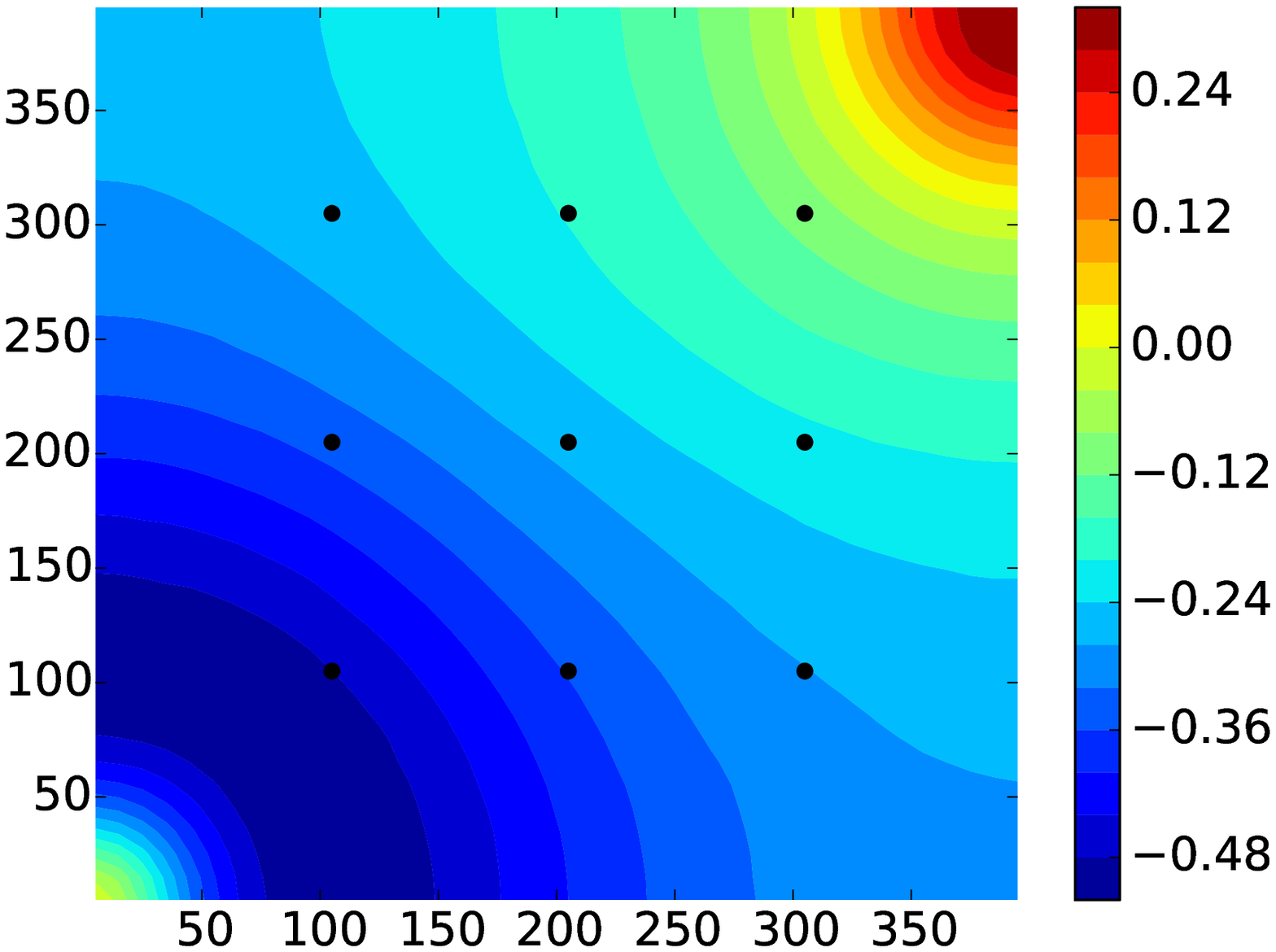}
\caption{Coefficients $u^{\bA(\bx)}_{i\epsilon_1}$, $i = 0,1,2$
  of the second-order one-dimensional Gaussian adaptation. \label{fig:adapted_coeffs_elliptic}}
\end{figure}

\begin{figure}[h]
\centering
\includegraphics[width = 0.70\textwidth]{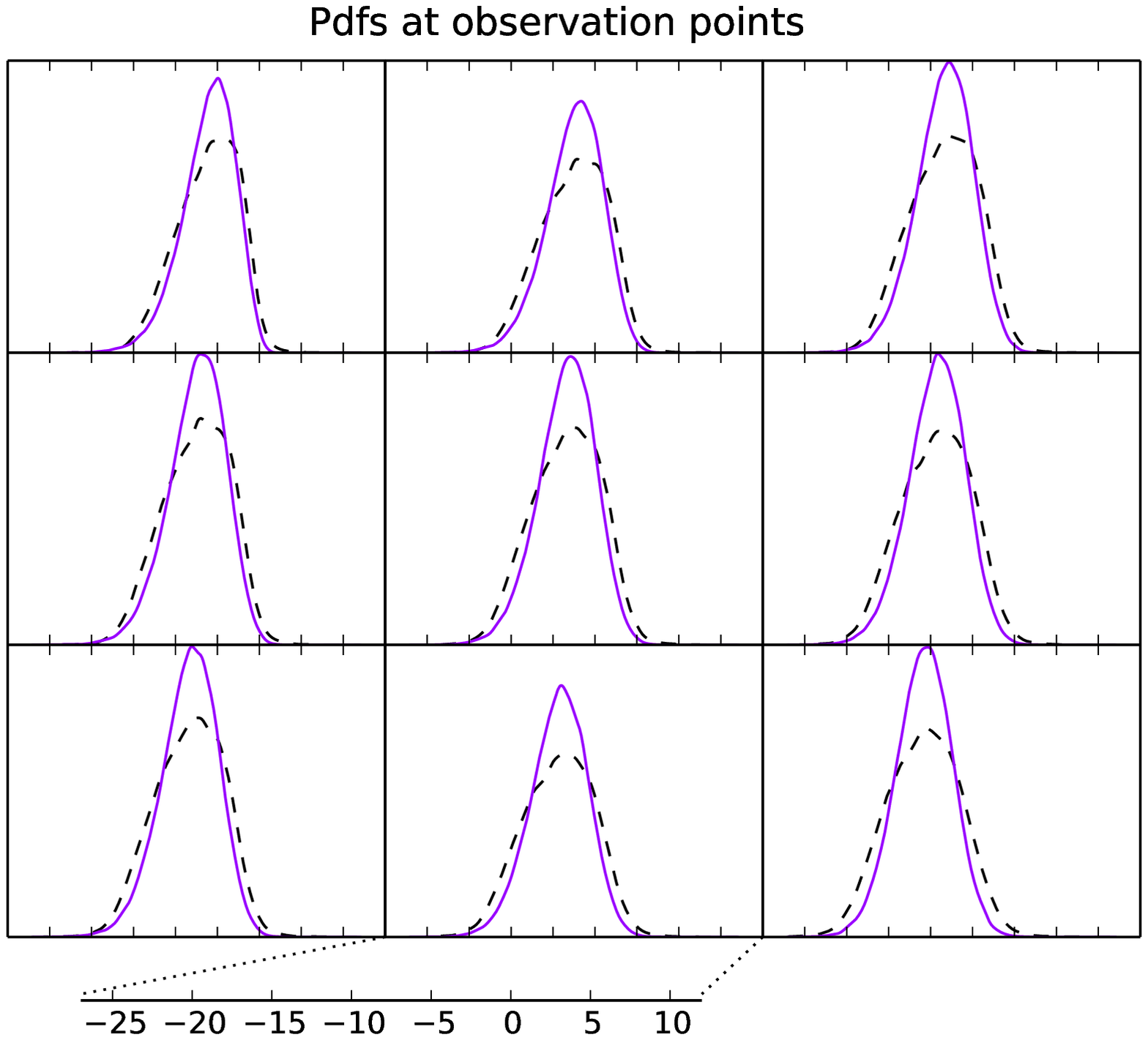}
\caption{Gaussian adaptation: Comparison of the pdfs of $\{u(\bx_i, \bxi)\}_{i=1}^{9}$ and
  $\{u^{\bA(\bx_i)}(\eta)\}_{i=1}^{9}$, where $\bx_i$, $i =
  1,...,9$ are the points of interest. The black dashed line
  corresponds to the original chaos expansion $u(\bx, \bxi)$, while the purple line
  indicates the adapted chaos exansion $u^{\bA(\bx)}(\eta)$.\label{fig:pdfs_gauss}}
\end{figure}

\begin{figure}[H]
\centering
\includegraphics[width =0.49\textwidth]{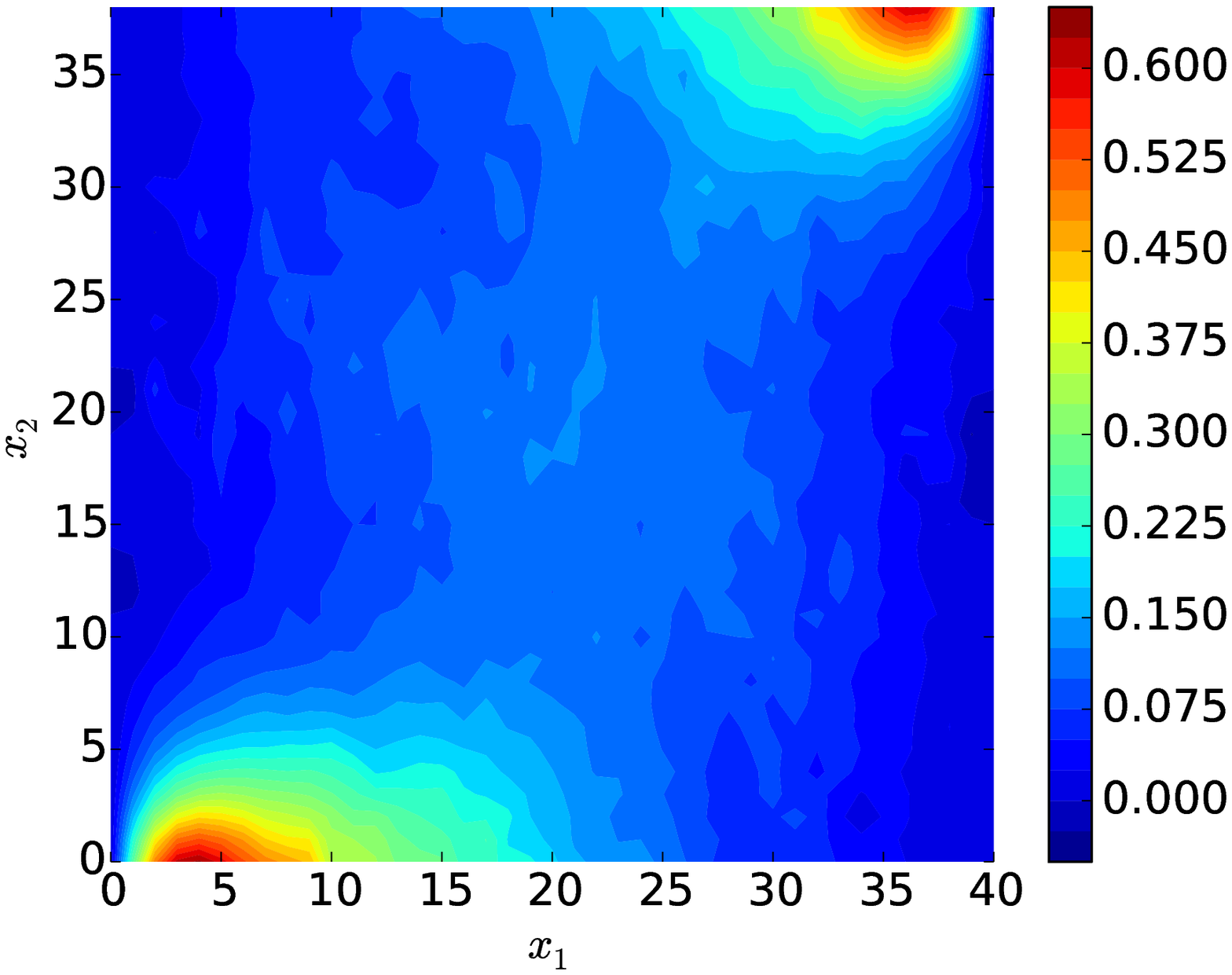}
\includegraphics[width =0.49\textwidth]{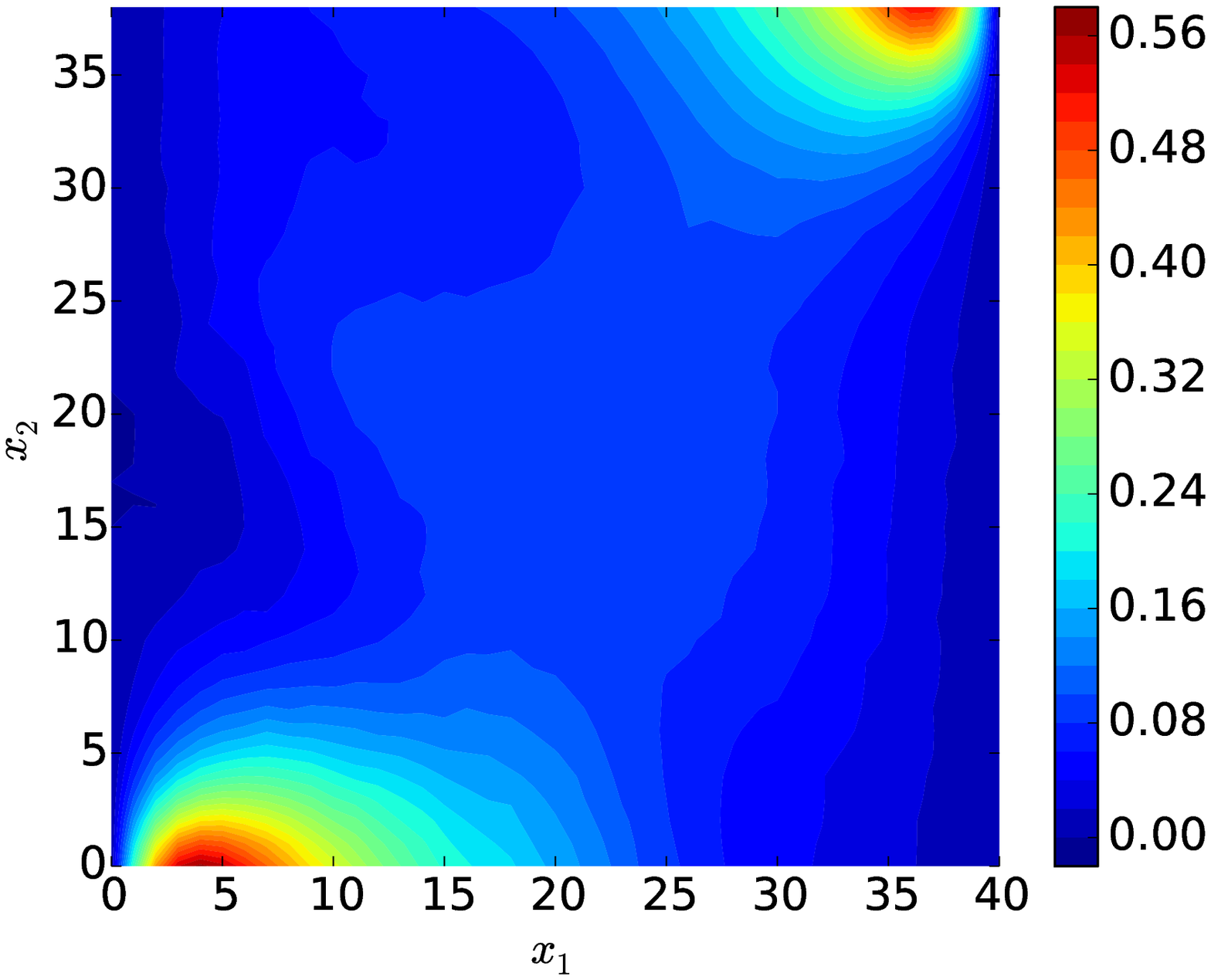}
\includegraphics[width =0.49\textwidth]{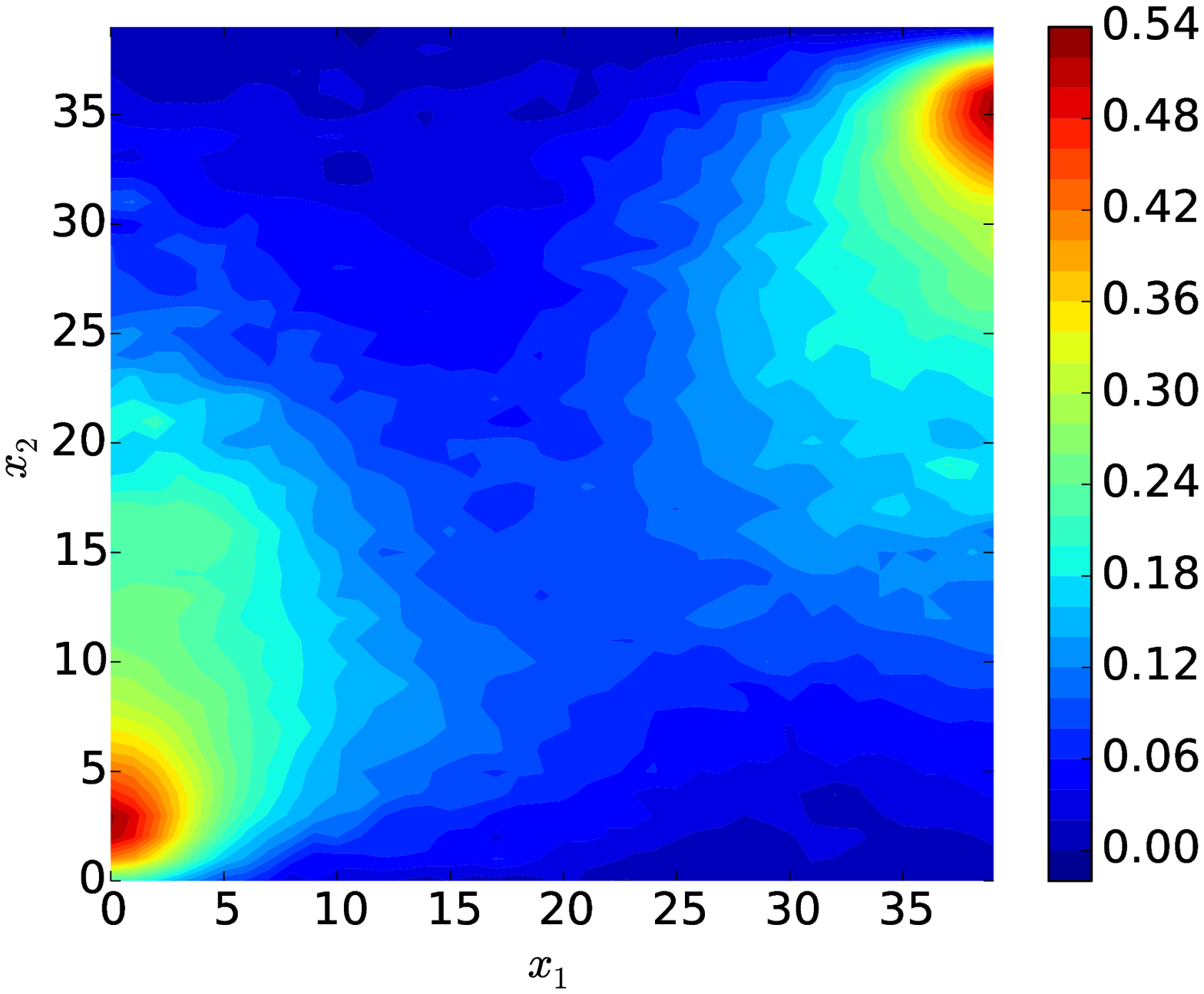}
\includegraphics[width =0.49\textwidth]{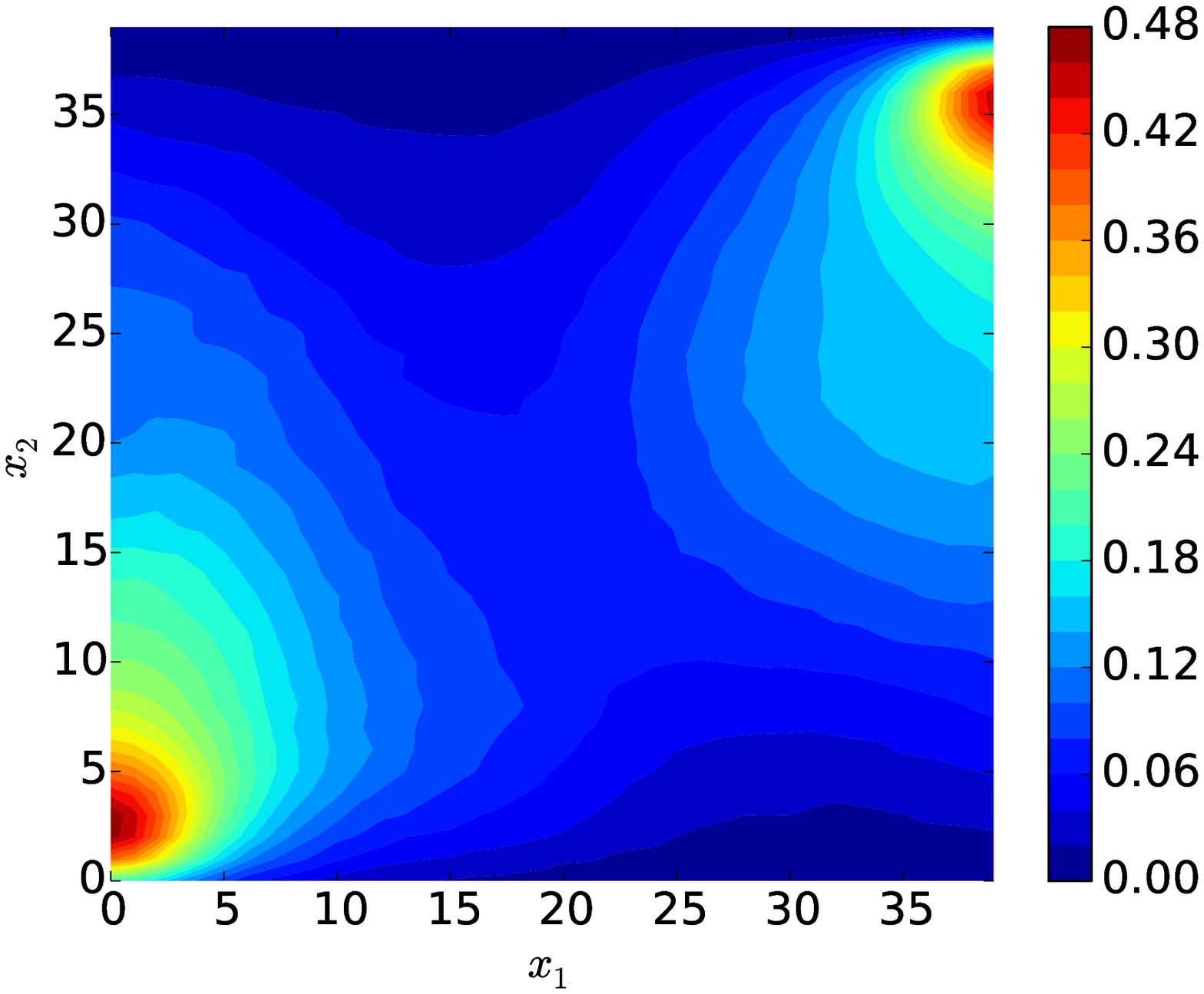}
\caption{Sample of velocity fields $\bv$ corresponding to $u(\bx,
  \bxi)$ (left column) and $u^{\bA(\bx)}(\eta)$ (right column). Top
  row shows $\bv_{x_1}$ and bottom row shows $\bv_{x_2}$. \label{fig:velocity_fields}}
\end{figure}

The coefficients in expression (\ref{eq:gauss_pce}) are shown in
Fig. \ref{fig:adapted_coeffs_elliptic}. As it seems by construction,
the finer scales of fluctuation that can be seen in the coefficients
of the full expansion, are merged within $\eta$ and are captured by
its distribution and its covariance kernel while the coefficients of
the adapted expansion display only the coarse
behavior. Analytically, it can be seen for instance (see Eq
\ref{eq:first_order}) that the first order coefficient is nothing but
the norm of the first order coefficients of the full expansion. The black dots indicate $9$
locations where a comparison of the
probability densities of $\{u^{\bA(\bx_i)}(\eta)\}_{i=1}^{9}$ and
  $\{u^{\bA(\bx_i)}(\eta)\}_{i=1}^{9}$ was
  performed, the results of which are shown in
  Fig. \ref{fig:pdfs_gauss}. The density functions of the two chaos
  expansions demonstrate good agreement among the two random
quantities, with those of the adapted expansions being slightly
more \emph{peaked} and with lighter tails, due to the relatively large
number of terms being essentially neglected via projection. Note
that while the initial series consists of $1771$ terms, the adapted
series consists of only $3$! At last, Fig. \ref{fig:velocity_fields} shows an example of
realizations of the velocity fields
\begin{equation}
\bv  = -\bk(\bx, \bxi) \nabla u(\bx, \bxi), \ \ \bx\in D
\end{equation}
computed for both $u(\bx,\bxi)$ and $u^{\bA(\bx)}(\eta)$.

\begin{figure}[h]
\centering
\includegraphics[width =0.32\textwidth]{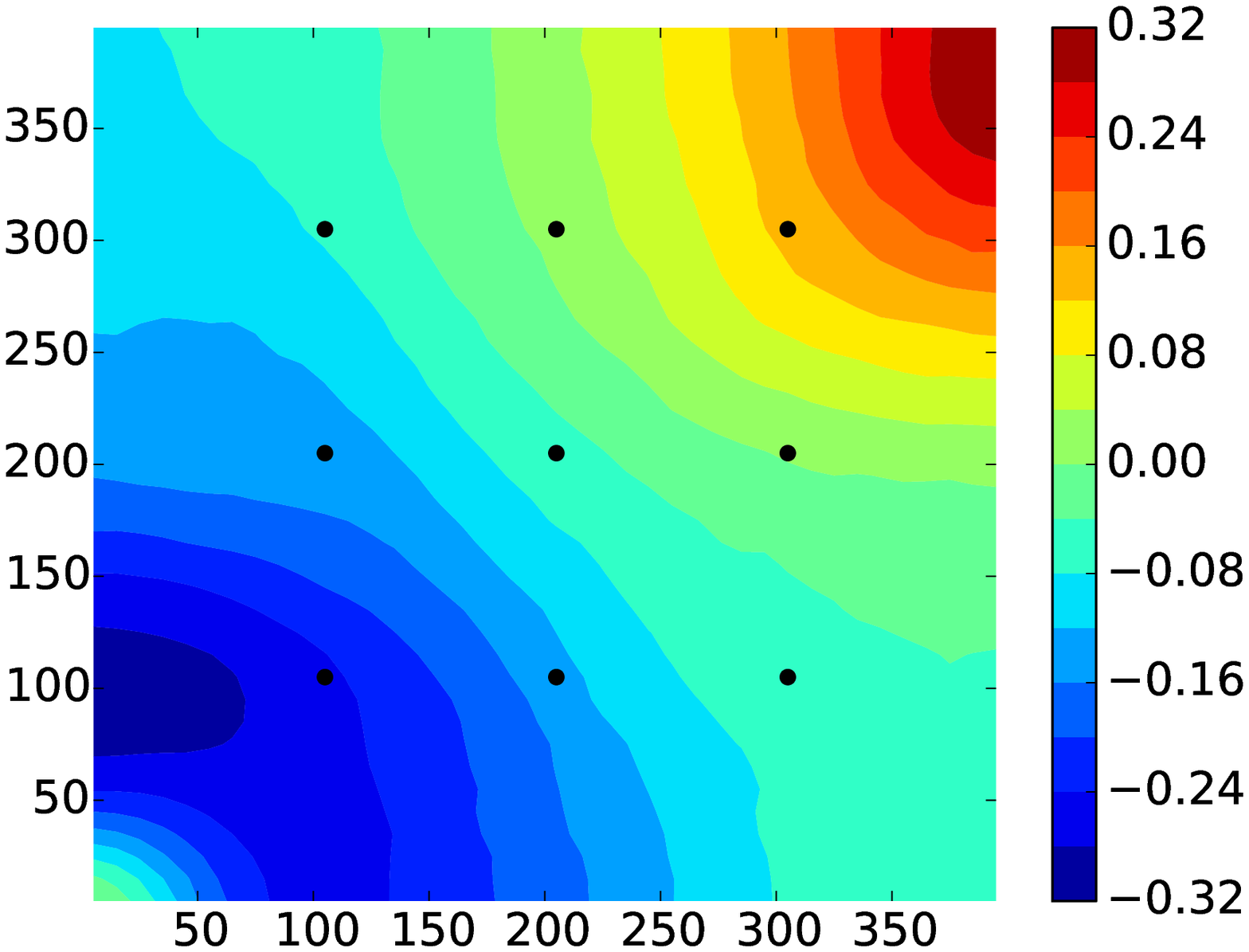}
\includegraphics[width =0.32\textwidth]{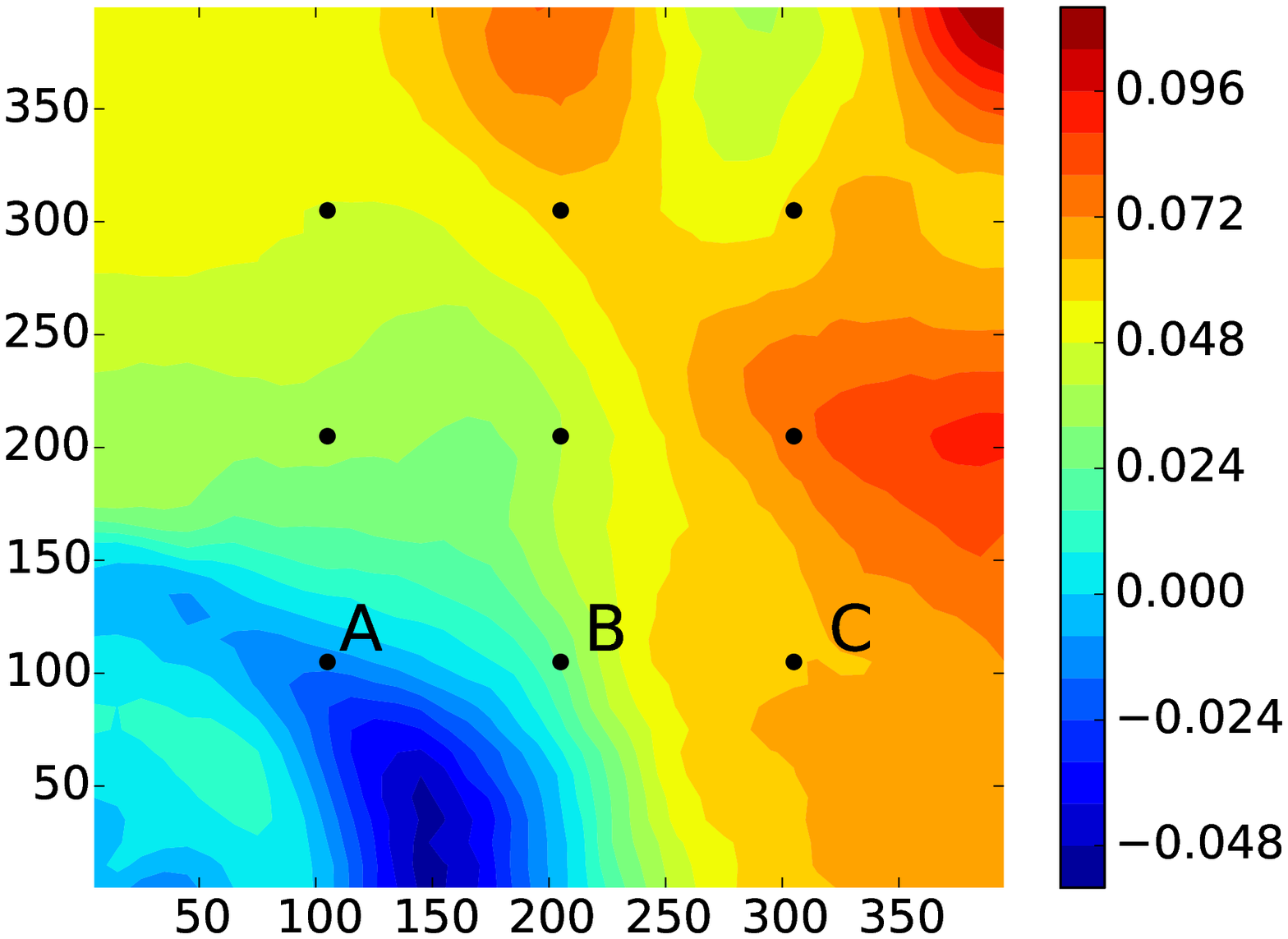}
\includegraphics[width =0.32\textwidth]{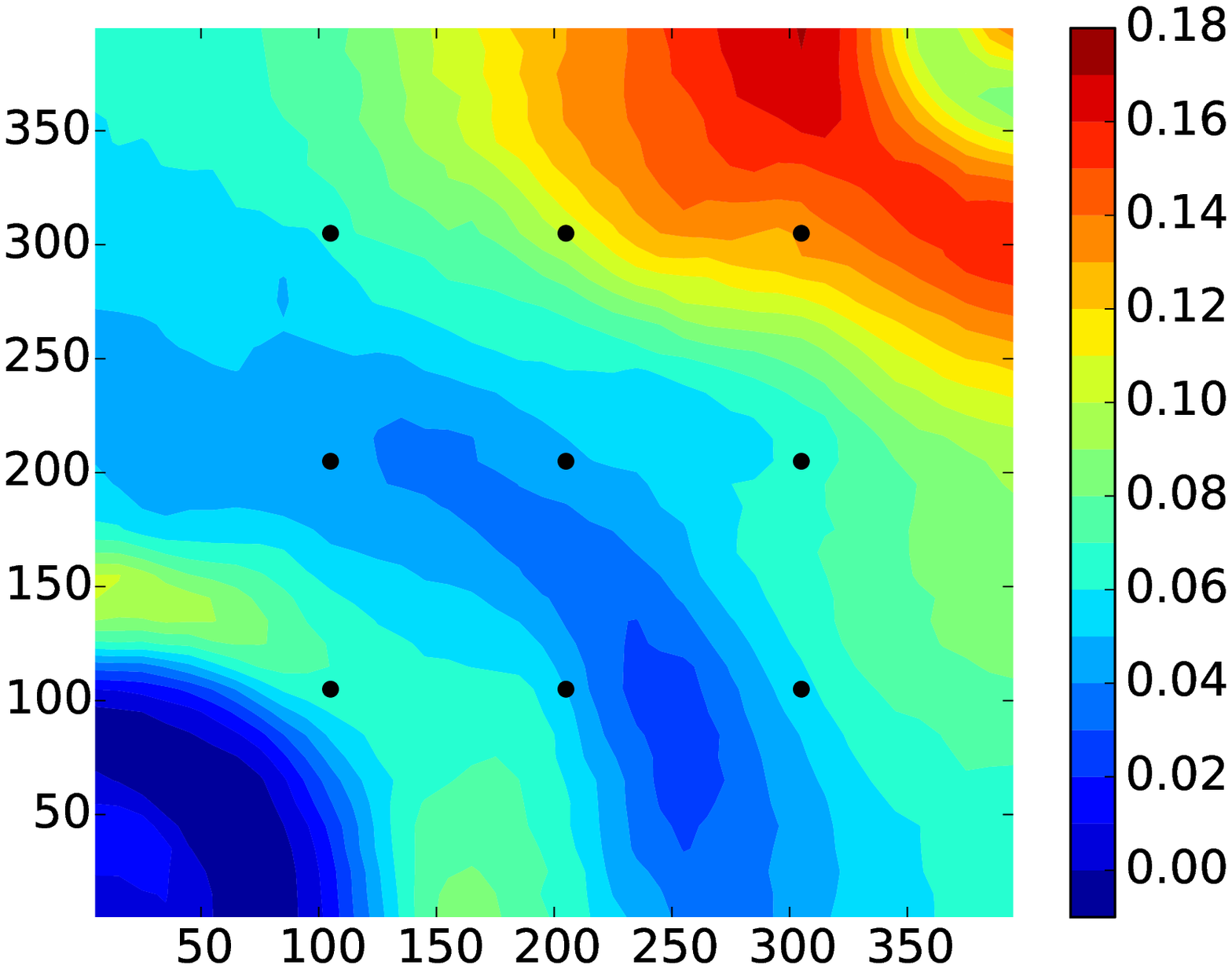}
\includegraphics[width =0.32\textwidth]{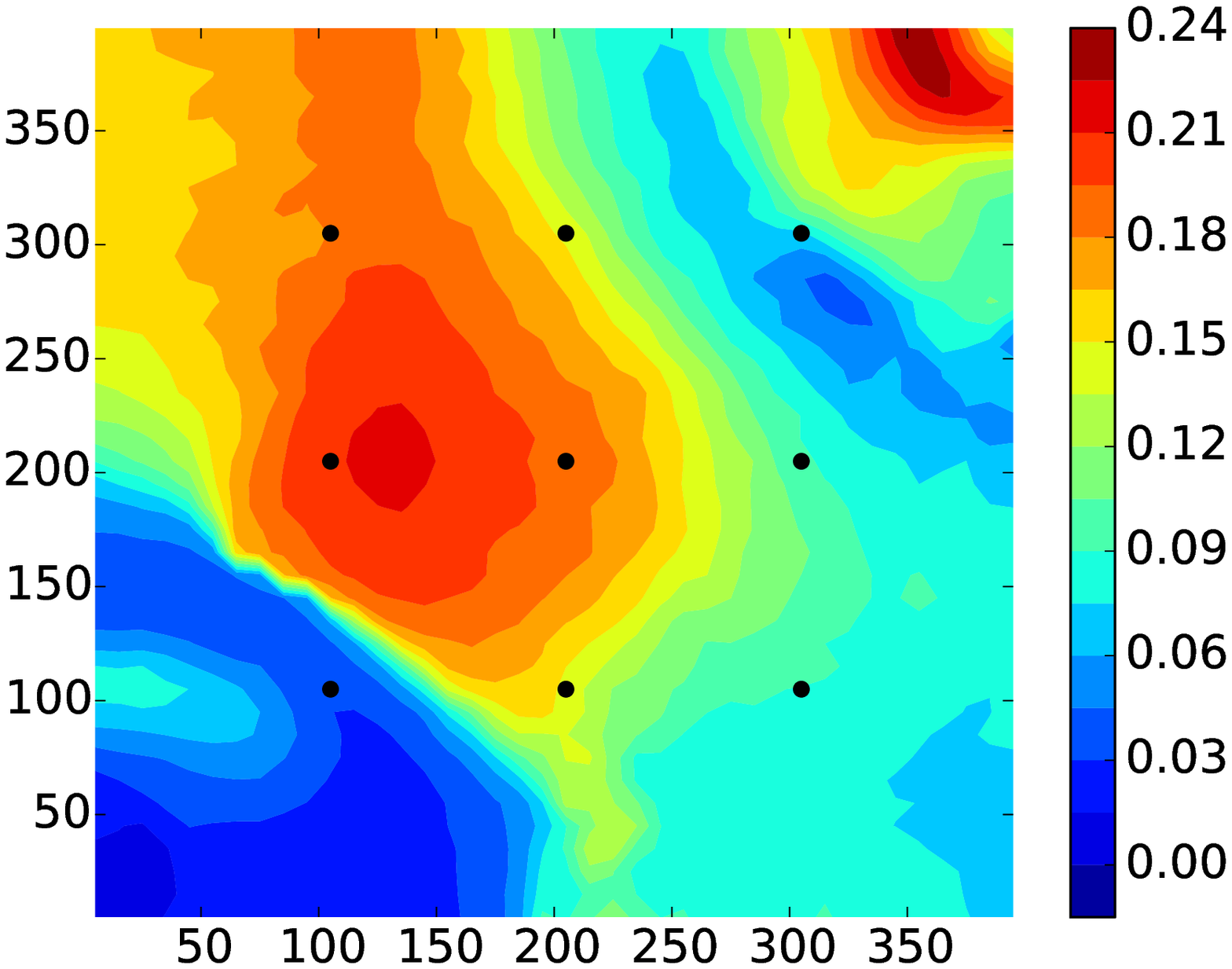}
\includegraphics[width =0.32\textwidth]{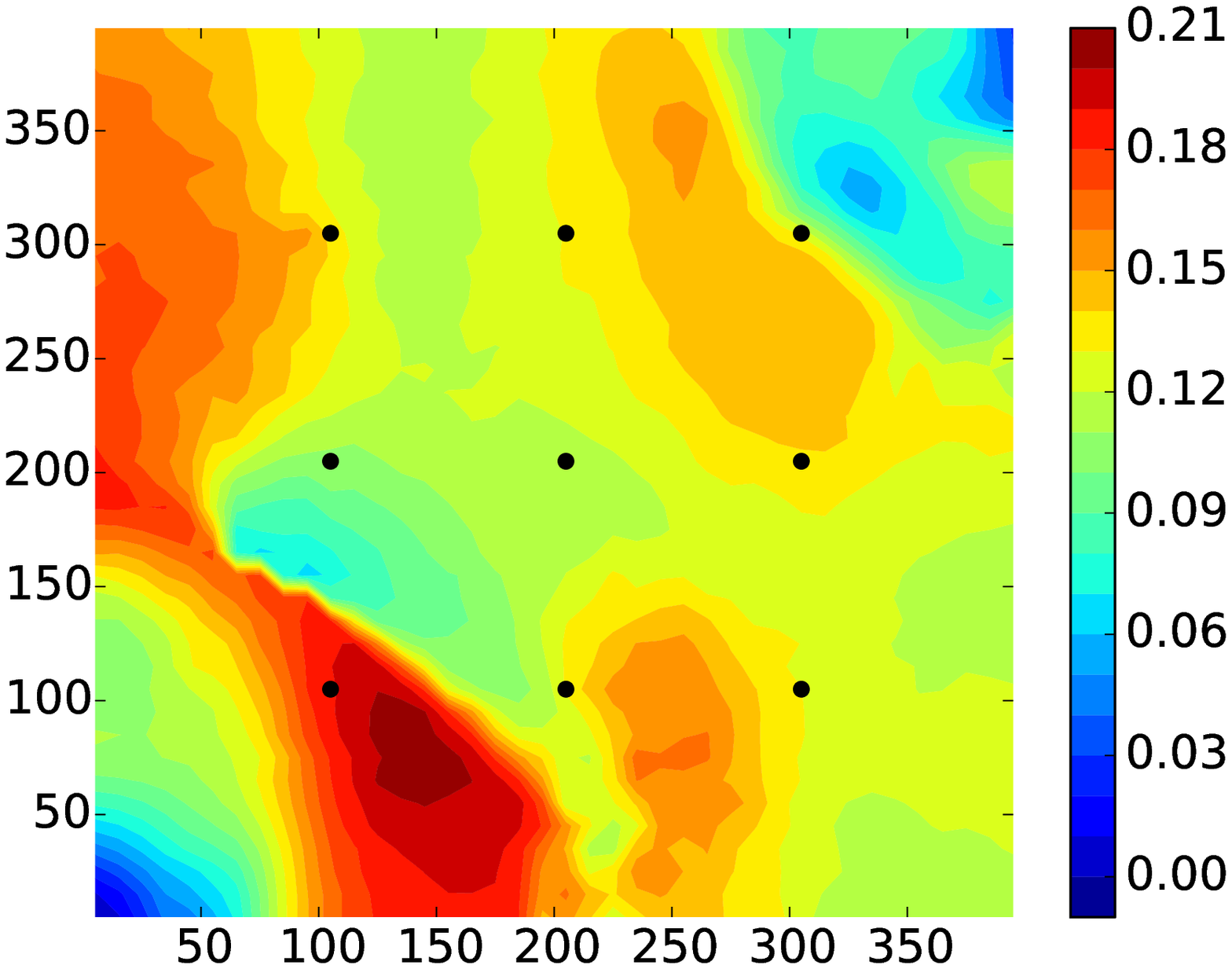}
\caption{Coefficients $u_{ii}^{\bA(\bx)}$, $i = 1,...,5$ of the
  second-order $5$-dimensional quadratic adaptation. \label{fig:coeffs_quadr}}
\end{figure}

\begin{figure}[t]
\centering
\includegraphics[width = 0.70\textwidth]{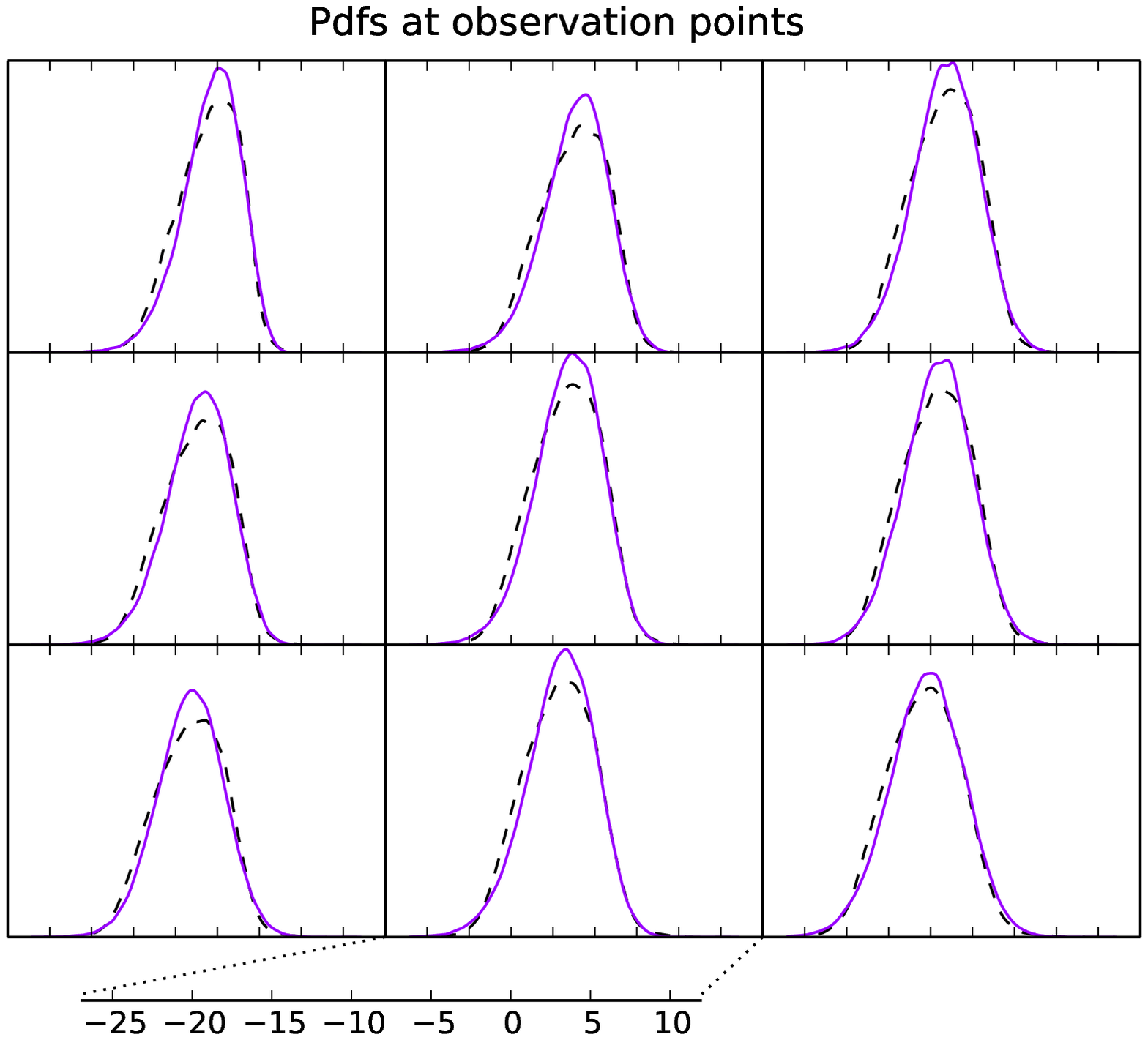}
\caption{Quadratic adaptation: Comparison of the pdfs of $\{u(\bx_i, \bxi)\}_{i=1}^{9}$ and
  $\{u^{\bA(\bx_i)}(\eta)\}_{i=1}^{9}$, where $\bx_i$, $i =
  1,...,9$ are the points of interest. The black dashed line
  corresponds to the original chaos expansion $u(\bx, \bxi)$, while the purple line
  indicates the adapted chaos exansion $u^{\bA(\bx)}(\eta)$.\label{fig:pdfs_quadr}}
\end{figure}

\subsubsection{Quadratic adaptation}

\begin{figure}[h]
\centering
\includegraphics[width = 0.49\textwidth]{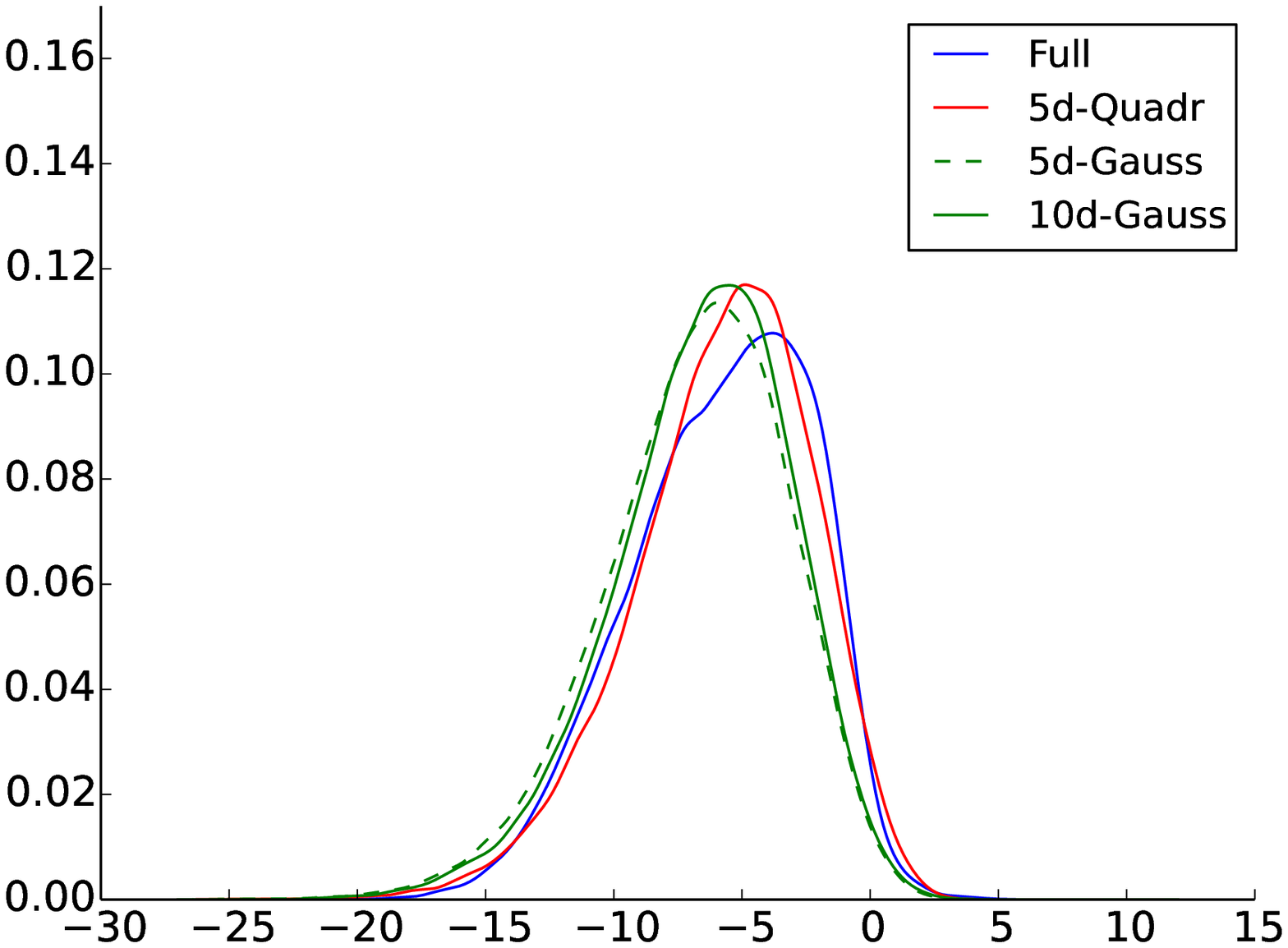}
\includegraphics[width = 0.49\textwidth]{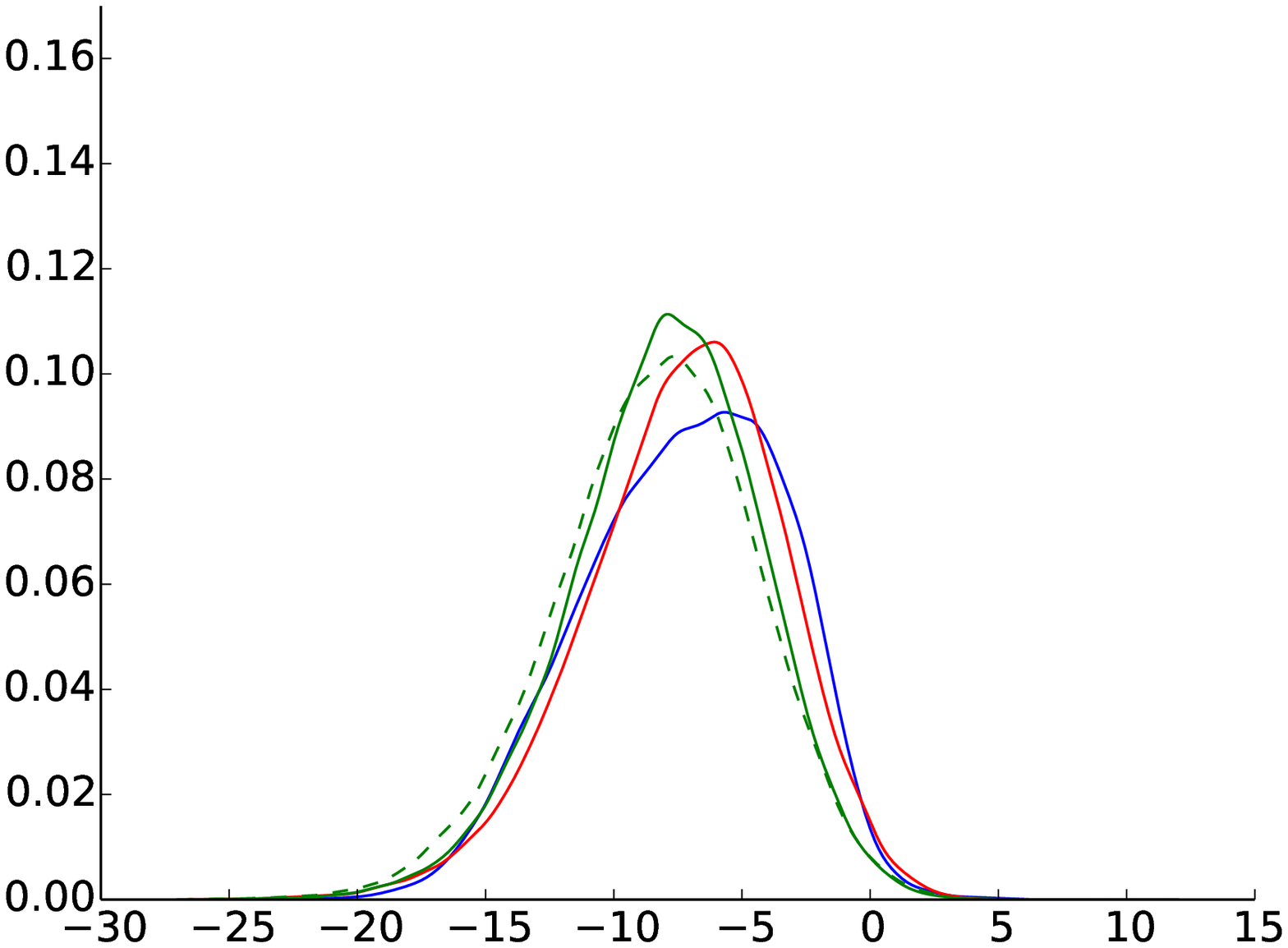}
\includegraphics[width = 0.49\textwidth]{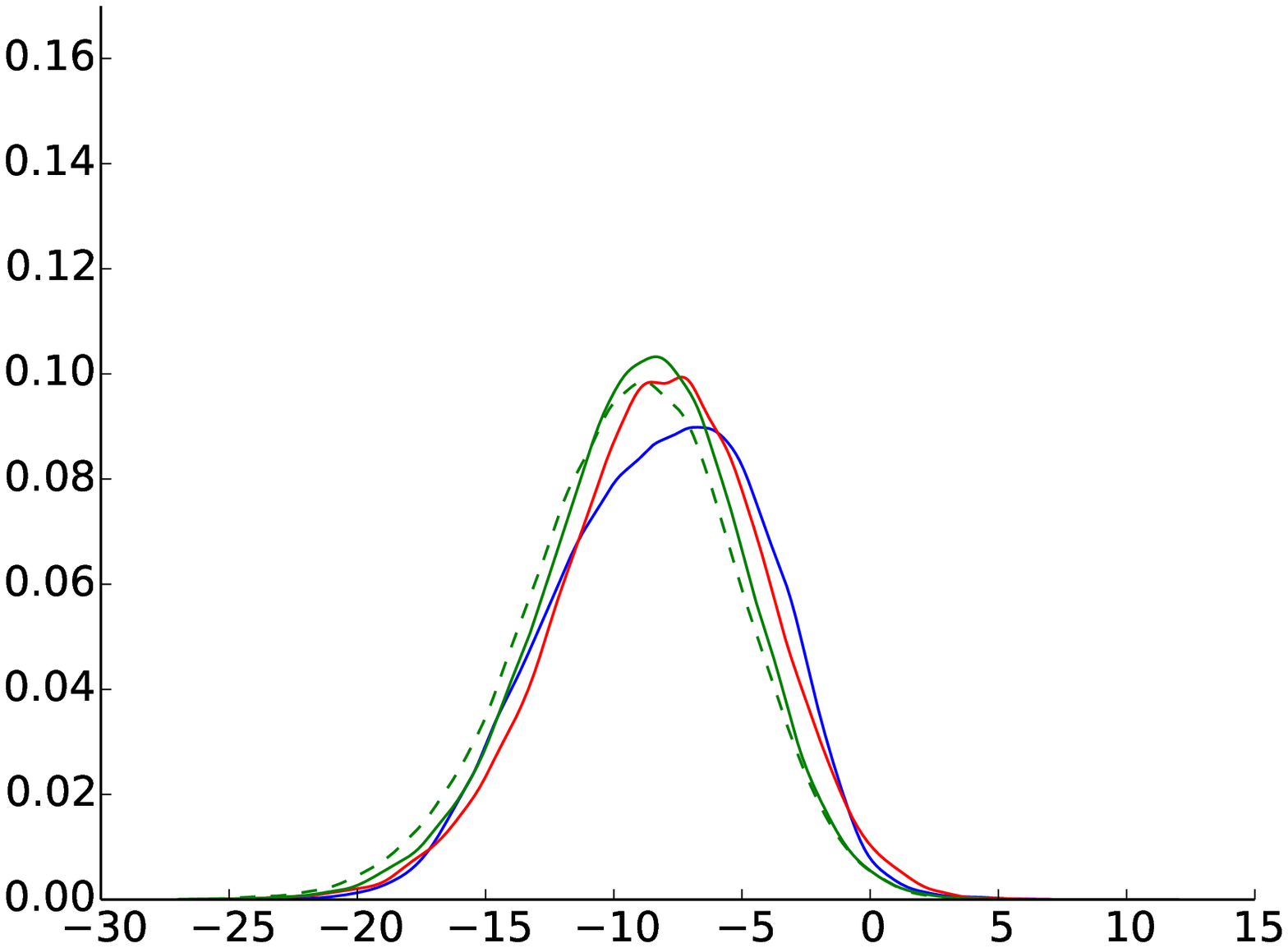}
\caption{Comparison of the Gaussian and quadratic adaptation for
  different choices of the dimension at the three locations A (up
  left), B (up right) and C (bottom). \label{fig:comparison}}
\end{figure}

Next we construct a $5$-dimensional quadratic adaptation, that is
\begin{equation}
\label{eq:quadr_pce}
u^{\bA(\bx)}(\eta) = u_0^{\bA(\bx)} + \sum_{i=1}^5u_i^{\bA(\bx)}
\eta_i + \sum_{i=1}^5 u_{ii}^{\bA(\bx)}\frac{(\eta_i^2 - 1)}{\sqrt{2}},
\end{equation}
where $\bA$ is constructed such that is satisfies
Eq. (\ref{eq:quadr_iso}). The quadratic adaptation can be seen
\cite{tipireddy} to have exactly the same sum of the polynomial terms
up to order two with those of the full expansion without essentially discarding
any terms via projection and the second order
coefficients $u_{ii}^{\bA(\bx)}$ are
proportional to the eigenvalues of $\bS$ (shown in
Fig. \ref{fig:coeffs_quadr}). Due to the small order of
our full expansion, this might be expexted to adapt better than the
Gaussian adaptation, given also that we include an expansion with higher dimensionality
than the 1-dimensional Gaussian adaptation. Comparison of the density functions at
$9$ locations with those of the full expansion can be seen in Fig. \ref{fig:pdfs_quadr} which
verifies our argument and shows particularly a better agreement
between the tails of the two pdfs. The two adaptations are
also compared with themselves at three locations, labeled A, B and C (shown in
$u_{22}^{\bA(\bx)}$ - Fig. \ref{fig:coeffs_quadr}) and the results are shown
in Fig. \ref{fig:comparison} where this time the distributions of
a $5$- and $10$-dimensional Gaussian adaptations are plotted together
with the $5$-dimensional quadratic adaptation. Again, good agreement
can be seen between the $3$ pdfs with the quadratic
adaptation being slightly closer to the true distribution. Another
interesting characteristic here is that as we keep increasing the
dimensionality of the expansion by adding only terms of
$1$-dimensional series, that is, dropping polynomial terms that depend
jointly on two or more $\xi$'s, the contribution is small and it seems
that the joint terms are essential in achieving a full distributional
equality (in fact the equality will be almost surely). However, the
agreement shown here can be considered sufficient for estimating
various statistics of interest. Further investigation in order to
identify the suitable rotations to optimally adapt the expansion while
maintaining low dimensionality could be pursued by minimizing an error
function of the form
(\ref{eq:projection_error}),(\ref{eq:projection_error_norm}) or within
the context of active subspaces \cite{constantine} and is beyond the scope of this work.

\subsubsection{Adaptation on expansion with random coefficients}

\begin{figure}[H]
\centering
\includegraphics[width = 0.32\textwidth]{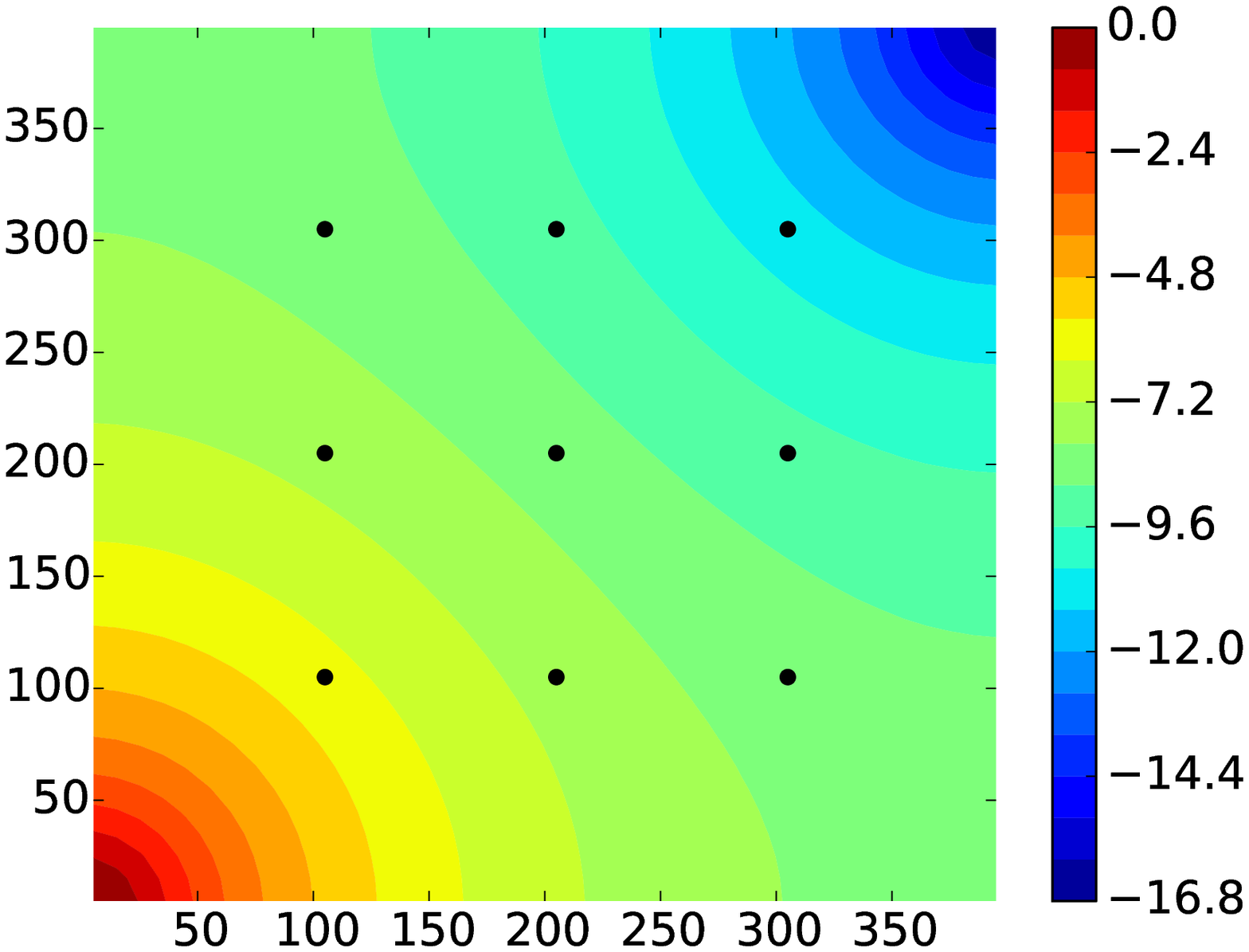}
\includegraphics[width = 0.32\textwidth]{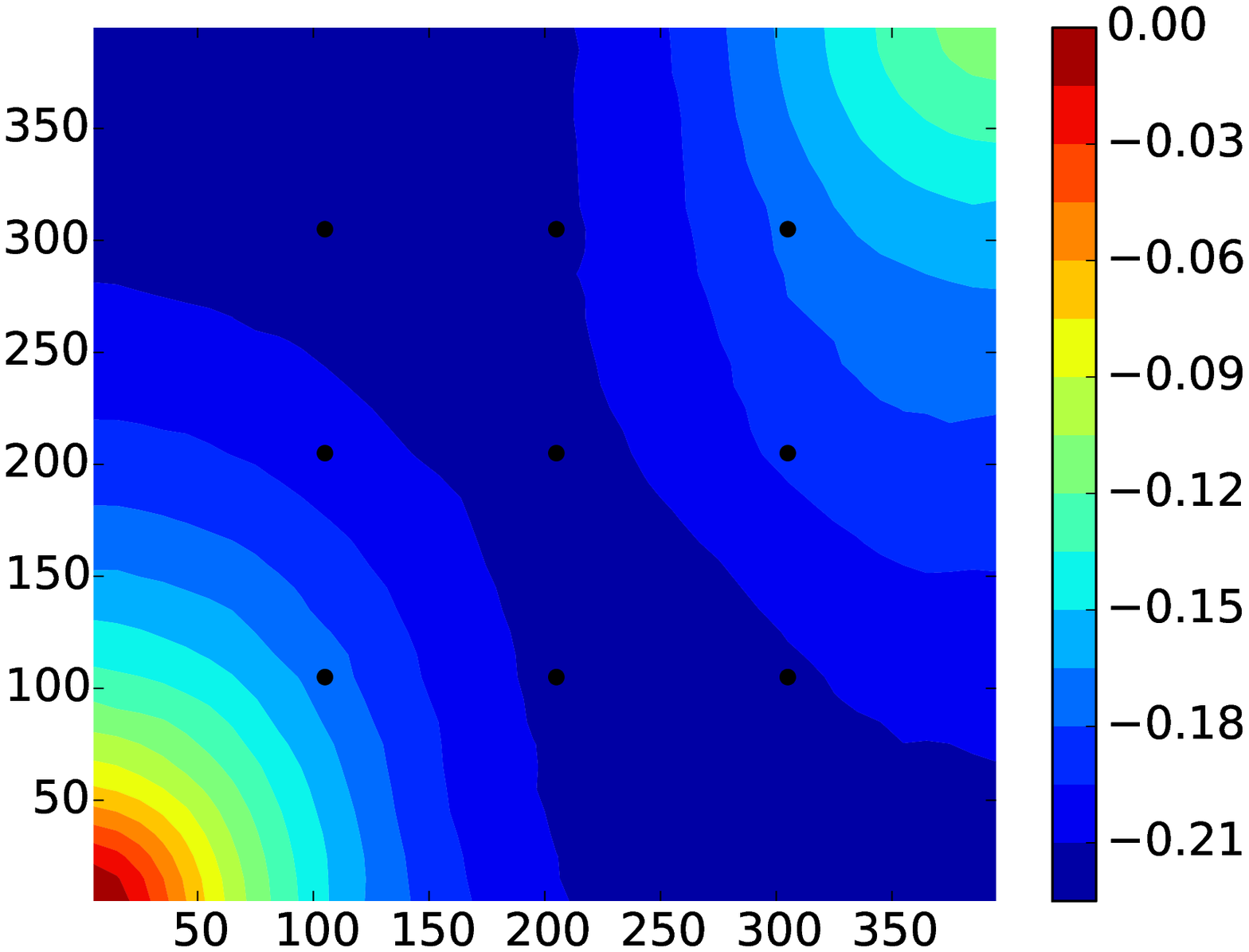}
\includegraphics[width = 0.32\textwidth]{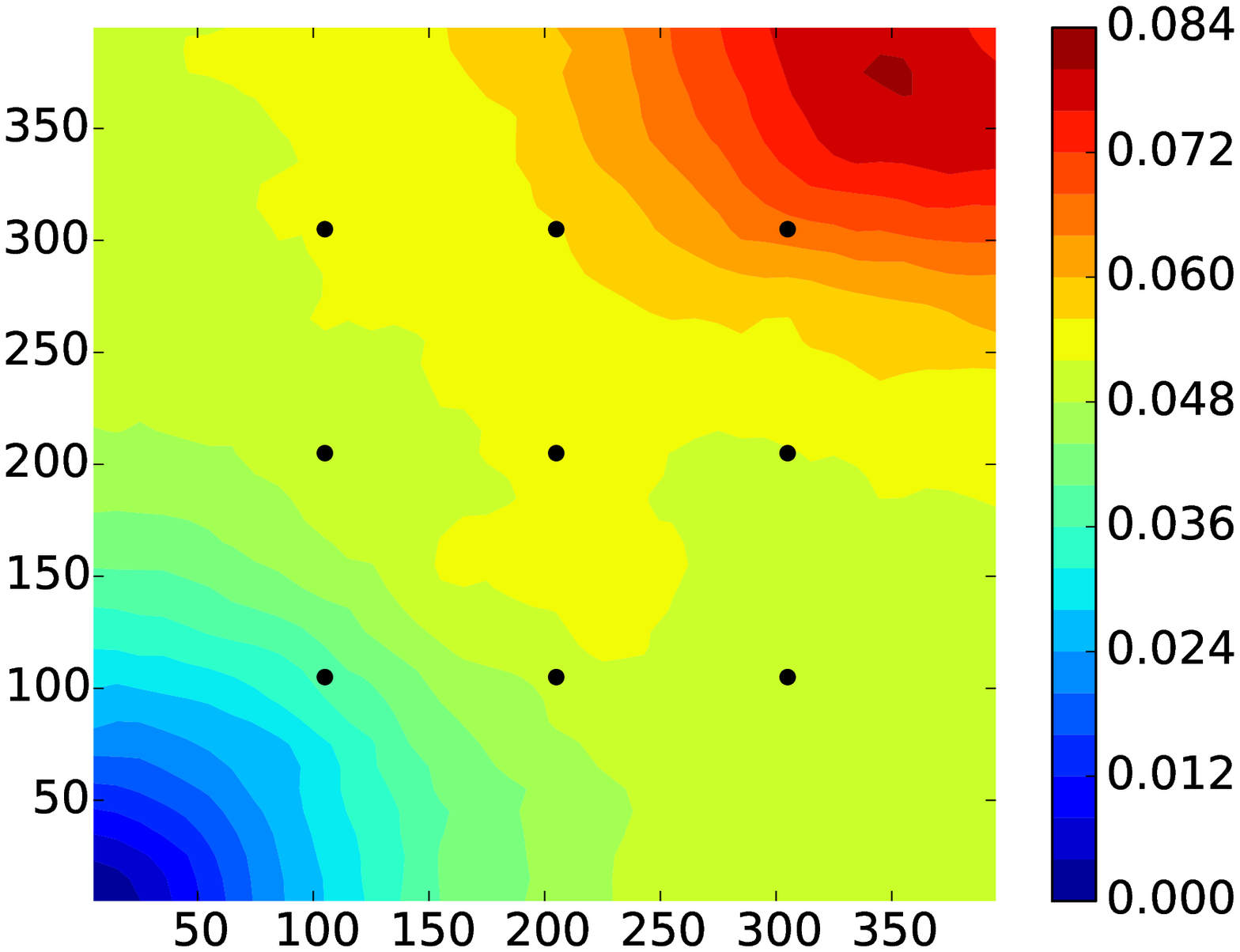}
\caption{Expectation of the random coefficients $\E\left[\bU^{\bA(\bx, \bzeta)}_{i}\right]$, $i = 0,1,2$
  of the second-order one-dimensional Gaussian adaptation of $u(\bx,
  \hat{\bxi}, \hat{\bzeta})$
  coefficients. As expected $\E\left[\bU^{\bA(\bx, \bzeta)}_{0}\right]
  = u_0^{\bA(\bx)} = u_{0}(\bx)$.\label{fig:adapted_rand_coeffs}}
\end{figure}

\begin{figure}[t]
\centering
\includegraphics[width = 0.70\textwidth]{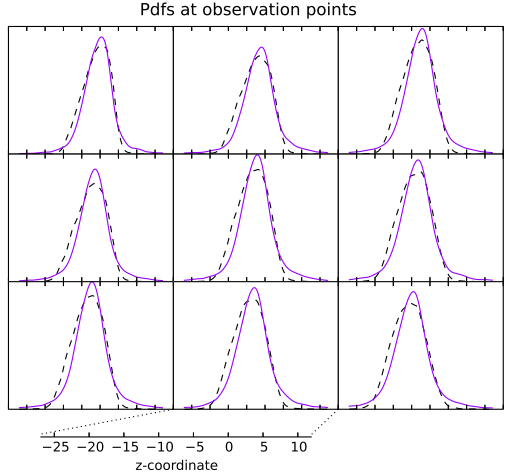}
\caption{Adaptation with random coefficients: Comparison of the pdfs of $\{u(\bx_i, \bxi)\}_{i=1}^{9}$ and
  $\{u^{\bA(\bx_i)}(\eta)\}_{i=1}^{9}$, where $\bx_i$, $i =
  1,...,9$ are the points of interest. The black dashed line
  corresponds to the original chaos expansion $u(\bx, \bxi)$, while the purple line
  indicates the adapted chaos exansion $u^{\bA(\bx)}(\eta)$.\label{fig:random_pdfs}}
\end{figure}

At last, we test our approach on a reduced chaos expansion with random
coefficients as given in Eq. (\ref{eq:reduced_chaos_trunc}) where we
have arbitrarily chosen $\hat{\bxi} = (\xi_1, \xi_2,\xi_3, \xi_4)$
and $\hat{\bzeta} = (\xi_5, ..., \xi_{20})$. Although this can be seen
as a way to separate fine and coarse random fluctuations (and this is
in fact our motivation behind this construction, as was introduced in
\cite{soize_reduced}), we do not claim this to be the case here since
the influence of the first four $\xi$'s is not necessarily significantly dominating
in this particular permeability model due to the relatively low
correlation lengths $\ell_1$, $\ell_2$. We restrict ourselves in
presenting only how the adaptation methodology applies in such a case and
leave the costruction of a more illustrating example for another paper. The
$4$-dimensional third-order expansion with respect to $\hat{\bxi}$
with the coefficients being dependent on $\hat{\bzeta}$ is, 
\begin{eqnarray}
\begin{array}{c} u(\bx,\hat{\bxi}, \hat{\bzeta}) = \bU_0(\bx, \hat{\bzeta}) +
\sum_{i=1}^{16}\bU_{i\epsilon_1}(\bx, \hat{\bzeta})\psi_{i\epsilon_1}(\xi_i)
+ \\ + \sum_{\balpha,|\balpha| = 2} \bU_{\balpha}(\bx, \hat{\bzeta})
\psi_{\balpha}(\hat{\bxi}) + \sum_{\balpha, |\balpha|=3}
\bU_{\balpha}(\bx, \hat{\bzeta}) \psi(\hat{\bxi}) \end{array}
\end{eqnarray}
where  $\bU_{\balpha}$ are given in Eq. (\ref{eq:random_coeffs}). We
use again the Gaussian adaptation scheme to construct a
$1$-dimensional second order expansion
\begin{equation}
u^{\bA(\bzeta)}(\hat{\bxi}, \eta) = \bU^{\bA(\bx,\bzeta)}_{0} +
\bU^{\bA(\bx, \bzeta)}_{1}\eta + \bU_2^{\bA(\bx, \bzeta)} \frac{\eta^2
- 1}{\sqrt{2}}
\end{equation}
Note here that only the $4$-dimensional $\hat{\bxi}$ has been merged
into a $1$-dimensional $\eta$ while the influence of all
dimensions incorporated in $\hat{\bzeta}$ is present both in the
coefficients and in the polynomials through the isometry $\bA(\bx,
\bzeta)$. The estimated expected values of the adapted
coefficients $\bU_i^{\bA(\bx, \bzeta)}$ are shown in
Fig. \ref{fig:adapted_rand_coeffs}. The density functions shown in
Fig. \ref{fig:random_pdfs} are constructed by simultanesously sampling from $\hat{\bzeta}$ and
$\hat{\bxi}$, then evaluating $\bU_{\balpha}$
and $\bA$ based on the values of $\hat{\bzeta}$ and subsequently computing the coefficients of the adapted
expansion that at last are evaluated on $\hat{\bxi}$. Again very good
agreement is observed when compared to the pdfs of the full
expansions. Since we have only applied the basis rotation on $4$
dimensions, upon re-expanding the series, this is a $17$-dimensional
expansion.

%% file: example_geometric.tex
\subsection{Infinite dimensional expansion with geometric series
  coefficients}

We consider a simple random process that is written as a function of an
infinite number of Gaussians given as
\begin{equation}
\label{eq:initial_chaos}
u(x, \bxi) = \sum_{n = 1}^{\infty} b_n(x) \xi_n +
\left(\sum_{n=1}^{\infty} b_n(x) \xi_n\right)^2
\end{equation}
where 
\begin{equation}
b_n(x) = x^{(n-1) / 2}, \ \ x \in (-1, 1).
\end{equation}
Since the sum of coefficients $b_n$ is square summable with
$\sum_{n=1}^{\infty} b_n^2 = \frac{1}{1 - x}$, then $u(x,\xi) <
+\infty$ a.s. for $|x| < 1$ with $\sum b_n\xi_n \sim
\calN(0, \frac{1}{1-x})$ and the variance of the summand blows up for
$x \to \pm 1$. We apply the $1$-dimensional Gaussian adaptation which
consists of transforming $\bxi$ to 
\begin{equation}
\eta = \frac{1}{\left(\sum_{n=1}^{\infty}b_n(x)^2\right)^{1/2}}\sum_{n=1}^{\infty} b_n(x) \xi_n
\end{equation}
and using expressions (\ref{eq:first_order}) and (\ref{eq:second_order}) we
take 
\begin{equation}
\label{eq:true_adapted}
u(x, \eta)  = u_1(x) \eta + u_2(x)\frac{(\eta^2 - 1)}{\sqrt{2}}
\end{equation}
where 
\begin{eqnarray}
\begin{array}{l}
u_1(x) = \frac{1}{\sqrt{1 - x}}\\
u_2(x) = \frac{1}{1+x} +
  \sqrt{2}\frac{x}{1 - x^2}.
\end{array}
\end{eqnarray}

\begin{figure}[t]
%\centering
\includegraphics[width = 0.32\textwidth]{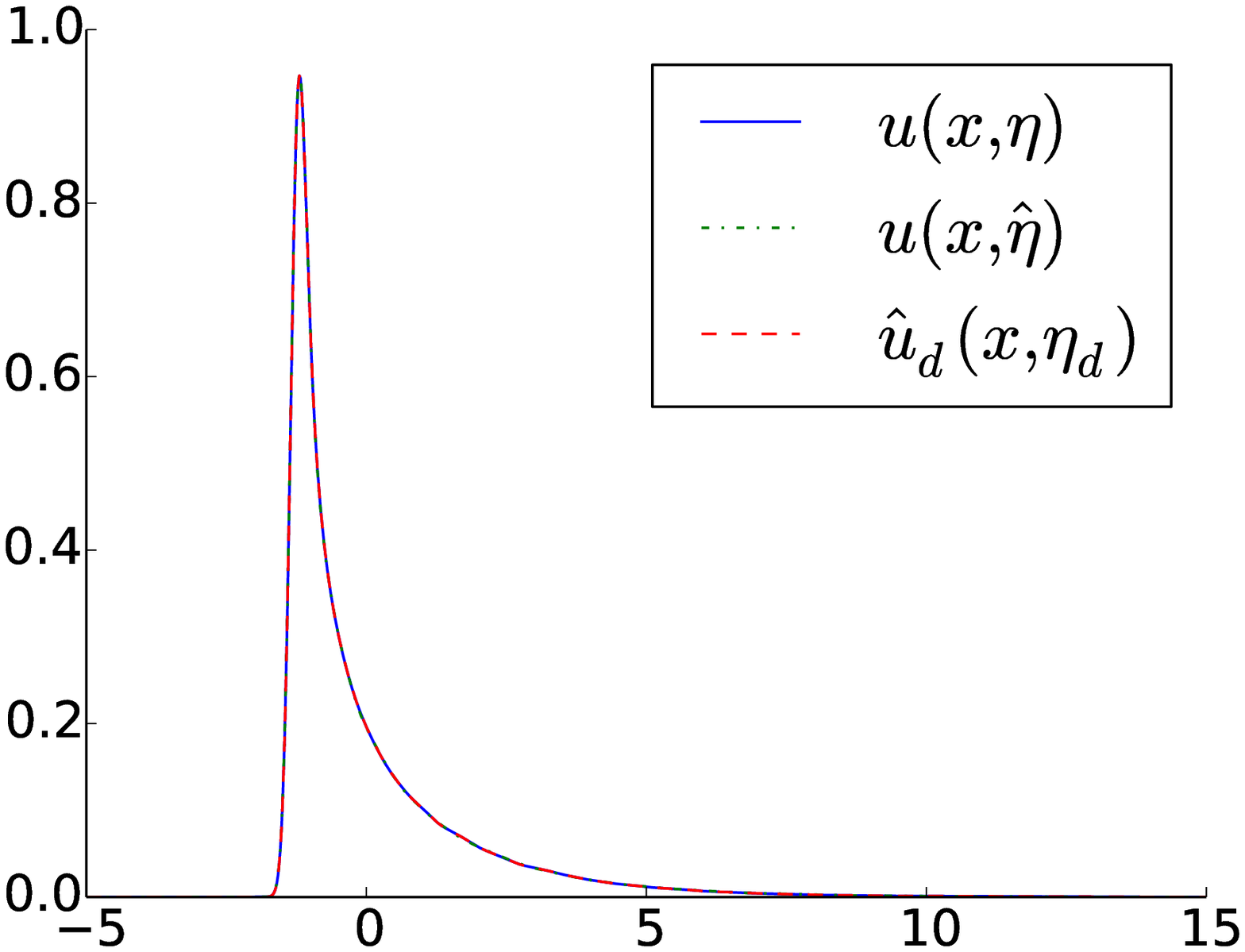}\\
\includegraphics[width = 0.32\textwidth]{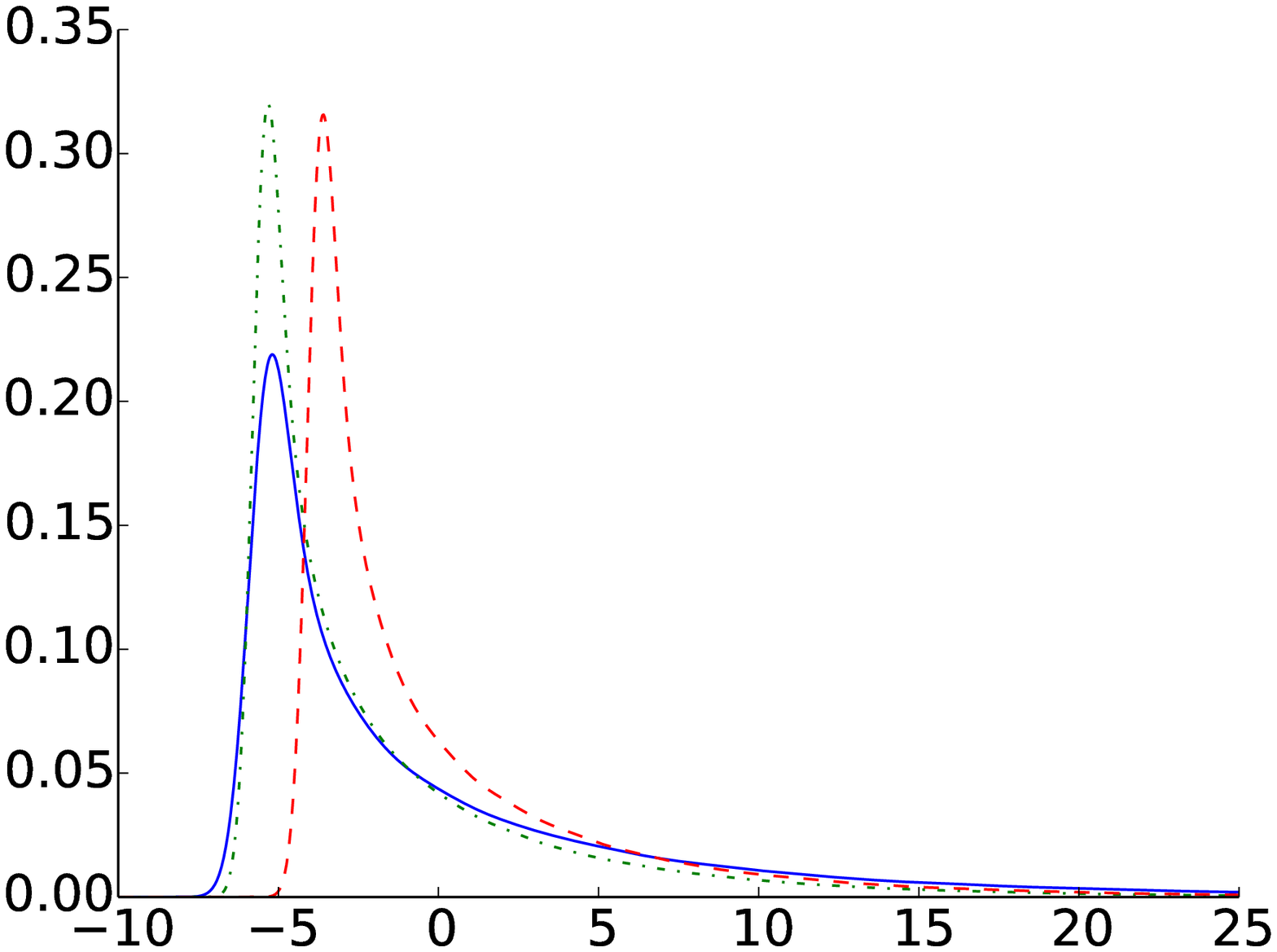}
\includegraphics[width = 0.32\textwidth]{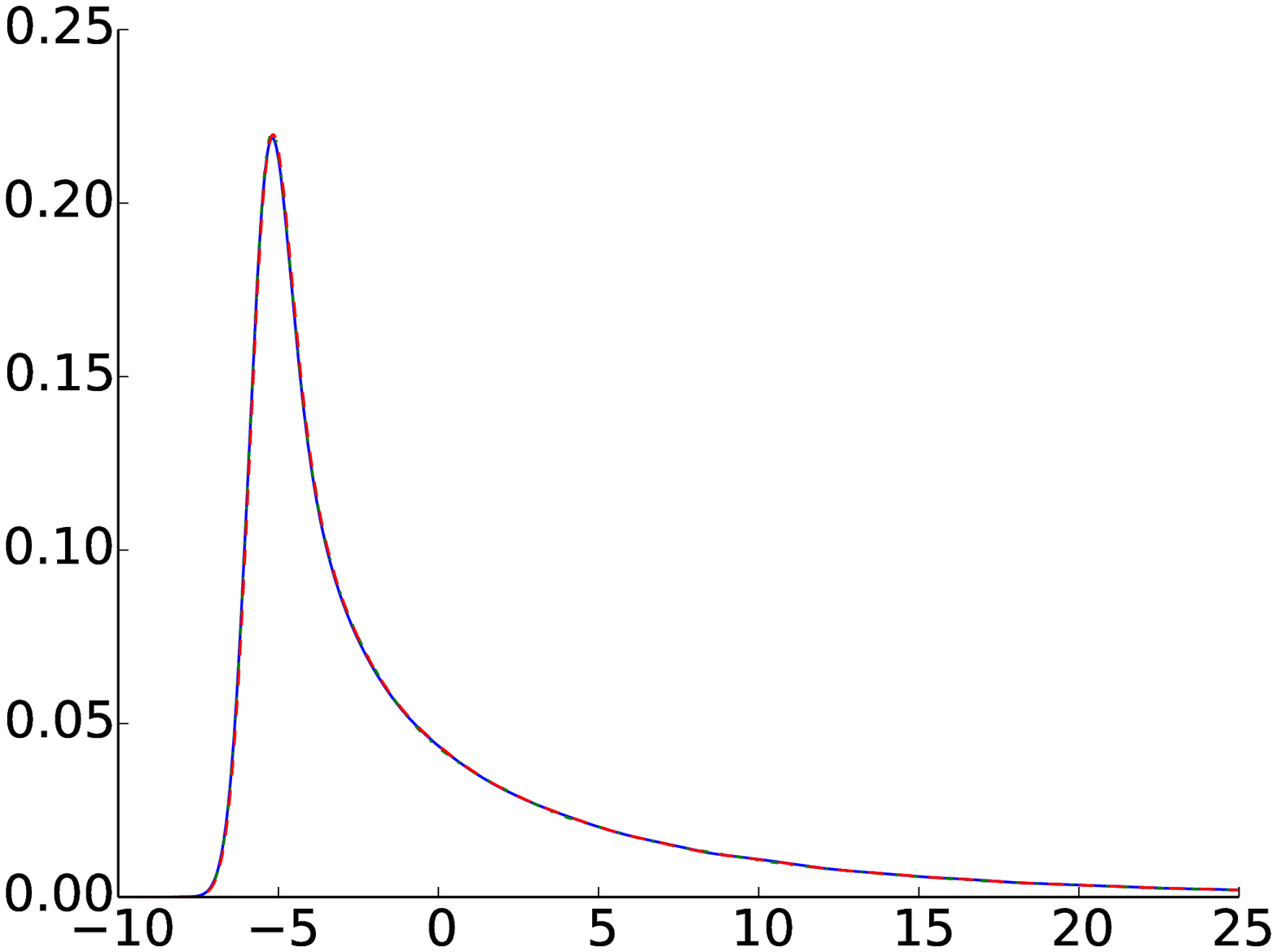}\\
\includegraphics[width = 0.32\textwidth]{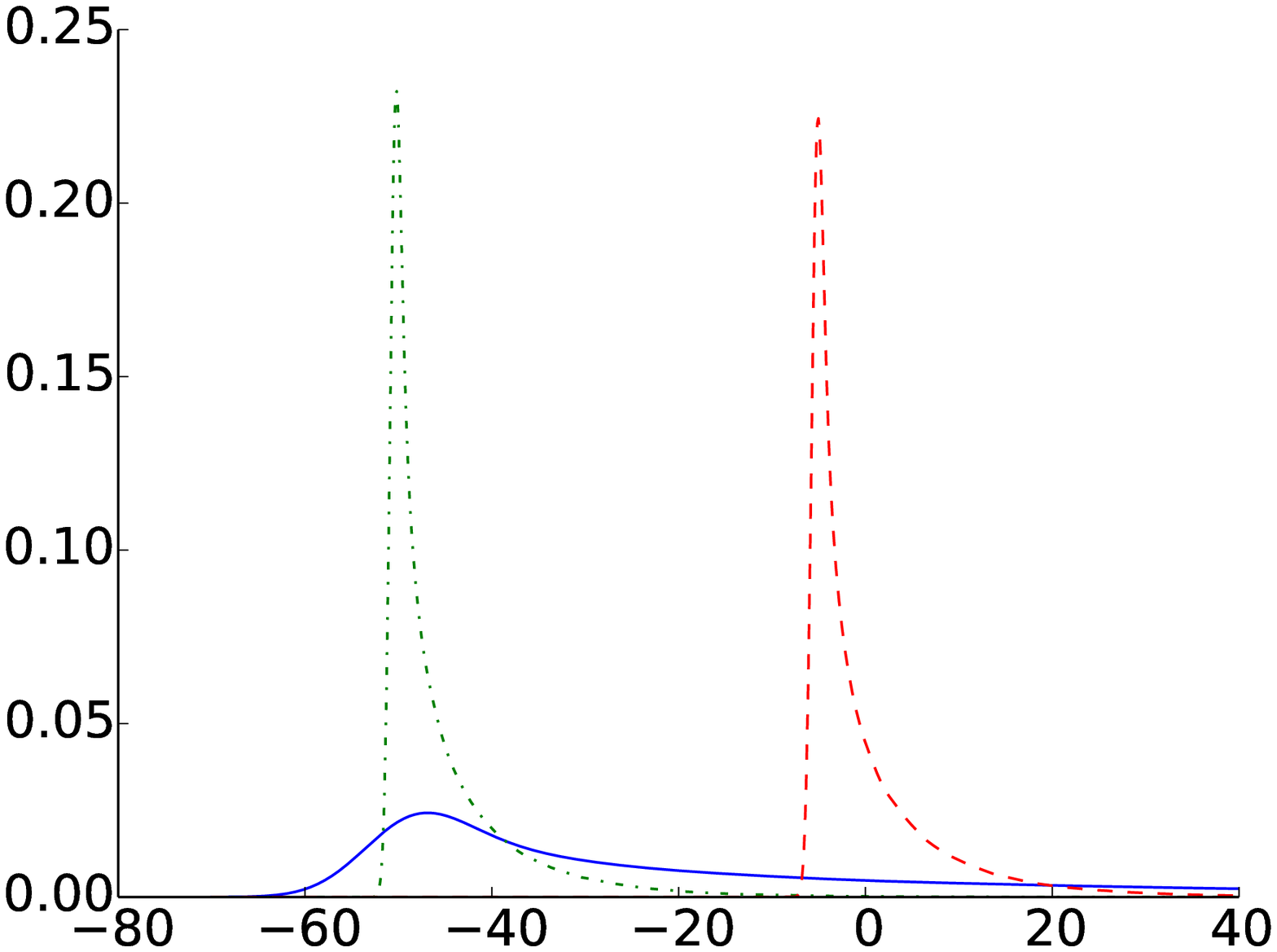}
\includegraphics[width = 0.32\textwidth]{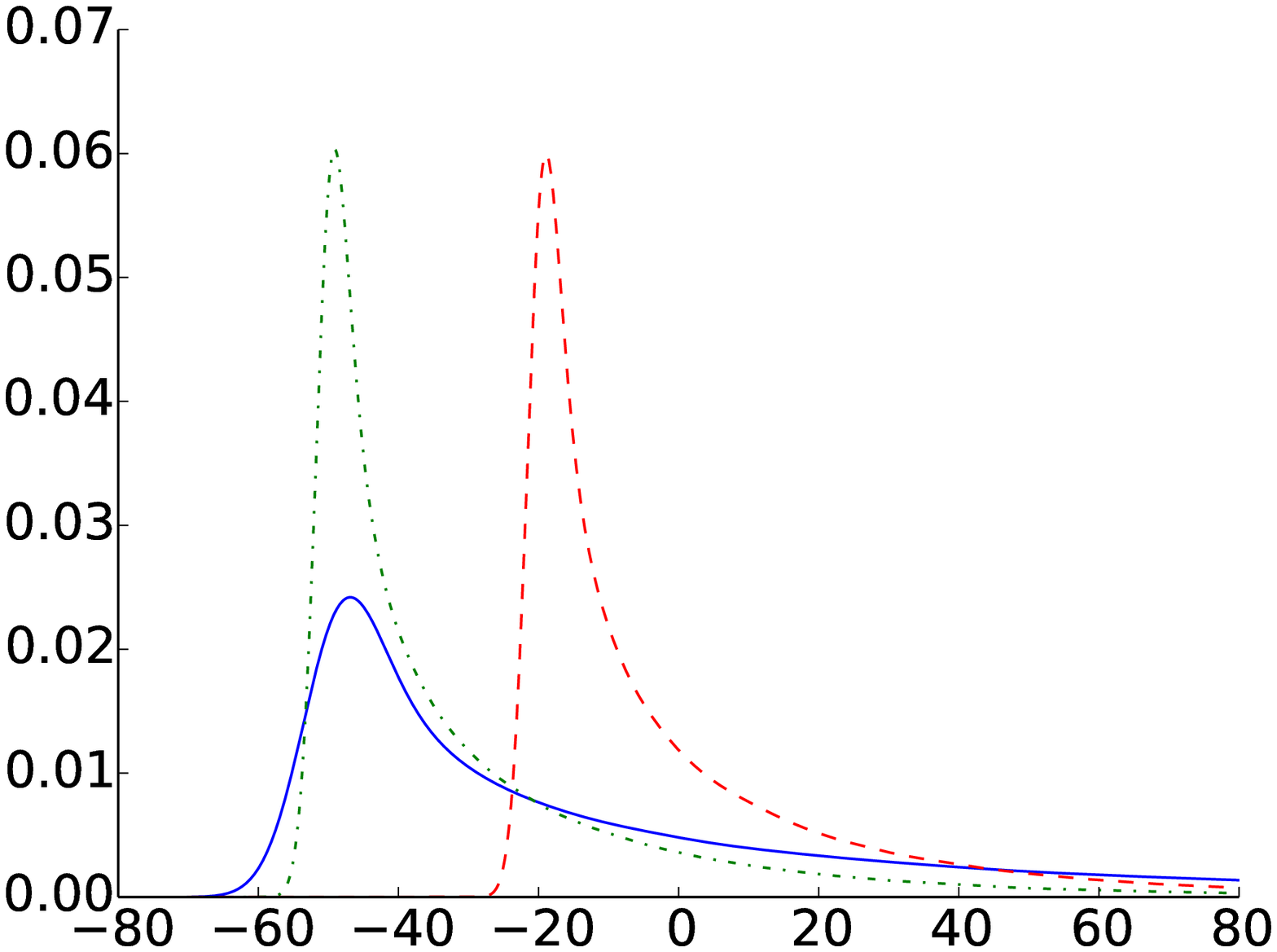}
\includegraphics[width = 0.32\textwidth]{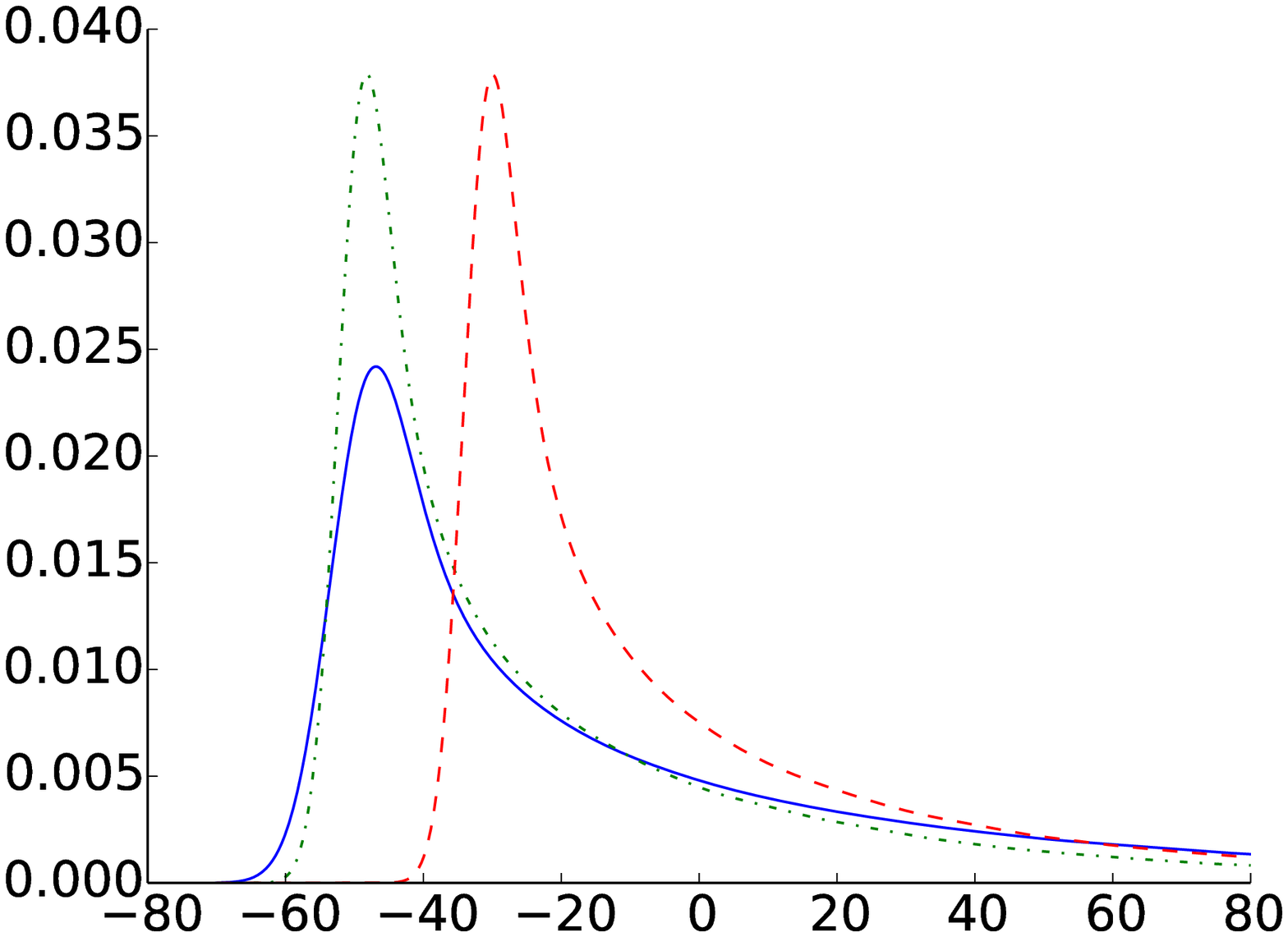}
\caption{Top: Adaptation $u(x,\eta)$ at $x = 0.3$ and its truncations
  using $d = 10$. Middle: Adaptation at $x = 0.9$ and its truncations
  using $d = 10$ (left) and $d = 50$ (middle). Bottom: Adaptation at
  $x = 0.99$ and its truncations using $d = 10$ (left), $d = 50$
  (middle) and $d = 100$ (right). \label{fig:geometric_pdfs}}
\end{figure}

Our goal is to compare the above analytical 1-dimensional
adaptation with two truncated versions. First, the summations in the initial
representation (Eq. (\ref{eq:initial_chaos})) are truncated at $d$ terms 
\begin{equation}
u_{d}(x, \xi) =  \sum_{n = 1}^{d} b_n(x) \xi_n +
\left(\sum_{n=1}^{d} b_n(x) \xi_n\right)^2, 
\end{equation}
which after adaptation gives 
\begin{equation}
\hat{u}_d(x, \eta_d ) =\hat{u}_1\eta_d + \hat{u}_2(x)\frac{\left(\eta_d^2 - 1\right)}{\sqrt{2}},
\end{equation}
where 
\begin{eqnarray}
\begin{array}{l}
\hat{u}_1(x) =  \left(\frac{1-x^d}{1-x}\right)^{1/2} \\
\hat{u}_2(x) = \frac{1 - x^{2d}}{(1-x^d)(1+x)} +
  \frac{\sqrt{2}}{1-x^d}\left(\frac{x(1-x^{2d})}{1-x^2} -
    \frac{x^d(1-x^d)}{1-x}\right)
\end{array}
\end{eqnarray}
and
\begin{eqnarray}
\eta_d = \frac{1}{\left(\sum_{n=1}^{d}b_n(x)^2\right)^{1/2}}\sum_{n=1}^{d} b_n(x) \xi_n.
\end{eqnarray}
Note here that $\hat{u}_i(x) \to u_i(x)$, $i = 1,2$ as $d \to \infty$. 
Second, the adapted expansion (\ref{eq:true_adapted}) is replaced
by one where only the input $\eta$ is truncated, depending only on $d$
terms, that is 
\begin{equation}
u(x, \hat{\eta})  = u_1(x) \hat{\eta} + u_2(x)\frac{(\hat{\eta}^2 - 1)}{\sqrt{2}},
\end{equation}
with
\begin{equation}
\hat{\eta} = \frac{1}{\left(\sum_{n=1}^{\infty}b_n(x)^2\right)^{1/2}}\sum_{n=1}^{d} b_n(x) \xi_n.
\end{equation}
Note that in the first truncation, although the dimensionality is
initially reduced to
$d$ terms, the adaptation procedure enforces $\eta$ to be standard
normally distributed by construction while in the second truncation
the truncated $\hat{\eta}$ is no longer standard normal (in fact it is
$\calN(0, 1-x^d)$) but it shares the same coefficients
with (\ref{eq:true_adapted}). 

The pdfs of the three expansions are shown in
Fig. \ref{fig:geometric_pdfs} for various choices of $x$ and the
truncation order $d$. Although for small choices of $x$ the terms $b_n(x)$
decay fast and both approximations behave well, as $x$ approaches $1$,
the discrepancy of $\hat{u}_d(x,\eta_d)$ from $u(x,\eta)$ increases
dramatically while $u(x,\hat{\eta})$ remains sufficiently close, thus
making a better approximation. This illustrates the fact
that a finite order truncation of the polynomial chaos expansion prior
to any adaptation can behave poorly compared to a
truncation that takes place after adapting the expansion. Note also
that in this example the coefficients decrease geometrically and
therefore their influence in the probability density of the QoI
$u(x,\eta)$ vanishes rapidly. The consequences of such truncations can
be even more severe in a case where all coefficients are of
significant importance.

%% file: conclusions.tex
\section{Conclusions}

We have presented a new formulation of random processes and random
fields using as starting point a homogeneous chaos expansion which
allows merging the dimensions of the
initial functional without deformation of its probability density
structure. The tranformed input variables can be seen as an input
random field with richer information about the quantity of interest
than the simple standard Gaussian inputs, that we think of as an
intermediate scale between the input and the chaos expansion. This novel
represention has significant potential as a dimensionality reduction
technique and can allow the exploration of higher dimensional
polynomial chaos expansions that appear in physical systems, an area
that undoubtedly has suffered a lot by the curse of dimensionality. 

%% file: appendix_A.tex
\section{Computation of the coefficients $q^{\bA}_{\bbeta}(\bx)$}
\label{sec:coeff_formulas}
\subsection{Derivation of the general formula}

Our goal is to derive an explicit expression for the coefficients
$u^{\bA}_{\bbeta}$ defined as 
\begin{equation}
u^{\bA}_{\bbeta} = \sum_{\balpha \in \calJ_p}u_{\balpha}
\lang\psi_{\balpha}(\bxi), \psi_{\bbeta}(\bA \bxi) \rang, \ \ \ \bbeta
\in \calJ_p.
\end{equation}
In order to prove our main result, we introduce some necessary tools
that will allow us to proceed. Let $\pi_n : L^2(\Omega) \to \calG^{\wick n
}$ be the orthogonal projection of $L^2(\Omega)$ onto $\calG^{\wick
  n}$. The \emph{Wick product} for Gaussian variables $\xi_i, ...,
\xi_n$ denoted with $\wick$, is 
\begin{equation}
\xi_1 \wick \cdots \wick \xi_n = \pi_{n}(\xi_1 \cdots\xi_n)
\end{equation} 
that is the projection of the ordinary product $\xi_1 \cdots \xi_n$ onto
$\calG^{\wick n}$. For the case where $\xi_1 = .... = \xi_n$ we write
$\xi^{\wick n} = \xi_1 \wick \cdots \wick \xi_n$. It is easy to see
\cite{janson}, for instance, that for $\xi \sim \calN(0,1)$, we have $\xi^{\wick n} =
h_n(\xi)$ and that for any $\{\xi_i\}_{i=1}^d$ orthonormal basis in
$\calG$, $\balpha \in \calJ$,
\begin{equation}
\xi_1^{\wick \alpha_1} \wick \cdots\wick \xi_d^{\wick \alpha_d} =
\prod_{i=1}^d h_{\alpha_i}(\xi_i) =  h_{\balpha}(\bxi).
\end{equation}
A \emph{Feynman diagram} $\gamma$ of order $n$ and rank $r$ is a graph
consisting of $n$ vertices and $r$ edges such that no two edges
share a common vertex. That means that there are always $2r$ paired
vertices and $n-2r$ unpaired ones. The diagram is called
\emph{complete} when $r = n/2$. A graph where each vertex is labelled
with a Gaussian random variable $\xi_i$, $i = 1,...,n$ is said to have
\emph{value}
\begin{equation}
v(\gamma) = \prod_{k=1}^r \lang\xi_{i_k}, \xi_{j_k} \rang \prod_{i\in C} \xi_i
\end{equation}
where $(\xi_{i_k}, \xi_{j_k})$, $k = 1,..., r$ are the pairs of
vertices and $C$ is the set of unpaired ones. Clearly, when $\gamma$
is complete, $C$ is empty and $v(\gamma)$ is a constant. Given the above definitions,
we can present the following (\cite{janson}, Th. $3.12$):

\textbf{Proposition 1.} \emph{Let $\{\zeta_{ij}\}_{1\leq i\leq k,1\leq j \leq l_i}$ be
  real jointly Gaussian random variables and define $Y_i =
  \zeta_{i_1}\wick \cdots \wick \zeta_{i_{l_i}}$, then 
\begin{equation}
\E\left[Y_1\cdots Y_k\right] = \sum_{\gamma} v(\gamma)
\end{equation} 
where the sum is taken over all complete Feynman diagrams such that no
edge joins any $\zeta_{i_1 j_1}$, $\zeta_{i_2 j_2}$ with $i_1 = i_2$.
}

This is also known as \emph{Wick's theorem} \cite{wick}. Taking this
into account, our main result follows:

\textbf{Proposition 2.} \emph{Let $\{\xi_i\}_{i=1}^d$ be an orthonormal
  basis in $\calG$, $A: \R^d \to \R^d $ be an isometry and
 take any $\balpha, \bbeta \in \calJ$. Let also $\{\eta_i\}_{i=1}^d$
 be such that $\bfeta = A\bxi$. Then
\begin{eqnarray}
\lang h_{\balpha}(\bxi), h_{\bbeta}(\bfeta) \rang =
\left\{\begin{array}{c}\sum_{\frak{A}_n} \prod_{k = 1}^{n}
    a_{i_k,j_k}, \ \ |\balpha| = |\bbeta| \\ 0, \ \ \ \ \ \ \ \ \ \
    \ \ \ \ \ \ \ \ \ \ |\balpha|
    \neq |\bbeta| \end{array} \right.
\end{eqnarray}
where $a_{i_k,j_k}$ are entries of $\bA$ and the sum is taken over
$\frak{A}_n$, which is the number of possible ways to choose $n$
entries of $\bA$ such that exactly $\alpha_i$ of them are in the $i$th
column and  $\beta_i$ of them are $i$th row, simultaneously, for all $i = 1, ..., d$.
}

\textbf{Proof.} Define $\{\zeta_{ij}\}_{1\leq i\leq k,1\leq j \leq l_i}$ with $k = 2$,
$l_1 = |\balpha| := n$, $l_2 = |\bbeta| := m$ where 
\begin{equation}
\{\zeta_{1j}\}_{j=1}^{n} := \left\{ \underbrace{\xi_1, ...,
    \xi_1}_{\alpha_1}, ...,
  \underbrace{\xi_d,...,\xi_d}_{\alpha_d}\right\},
\end{equation}
\begin{equation}
\{\zeta_{2j}\}_{j=1}^{m} := \left\{ \underbrace{\eta_1,...,
    \eta_1}_{\bbeta_1}, ..., \underbrace{\eta_d, ...,
    \eta_d}_{\bbeta_d}\right\}.
\end{equation}
Then for $Y_1 := \xi_1^{\wick \alpha_1} \wick \cdots\wick \xi_d^{\wick
  \alpha_d} = h_{\balpha}(\bxi)$ and $Y_2 := \eta_1^{\wick \alpha_1} \wick \cdots\wick
\eta_d^{\wick \alpha_d} = h_{\bbeta}(\bfeta) $, Prop. 1 gives that 
\begin{equation}
\lang h_{\balpha}(\bxi), h_{\bbeta}(\bfeta) \rang = \E\left[Y_1
  Y_2\right] = \sum_{\gamma} v(\gamma)
\end{equation}
where the sum is taken over all complete Feynman diagrams with edges
that connect $\{\zeta_{1j}\}_{j=1}^{n}$ with
$\{\zeta_{2j}\}_{j=1}^{m}$. Clearly for $n \neq m$ there is no such
complete Feynman diagram and the sum is zero. For $n = m$, any such
diagram $\gamma$ can be represented by its pairs $\left\{(\zeta_{1j},
  \zeta_{2l_j})\right\}_{j=1}^n$ and has value 
\begin{equation}
v(\gamma) =  \prod_{j = 1}^n \lang \zeta_{1j}, \zeta_{2l_j} \rang =
\prod_{j=1}^{\alpha_1} \lang \xi_1, \zeta_{2l_j}\rang \prod_{j=\alpha_1+1}^{\alpha_1+\alpha_2} \lang \xi_2,
\zeta_{2l_j}\rang \cdots \prod_{j
  =n - \alpha_d}^{n} \lang \xi_d, \zeta_{2l_j} \rang.
\end{equation}
Observe that any $\zeta_{2l_j} \in \{\eta_s\}_{s=1}^d$ and that for
any $\eta_s$
\begin{equation}
\lang \xi_j, \eta_s \rang = \lang \xi_j,  \sum_{r=1}^d
a_{sr}\xi_r\rang = a_{s,j}
\end{equation}
that follows from $\bfeta = \bA\bxi$ and $a_{s,j}$ is the $(s,j)$th
entry of $\bA$. Therefore, substituting in the above equation and noting
that exactly $\alpha_i$ of the products $\lang\cdot, \cdot\rang$
include $\xi_i$ and exactly $\beta_i$ include $\eta_i$ we
obtain that exactly $\alpha_i$ and $\beta_i$ entries of $\bA$ will be taken from the
$i$th column and $i$th row respectively, which completes the
proof. \hspace{8cm} $\square$

An immediate consequence of the above proposition, when one wants to compute
the coefficients of a chaos expansion with respect to a rotated basis
$\bfeta = \bA\bxi$, is that the sum is reduced to
\begin{equation}
u^{\bA}_{\bbeta} = \sum_{\balpha \in \calJ_p}u_{\balpha}
\lang\psi_{\balpha}(\bxi), \psi_{\bbeta}(\bA \bxi) \rang =
\sum_{\balpha, |\balpha| = |\bbeta|} u_{\balpha} \lang\psi_{\balpha}(\bxi), \psi_{\bbeta}(\bA \bxi) \rang .
\end{equation}
The above formula can be further simplified for the case of
$1$-dimensional polynomials:

\textbf{Corollary 1.} \emph{For any $n \in\N$, $\balpha \in \calJ$
  with $|\balpha| = n$ and $i = 1,...,d$, we have
\begin{equation}
\lang h_{\balpha}(\bxi), h_{n}(\eta_i) \rang = n!\prod_{k = 1}^d a_{i,k}^{\alpha_k}
\end{equation}
}

\textbf{Proof.} Let $\bbeta \in \calJ$ with $\bbeta =
  n\bepsilon_i = (0,...,n,...,0)$ and by working as in the proof of Prop. 2
  with
\begin{equation}
\{\zeta_{2j}\}_{j=1}^n = \left\{\underbrace{\eta_i, ..., \eta_i}_{n}\right\},
\end{equation}
it is easy to see that all complete Feynman diagrams take the
same value, that is 
\begin{equation}
v(\gamma) = \prod_{j=1}^{\alpha_1} \lang\xi_1, \eta_i \rang \cdots
\prod_{j=1}^{\alpha_d} \lang\xi_d, \eta_i \rang = a_{i,1}^{\alpha_i}
\cdots a_{i,d}^{\alpha_d} = \prod_{k = 1}^d a_{i,k}^{\alpha_k}
\end{equation}
and the total number of such diagrams is $n!$. \hspace{3.6cm} $\square$

\subsection{Coefficients for $1$-dimensional subspaces}

Taking into account Corollary 1 from the previous paragraph, we are now
ready to derive explicit formulas for the coefficients along
$1$-dimensional subspaces of chaos expansion
$u^{\bA}(\bfeta)$. Namely, for
any $\bbeta \in \calJ$ with $\bbeta = n\bepsilon_i = (0,...,n,...,0)$,
$i = 1,..., d$ and $n \in \N$, we have 
\begin{eqnarray}
%\begin{array}{r}
u^{\bA}_{\bbeta} &  = & \sum_{\balpha, |\balpha| = n} u_{\balpha} \lang 
\psi_{\balpha}(\bxi) , \psi_{n}(\eta_i) \rang = \\ & = & \sum_{\balpha,
  |\balpha| = n} \frac{u_{\balpha}}{\sqrt{\balpha !}\sqrt{n!}} \lang
h_{\balpha}(\bxi), h_n(\eta_i) \rang = \\ & = & \sqrt{n!} \sum_{\balpha,
|\balpha| = n} \frac{u_{\balpha}}{\sqrt{\balpha!}} \prod_{k=1}^d
a_{i,k}^{\alpha_k}.
%\end{array}
\end{eqnarray}
The coefficients of $\psi_{\bbeta}(\bfeta)$ of order up to $3$ are
given by:
\begin{eqnarray}
u_0^{\bA} & = & u_0 \\
\label{eq:first_order}
u_{\epsilon_i}^{\bA} & = & \sum_{k = 1}^d a_{i,k}u_{\epsilon_k} \\
\label{eq:second_order}
u_{2\epsilon_i}^{\bA} & = & \sum_{k=1}^d u_{kk}a_{i,k}^2 + \sqrt{2}\sum_{\substack{k =
  1\\ j > k}}^d u_{kj} a_{i,k} a_{i,j} \\
u_{3\epsilon_i}^{\bA} & = & \sum_{k = 1}^d u_{kkk} a_{i,k}^3 +
\sqrt{3}\sum_{\substack{k = 1 \\j > k}}^d u_{kkj} a_{i,k}^2 a_{i,j} +
\\ & +  &\sqrt{6}
\sum_{\substack{k = 1\\ j >k\\ l > j}}^d u_{kjl}a_{i,k}a_{i,j}a_{i,l}
\end{eqnarray}

%For notational simplicity we write $q_n^{\bA}(\bx)$ instead of
%$q^{\bA}_{n\epsilon_1}(\bx)$. Clearly we have
%\begin{eqnarray*}
%q^{\bA}_0(x) & = & \sum_{\alpha} q_{\alpha}(\bx) \lang
%\psi_{\alpha}(\bxi), 1 \rang  = q_0(\bx),\\
%q^{\bA}_1(\bx) & = &  \sum_{\alpha} q_{\alpha}(\bx) \lang
%\psi_{\alpha}(\bxi), \eta_1 \rang = \sum_{\alpha} q_{\alpha}(\bx) \lang
%\psi_{\alpha}(\bxi), \sum_{i=1}^da_i(\bx)\xi_i \rang \\ & = &
%\sum_{i=1}^d a_i(\bx)\sum_{\alpha} q_{\alpha}(\bx) \lang
%\psi_{\alpha}(\bxi), \xi_i \rang  = \sum_{i=1}^d a_i(\bx)
%q_{\epsilon_i}(\bx) \lang \xi_i^2 \rang = \left(\sum_{i=1}^d q_{i\epsilon_i}^2(\bx)\right)^{1/2}
%\end{eqnarray*}
%For $n \geq 2$, using the recurrence relation for
%Hermite polynomials 
%\begin{equation*}
%h_{n}(\eta) = \eta h_{n-1} - (n-1) h_{n-2}(\eta), \ \ \ n
%\geq 2,
%\end{equation*}
%we have
%\begin{eqnarray*}
%q^{\bA}_{n}(\bx)  & = & \sum_{\alpha} q_{\alpha}(\bx) \lang
%\psi_{\alpha}(\bxi), \psi_{n}(\eta_1) \rang = \frac{1}{\sqrt{n!}}\sum_{\alpha} q_{\alpha}(\bx) \lang
%\psi_{\alpha}(\bxi), h_{n}(\eta_1) \rang \\ & = & \frac{1}{\sqrt{n!}}\sum_{\alpha}
%q_{\alpha}(\bx) \lang \psi_{\alpha}(\bxi) ,
%\eta_1h_{n-1}(\eta_1) \rang - \frac{(n-1)}{\sqrt{n!}} \sum_{\alpha}
%q_{\alpha}(\bx) \lang \psi_{\alpha}(\bxi) , h_{n-2}(\eta_1) \rang
%\\ & = & \frac{1}{\sqrt{n!}} \sum_{\alpha} q_{\alpha}(\bx) \lang \psi_{\alpha}(\bxi) ,
%\eta_1h_{n-1}(\eta_1) \rang - \sqrt{\frac{n-1}{n}} q_{n-2}^{\bA}(\bx).
%\end{eqnarray*}

%% file: appendix_B.tex
\section{Proof of Theorem 1}
\label{sec:theorem_1}

We have
\begin{eqnarray*}
\int_D\int_D |k_i(\bx, \by)|^2 d\bx d\by = \\ =  \sum_{j,k = 1}^d
\int_Da_{ij}(\bx)a_{ik}(\bx) d\bx \int_D a_{ij}(\by) a_{ik}(\by)d\by \leq
\\  \leq  \sum_{j,k = 1}^d ||a_{ij}||^2_{L^2(D)}
||a_{ik}||^2_{L^2(D)} < +\infty
\end{eqnarray*}
where the second row is derived after applying the Cauchy-Schwarz
inequality.

\section{Proof of Lemma 1}
\label{sec:lemma_1}

Clearly by definition $\eta(\bx) \in \calG$ for all
$\bx \in D$ since $\{\xi_i\}_{i\in\N}$ forms a basis in $\calG$,
therefore $span\{\eta(\bx)\}_{\bx\in D} \subset \calG$. On the oher
hand, for any $\xi \in \calG$ there exists $\chi \in H$ such that 
\begin{equation}
\xi = \lang \chi, \bxi \rang_H = \sum_{i} \chi_i \xi_i
\end{equation}
where $\chi_n = \langle \chi, \be_n\rangle_H$ with $\{\be_n\}_{n\in \N}$ some
basis in $H$. Set $\bx_1 \in \calA^{-1}\left(\{\be_n\}_{n\in\N}\right)$ and for $n
\geq 2$ choose $\bx_n$ such that $\be_n = \hat{\be}_1(\bx_n)$ where
$\hat{\be}_1(\bx_n)$ is the first basis element of $\calA(\bx_n)$. This is
possible since we can continuously rotate any basis until its $n$-th
element becomes the first element of another basis. Then 
\begin{eqnarray*}
\xi & = & \lang \chi, \bxi \rang_H = \sum_i \lang \chi_i, \be_i
\rang_H \lang \be_i, \bxi \rang_H \\ & = & \sum_i \chi_i \lang
\be(\bx_i), \bx \rang_H = \sum_i \chi_i \eta(\bx_i),
\end{eqnarray*}
therefore $\xi \in span\{\eta(\bx)\}_{\bx\in D}$ and $\calG \subset
span\{\eta(\bx)\}_{\bx\in D}$ which completes the proof.

\section{Proof of Theorem 2}
\label{sec:theorem_2}

It is known (\cite{janson}, Theorem $8.15$) that the
linear mapping $R_{\eta}(\cdot)(\bx)$ is an isometry from
$span\{\eta(\bx)\}_{\bx\in D}$ to $\calR_{\eta}(\calG)$ and by using
Lemma 1 we have that $\calG$ and $\calR_{\eta}(\calG)$ have the same
dimension. Also
(\cite{janson}, Corollary $8.16$) $\calR_{\eta}(\calG)$ is spanned by
the covariance kernels 
\begin{equation}
k_{\by}(\bx) = \calR_{\eta}(\eta(\by))(\bx) = \E[\eta(\by) \eta(\bx)], \ \ \
\by \in D
\end{equation}
and (\cite{janson}, Theorem $8.22$) $\eta(\bx)$ admits a representation 
\begin{equation}
\eta(\bx) = \sum_{i = 1}^{\infty} \rho_i(\bx)\xi_i
\end{equation}
where $\{\rho_i\}_{i\in \N}$ is a basis in $\calR_{\eta}(\calG)$ and
$\{\xi_i\}$ a basis in $span\{\eta(\bx)\}_{\bx\in D} = \calG$ and the limit is taken in $L^2$. Let
$\{\by_i\}_{i\in \N}$ such that $\rho^i(\bx) = k_{\by_i}(\bx)$. From
the proof of Lemma 1 we can see that it is possible to choose
$\by_i$ such that $\eta(\by_i) = \xi_i$, all $i\in \N$. That is due to
the fact that the isometry $R_{\eta}(\cdot)(\bx)$ will map the basis $\{\eta(\by_i)\}_{i\in\N}$ to
a basis $\{k_{\by_i}(\bx)\}_{i\in\N}$ in $\calR_{\eta}(\calG)$. Then 
\begin{eqnarray*}
\eta(\bx) & = & \sum_i \rho_i(\bx) \xi_i = \sum_i k_{\by_i}(\bx) \xi_i
 =  \sum_i \E[\xi_i \eta(\bx)] \xi_i \\ & = & \sum_i \lang \xi_i, \eta(\bx)
\rang_\calG \xi_i = \sum_i \sum_j e^j(\bx) \lang \xi_i, \xi_j
\rang_\calG \xi_i \\ & = & \sum_i e^i(\bx) \xi_i,
\end{eqnarray*}
from where we obtain $\rho^i(\bx) = e^i(\bx)$.

%\section{Projection error for $1$-dimensional adaptations}
%\label{sec:error}

%\subsection{Structure of $\bC$}

%Using Proposition $2$ from
%\ref{sec:coeff_formulas} we observe that the entries $\bC_{\balpha,
 % \bbeta}$ with $|\balpha| \neq |\bbeta|$ are zero resulting a
%$\bC$ being a block diagonal matrix. 
%\begin{equation*}
%\bC = \left[\begin{array}{cccc} \bC_0 & & \cdots & \mathbb{O} \\ &
 %   \bC_1 & &  \\ \vdots & & \ddots & \vdots
  %  \\ \mathbb{O} & & \cdots & \bC_p\end{array} \right].
%\end{equation*}
%and particularly it is easy to verify that $\bC_0 = 0$ and in the case
%$\calI = \calJ_p$ we have $\bC_1 = \bA^T$.
%Then taking into account that the columns of $\bC$ corresponding to $\bbeta \in
%\calJ_p \setminus \calI$ are zero, we observe that additional columns in
%$\bC_i$, $i = 0, ...,p$ are zero as well. 

%% file: notes.bbl
\begin{thebibliography}{10}

\bibitem{aarnes}
J.~Aarnes, T.~Gimse, and K.-A. Lie.
\newblock An introduction to the numerics of flow in porous media using matlab.
\newblock {\em Geometric Modelling, Numerical Simulation and Optimization},
  pages 265--306, 2007.

\bibitem{arnst_ghanem}
M.~Arnst, R.~Ghanem, and C.~Soize.
\newblock Identification of bayesian posteriors for coefficients of chaos
  expansions.
\newblock {\em Journal of Computational Physics}, 229:3134--3154, 2010.

\bibitem{cameron}
R.~Cameron and W.~Martin.
\newblock The orthogonal development of nonlinear functionals in series of
  fourier-hermite functionals.
\newblock {\em Annals of Mathematics}, 48:385--392, 1947.

\bibitem{constantine}
P.~G. Constantine, E.~Dow, and Q.~Wang.
\newblock Active subspace methods in theory and practice: applications to
  kriging surfaces.
\newblock {\em SIAM Journal on Scientific Computing}, 36:A1500--A1524, 2014.

\bibitem{das_etal}
S.~Das, R.~Ghanem, and J.C. Spall.
\newblock Asymptotic sampling distribution for polynomial chaos representation
  from data: a maximum entropy and fisher information approach.
\newblock {\em SIAM Journal on Scientific Computing}, 30:2207--2234, 2008.

\bibitem{desceliers_etal}
C.~Desceliers, R.~Ghanem, and C.~Soize.
\newblock Maximum likelihood estimation of stochastic chaos representations
  from experimental data.
\newblock {\em International Journal for Numerical Methods in Engineering},
  66:978--1001, 2006.

\bibitem{doostan_icc}
A.~Doostan and G.~Iaccarino.
\newblock A least-squares approximation of partial differential equations with
  high-dimensional random inputs.
\newblock {\em Journal of Computational Physics}, 228:4332--4345, 2009.

\bibitem{doostan_owhadi}
A.~Doostan and H.~Owhadi.
\newblock A non-adapted sparse approximation of pdes with stochastic inputs.
\newblock {\em Journal of Computational Physics}, 230:3015--3034, 2011.

\bibitem{gelfand}
I.M. Gelfand and N.~Ya. Vilenkin.
\newblock {\em Generalized functions, Vol 4: Applications to Harmonic
  Analysis}.
\newblock Academic Press, 1964.

\bibitem{ghanem_wrr}
R.~Ghanem.
\newblock Scales of fluctuation and the propagation of uncertainty in random
  porous media.
\newblock {\em Water Resources Research}, 34:2123--2136, 1998.

\bibitem{ghanem}
R.~Ghanem.
\newblock Ingredients for a general purpose stochastic finite elements
  implementation.
\newblock {\em Computer Methods in Applied Mechanics and Engineering},
  168:19--34, 1999.

\bibitem{ghanem_dham}
R.~Ghanem and S.~Dham.
\newblock Stochastic finite element analysis for multiphase flow in
  heterogeneous porous media.
\newblock {\em Transport in Porous Media}, 32:239--262, 1998.

\bibitem{ghanem_doostan}
R.~Ghanem and R.~Doostan.
\newblock Characterization of stochastic system parameters from experimental
  data: A bayesian inference approach.
\newblock {\em Journal of Computational Physics}, 217:63--81, 2006.

\bibitem{ghanem_redhorse}
R.~Ghanem and J.~Red-Horse.
\newblock Propagation of probabilistic uncertainty in complex physical systems
  using a stochastic finite element approach.
\newblock {\em Physica D: Nonlinear Phenomena}, 133:137--144, 1999.

\bibitem{ghanem_spanos}
R.~Ghanem and P.~Spanos.
\newblock {\em Stochastic finite elements: A spectral approach}.
\newblock Springer-Verlag, 1991.

\bibitem{ghanem_doostan_redhorse}
R.G. Ghanem, A.~Doostan, and J.~Red-Horse.
\newblock A probabilistic construction of model validation.
\newblock {\em Computer Methods in Applied Mechanics and Engineering},
  197:2585--2595, 2008.

\bibitem{huan}
X.~Huan and Y.M. Marzouk.
\newblock Simulation-based optimal bayesian experimental design for nonlinear
  systems.
\newblock {\em Journal of Computational Physics}, 232:288--317, 2013.

\bibitem{ito}
K.~It\^{o}.
\newblock An elementary approach to malliavin fields.
\newblock In {\em Asymptotic problems in probability theory: Wiener functionals
  and asymptotics}, pages 35--89, Essex, 1990. Sanda and Kyoto.

\bibitem{janson}
S.~Janson.
\newblock {\em Gaussian Hilbert spaces}.
\newblock Cambridge University Press, 1999.

\bibitem{karhunen}
K.~Karhunen.
\newblock \"{U}ber lineare methoden in der wahrscheinlichkeits-rechnung.
\newblock {\em Annals of Academic Science Fennicade Series A1, Mathematical
  Physics}, 37:3--79, 1946.

\bibitem{le_maitre_etal}
O.P. Le~Ma\^{i}tre, M.T. Reagan, H.N. Najm, R.G. Ghanem, and O.M. Knio.
\newblock A stochastic projection method for fluid flow: Ii. random process.
\newblock {\em Journal of Computational Physics}, 181:9--44, 2002.

\bibitem{loeve}
M.~Lo\'{e}ve.
\newblock {\em Probability Theory, D}.
\newblock Van Nostrand, Princeton, New Jersey, 1955.

\bibitem{marzouk_etal}
Y.~M. Marzouk, H.~N. Najm, and L.~Rahn.
\newblock Stochastic spectral methods for efficient bayesian solution of
  inverse problems.
\newblock {\em Journal of Computational Physics}, 224:560--586, 2007.

\bibitem{marzouk_najm}
Y.M. Marzouk and H.N. Najm.
\newblock Dimensionality reduction and polynomial chaos acceleration of
  bayesian inference in inverse problems.
\newblock {\em Journal of Computational Physics}, 228:1862--1902, 2009.

\bibitem{matthies_bucher}
H.G. Matthies and C.~Bucher.
\newblock Finite elements for stochastic media problems.
\newblock {\em Computer Methods in Applied Mechanics and Engineering},
  168:3--17, 1999.

\bibitem{mercer}
J.~Mercer.
\newblock Functions of positive and negative type, and their connection with
  the theory of integral equations.
\newblock {\em Philosophical Transactions of the Royal Society of London.
  Series A, containing papers of a mathematical or physical character},
  209:415--446, 1909.

\bibitem{najm}
H.N. Najm.
\newblock Uncertainty quantification and polynomial chaos techniques in
  computational fluid dynamics.
\newblock {\em Annual Review of Fluid Mechanics}, 41:35--52, 2009.

\bibitem{redhorse_2009}
J.~Red-Horse and R.~Ghanem.
\newblock Elements of a functional analytic approach to probability.
\newblock {\em International Journal for Numerical Methods in Engineering},
  80(6-7):689--716, 2009.

\bibitem{saad_ghanem}
G.~Saad and R.~Ghanem.
\newblock Characterization of reservoir simulation models using a polynomial
  chaos-based ensemble kalman filter.
\newblock {\em Water Resources Research}, 45:Art. W04417, 2009.

\bibitem{soize_physical}
C.~Soize and R.~Ghanem.
\newblock Physical systems with random uncertainties: chaos representations
  with arbitrary probability measure.
\newblock {\em SIAM Journal on Scientific Computing}, 26:395--410, 2004.

\bibitem{soize_reduced}
C.~Soize and R.~Ghanem.
\newblock Reduced chaos decomposition with random coefficients of vector-valued
  random variables and random fields.
\newblock {\em Computer Methods in Applied Mechanics and Engineering},
  198:1926--1934, 2009.

\bibitem{tipireddy}
R.~Tipireddy and R.G. Ghanem.
\newblock Basis adaptation in homogeneous chaos spaces.
\newblock {\em Journal of Computational Physics}, 259:304--317, 2014.

\bibitem{tsilifis}
P.~Tsilifis, R.G. Ghanem, and P.~Hajali.
\newblock Efficient bayesian experimentation using an expected information gain
  lower bound.
\newblock {\em arXiv pre-print, arXiv:1506.00053v2}, 2015.

\bibitem{wick}
G.C. Wick.
\newblock The evaluation of the collision matrix.
\newblock {\em Physical Review}, 80:268--272, 1950.

\bibitem{wiener}
N.~Wiener.
\newblock The homogeneous chaos.
\newblock {\em American Journal of Mathematics}, 60:897--936, 1938.

\bibitem{xiu_karniadakis}
D.~Xiu and G.E. Karniadakis.
\newblock The wiener--askey polynomial chaos for stochastic differential
  equations.
\newblock {\em SIAM Journal on Scientific Computing}, 24:619--644, 2002.

\bibitem{xiu_fluid}
D.~Xiu and G.E. Karniadakis.
\newblock Modeling uncertainty in flow simulations via generalized polynomial
  chaos.
\newblock {\em Journal of Computational Physics}, 187:137--167, 2003.

\end{thebibliography}
